\documentclass[twocolumn]{aastex631}

\usepackage{graphicx}
\usepackage{amsmath}

\newcommand{\rv}{$R_V$}
\newcommand{\ebv}{$E(B-V)$}
\newcommand{\av}{$A_V$}
\newcommand{\avir}{$\langle A^\mathrm{IR}_V \rangle$}

\newcommand{\izw}{I\,Zw\,18}
\newcommand{\sbs}{SBS\,0335$-$052}
\newcommand{\msunyr}{M$_\odot$\,yr$^{-1}$}
\newcommand{\zsun}{$Z_\odot$}
\newcommand{\zzsun}{$Z/Z_\odot$}
\newcommand{\hi}{H{\sc i}}
\newcommand{\hii}{H{\sc ii}}
\newcommand{\htwo}{H$_2$}

\newcommand{\mstar}{$M_*$}

\newcommand{\msun}{M$_\odot$}

\newcommand{\logoh}{12$+$log(O/H)}

\newcommand{\cmthree}{cm$^{-3}$}
\newcommand{\hst}{{HST}}
\newcommand{\jwst}{JWST}

\newcommand{\hua}{Hu$\alpha$}
\newcommand{\hub}{Hu$\beta$}
\newcommand{\hug}{Hu$\gamma$}
\newcommand{\pfa}{Pf$\alpha$}

\newcommand{\feii}{[Fe\,{\sc ii}]}
\newcommand{\ha}{H$\alpha$}
\newcommand{\hb}{H$\beta$}
\newcommand{\hd}{H$\delta$}
\newcommand{\heii}{He\,{\sc ii}}
\newcommand{\heiii}{He\,{\sc iii}}
\newcommand{\neii}{[Ne\,{\sc ii}]}

\newcommand{\neiii}{[Ne\,{\sc iii}]}

\newcommand{\oiv}{[O\,{\sc iv}]}
\newcommand{\fev}{[Fe\,{\sc v}]}
\newcommand{\nev}{[Ne\,{\sc v}]}

\newcommand{\siv}{[S\,{\sc iv}]}

\newcommand{\arii}{[Ar\,{\sc ii}]}
\newcommand{\ariii}{[Ar\,{\sc iii}]}

\newcommand{\siii}{[S\,{\sc iii}]}
\newcommand{\piii}{[P\,{\sc iii}]}

\newcommand{\spit}{\textit{Spitzer}}

\newcommand{\kms}{km\,s$^{-1}$}

\newcommand{\wmsq}{W\,m$^{-2}$}

\newcommand{\logu}{$\log\,\cal{U}$}

\newcommand{\fagn}{$f_\mathrm{AGN}$}

\begin{document}

\title{The Interstellar Medium in I\,Zw\,18 seen with JWST/MIRI: I. 
Highly Ionized Gas}

\correspondingauthor{Leslie Hunt}
\email{leslie.hunt@inaf.it}

\author[0000-0001-9162-2371]{L.~K. Hunt}
\affiliation{INAF -- Osservatorio Astrofisico di Arcetri, Largo E. Fermi 5, 50125 Firenze, Italy}

\author[0000-0003-4137-882X]{A. Aloisi}
\affiliation{Space Telescope Science Institute, 3700 San Martin Drive, Baltimore, MD 21218, USA}


\author[0000-0002-1860-2304]{M.~G. Navarro}
\affiliation{INAF -- Osservatorio Astronomico di Roma, Via di Frascati 33, 00040 Monteporzio Catone, Italy}

\author[0000-0001-9719-4080]{R.~J. Rickards Vaught}
\affiliation{Space Telescope Science Institute, 3700 San Martin Drive, Baltimore, MD 21218, USA}

\author[0000-0002-0846-936X]{B.~T. Draine}
\affiliation{Dept. of Astrophysical Sciences, Princeton University, Princeton, NJ 08544, USAA}

\author[0000-0002-8192-8091]{A. Adamo}
\affiliation{Department of Astronomy, The Oskar Klein Centre, Stockholm University, AlbaNova, SE-10691 Stockholm, Sweden}

\author[0000-0003-3758-4516]{F. Annibali}
\affiliation{INAF -- Osservatorio di Astrofisica e Scienza dello Spazio, Via Gobetti 93/3, 40129 Bologna, Italy}

\author[0000-0002-5189-8004]{D. Calzetti}
\affiliation{Department of Astronomy, University of Massachusetts Amherst, 710 North Pleasant Street, Amherst, MA 01003, USA}

\author[0000-0003-4857-8699]{S. Hernandez}
\affiliation{AURA for ESA, Space Telescope Science Institute, 3700 San Martin Drive, Baltimore, MD 21218, USA}

\author[0000-0003-4372-2006]{B.~L. James}
\affiliation{AURA for ESA, Space Telescope Science Institute, 3700 San Martin Drive, Baltimore, MD 21218, USA}

\author[0000-0003-2589-762X]{M. Mingozzi}
\affiliation{Space Telescope Science Institute, 3700 San Martin Drive, Baltimore, MD 21218, USA}
\affiliation{AURA for ESA, Space Telescope Science Institute, 3700 San Martin Drive, Baltimore, MD 21218, USA}

\author[0000-0001-9317-2888]{R. Schneider}
\affiliation{Dipartimento di Fisica, 'Sapienza' Universit{\`a} di Roma, Piazzale Aldo Moro 2, I-00185 Roma, Italy}
\affiliation{INAF -- Osservatorio Astronomico di Roma, Via di Frascati 33, I-00040 Monte Porzio Catone, Italy}
\affiliation{INFN, Sezione Roma1, Dipartimento di Fisica, 'Sapienza' Universit{\`a} di Roma, Piazzale Aldo Moro 2, I-00185 Roma, Italy}
\affiliation{Sapienza School for Advanced Studies, Viale Regina Elena 291, I-00161 Roma, Italy}

\author[0000-0002-0986-4759]{M. Tosi}
\affiliation{INAF -- Osservatorio di Astrofisica e Scienza dello Spazio, Via Gobetti 93/3, 40129 Bologna, Italy}

\author[0000-0001-9737-169X]{B. Brandl}
\affiliation{Leiden Observatory, Leiden University, PO Box 9513, 2300 RA Leiden, The Netherlands}
\affiliation{Faculty of Aerospace Engineering, Delft University of Technology, Kluyverweg 1, 2629 HS Delft, The Netherlands}

\author[0000-0002-0191-4897]{M.~G. del Valle-Espinosa}
\affiliation{Space Telescope Science Institute, 3700 San Martin Drive, Baltimore, MD 21218, USA}

\author[0000-0002-6460-3682]{F. Donnan}
\affiliation{Department of Physics, University of Oxford, Keble Road, Oxford, OX1 3RH, UK}

\author[0000-0002-2954-8622]{A.~S. Hirschauer}
\affiliation{Department of Physics \& Engineering Physics, Morgan State University, 1700 East Cold Spring Lane, Baltimore, MD 21251, USA}

\author[0000-0002-0522-3743]{M. Meixner}
\affiliation{Jet Propulsion Laboratory, California Institute of Technology, 4800 Oak Grove Dr., Pasadena, CA 91109, USA}

\author[0000-0001-6854-7545]{D. Rigopoulou}
\affiliation{Department of Physics, University of Oxford, Keble Road, Oxford OX1 3RH, UK}
\affiliation{School of Sciences, European University Cyprus, Diogenes street, Engomi, 1516 Nicosia, Cyprus}

\author[0000-0002-3703-0719]{C.~T. Richardson}
\affiliation{Department of Physics \& Astronomy, Elon University, 100 Campus Drive, Elon, NC 27244, USA}

\author[ 0009-0003-6156-8310]{J.~M. Levanti}
\affiliation{Department of Physics \& Astronomy, Elon University, 100 Campus Drive, Elon, NC 27244, USA}

\author[0000-0001-8525-4920]{A.~R. Basu-Zych}
\affiliation{NASA Goddard Flight Center, Code 662, Greenbelt, MD 20771, USA}




\begin{abstract}
We present \jwst/MIRI spectra from the Medium-Resolution Spectrometer of \izw,
a nearby dwarf galaxy with a metallicity of $\sim$3\% Solar.
Its proximity enables a detailed study of 
highly ionized gas that can be interpreted in the context of newly discovered
high-redshift dwarf galaxies.
We derive aperture spectra centered on eleven regions of interest;
the spectra show very low extinction, \av\,$\la\,0.1$, consistent with optical determinations.
The gas is highly ionized; we have detected 10 fine-structure lines, 
including \oiv\,25.9\,\micron\ with an ionization potential
(IP) of $\sim$\,55\,eV, and \nev\,14.3\,\micron\ with an IP of $\sim$\,97\,eV.
The ionization state of \izw\ falls at the extreme upper end of all of the line
ratios we analyzed, but not coincident with galaxies containing an accreting massive black hole
(active galactic nucleus).
Comparison of the line ratios with state-of-the-art photoionization and shock models suggests
that the high ionization state in \izw\ is not due to shocks.
Rather it can be attributed to metal-poor stellar populations with a 
self-consistent contribution of X-ray binaries or ultra-luminous X-ray sources.
It could also be partially due to a small number of hot low-metallicity Wolf-Rayet stars
ionizing the gas; a small fraction (a few percent) of the ionization could come from an intermediate-mass black hole.
Our spectroscopy also revealed four 14\,\micron\ continuum sources, $\ga\,30-100$\,pc in diameter,
three of which were not previously identified. 
Their properties are consistent with \hii\ regions ionized by 
young star clusters.
\end{abstract}



\section{Introduction} \label{sec:intro}

Dwarf galaxies (stellar mass \mstar\ $\la 10^9$\,\msun) are the smallest, faintest, least 
chemically evolved systems in the Universe,
and the most abundant galaxy population at any redshift
\citep[e.g.,][]{white91,babul96,bullock17,davidzon17,stefanon21}.
In the $\Lambda$Cold-Dark-Matter paradigm, dwarf galaxies resided in
early low-mass dark matter (DM) haloes \citep[$\la 10^{11}$\,\msun,][]{bullock17,behroozi19,behroozi20}
that were the sites of the first star formation (SF).
Due to their high ionizing photon escape fractions, 
the dwarf galaxies forming in low-mass DM halos are thought to provide most of the photons 
needed to reionize the Universe \citep[e.g.,][]{wise09,atek22,atek24}.
They are also expected to be extremely metal-poor, with a rapid transition from
virtually pristine (Population III) stars in mini-haloes to the few percent level of metal enrichment 
relative to Solar metallicity (\zsun)
that can be observed today \citep[e.g.,][]{bromm11,klessen23}. 

Before the advent of \jwst, it was virtually impossible to 
study dwarf galaxies in detail at or beyond the Epoch of Reionization.
Now, early observations have revealed a wealth of information about metal-poor dwarf galaxies at
those redshifts 
\citep{trump23,furtak23,heintz23,rhoads23,morishita24,curti24}.
These new observations find that the majority of galaxies at $z \ga 7$ are dwarfs, 
\mstar $\la 10^8$\,\msun\ \citep{carnall23,furtak23,nakajima23,endsley24}, 
with low metallicities \zzsun\,$\la$ 0.2 
\citep[][]{carnall23,curti23,trump23,rhoads23,morishita24,curti24}.
These high-$z$ dwarf galaxies also exhibit high star formation rates (SFRs) for their stellar mass, 
a consequence of the evolutionary shift of SFR/\mstar\ to higher values at higher redshifts
\citep[e.g.,][]{noeske07,speagle14,whitaker14,leslie20,conselice25},
and the tendency of low-mass galaxies at high redshift to have ``bursty' star-formation histories,
unlike their more massive counterparts \citep{looser25}.
Although these preliminary \jwst\ results are not definitive, they hint at extremely
intense SF activity in these dwarf galaxies at very early epochs.

Despite their cosmological importance,
it is not yet clear how metal-poor dwarfs fit within the current framework for massive star formation.
The interstellar medium (ISM) in these systems tends to be less dusty \citep[e.g.,][]{remy14,galliano21}, 
therefore with less shielding from the intense radiation field (RF) 
produced by massive stars.
Cooling is 
inefficient because of the low metal and dust content.
Overall, this results in a paucity of ISM diagnostics such as CO, the usual proxy for \htwo,
because it tends to be photo-dissociated in unshielded metal-poor environments
\citep[e.g.,][]{vandishoeck88,bolatto13}.
At low metallicity, there is also a deficit of dust features such as those emitted by 
Polycyclic Aromatic Hydrocarbons 
\citep[PAHs, e.g.,][]{engelbracht05,madden06,hunt10}.
The RF tends to be harder and more intense at low metallicity \citep[e.g.,][]{cormier15},
and able to effectively carve cavities in the surrounding neutral ISM.
This makes the ISM more porous and clumpy than in less extreme environments.
However, the resolved chemical and ionization structure of the ISM is virtually 
impossible to study in dwarf galaxies at high redshift,
requiring us to study examples in the Local Universe.

Objects like the recently-discovered \jwst\ high-$z$ dwarf galaxies do exist in the Local Universe,
namely Blue Compact Dwarf galaxies (BCDs), defined by high surface brightness and blue optical colors
\citep[e.g.,][]{gildepaz03}.
At a distance of 18.2\,Mpc \citep[88\,pc\,arcsec$^{-1}$,][]{aloisi07}, the most extreme BCD is \izw,
with a metallicity of  \logoh\,=\,7.2 \citep[$\sim$3\%\,\zsun,][]{izotov99,lebout13}\footnote{The 
O/H determination is discussed in Rickards Vaught et al., Paper\,III in this series.},
among the lowest in the nearby 
universe, a stellar mass of $\sim 10^6 - 10^7$\,\msun\ \citep{fumagalli10,madden14,jano17,nanni20},
and an SFR estimated from the radio free-free continuum of $\sim 0.2$\,\msunyr\ \citep{hunt05}.
Over the last 10\,Myr, the SFR is estimated to have been greater, 
$\la\,1$\,\msunyr\ \citep{annibali13,bortolini24}.
Its specific SFR (sSFR\,=\,SFR/\mstar) is
very high, $\sim 10^{-7} - 10^{-8}$\,yr$^{-1}$, making 
\izw\ a ``starburst'', lying well above the local SF main sequence. 
It is thus an ideal target to characterize with high spatial definition the ISM
and the massive star 
formation taking place in an almost pristine environment similar to that of the dwarf galaxies 
newly discovered at redshift $z\ga 7$ by \jwst.

\izw\ is extremely gas rich with $\sim\,1\times10^8$ \msun\ of \hi\
\citep{vanzee98,lelli12};
its main body is interacting 
with a fainter companion (\izw\ C, or ``Zwicky's flare'')
$\sim$\,1.5 kpc to the NW, embedded in the same \hi\ envelope \citep{vanzee98}.
The radio continuum at 5\,GHz in \izw\ is optically thin 
\citep{cannon05,hunt05}
indicating lack of self-absorption from dense (ionized) gas.
The lack of dense gas is also implied by the measured ionized gas density
of $\sim\,10-100$ \cmthree, 
both from optical spectra and inferred from the radio continuum \citep{izotov99,hunt05}.
SF in \izw\ is mostly concentrated in two prominent OB associations 
(NW, SE), each $\sim$ 200\,pc in diameter, separated by $\sim$ 450\,pc. 
Together they host a total of $\sim$\,2,000 O stars, 
\citep{hunter95,hunt05,cannon05},
similar to normal OB associations in the Milky Way. 

In the optical,
\izw\ emits high-ionization lines (\fev, \heii) with ionization potentials required to create the corresponding ions of 
$\geq\,$54.4\,eV,
indicating an extremely hard RF \citep[e.g.,][]{thuan05}.
The \heii\ emission cannot be explained with a ``standard'' massive stellar population,
including Wolf-Rayet stars \citep{kehrig15},
and has been attributed to Population III stars 
\citep[e.g.,][]{kehrig15,kehrig16},
or possibly luminous X-ray binaries 
\citep[XRBs,][]{schaerer19,rickards21},
because of the bright X-ray source associated with the NW OB complex
\citep{thuan04,kaaret13}.
There are also hints of faint extended diffuse X-ray emission \citep{thuan04},
that could be associated with superbubbles driven by massive stellar feedback
\citep[e.g.,][]{martin96}.

There is apparently little cool dust in \izw.
Optical and near-infrared H recombination lines indicate very low extinction
\citep[A$_V \sim 0-0.2$ mag,][]{izotov99,izotov16},
making \izw\ ostensibly similar to the pristine environments of primordial galaxies. 
The total dust mass in \izw\ constrained by fitting the spectral energy distribution (SED) 
over almost 6 orders of magnitude in wavelength 
from the UV to the radio amounts to $\sim 340$\,\msun\ \citep{hunt14};
estimated with a different dust model,
it could be as high as $\sim 1,800$\,\msun\ \citep{fisher14}.
\izw\ has been observed with \spit/IRS \citep{wu07},
but the coarse spatial resolution made it impossible to 
separate the individual OB complexes, 
and the spectral resolution was low.
Nevertheless, \citet{wu07} detected \oiv\,25.89,
consistent with the 
hard RF suggested by the optical high-ionization lines.

Here we present \jwst\ MIRI Medium Resolution Integral Field Unit Spectrometer (MRS)
observations of \izw\ 
at a spatial resolution $\sim 25-50$\,pc. 
Our \jwst\ observations reveal high-ionization gas, several \hi\ recombination lines, and
a full complement of \htwo\ transitions, but no strong emission from PAHs. 
In the following,
we first describe the observations, data processing, and analysis in Sect. \ref{sec:data}.
The ionized-gas lines detected in 
spectra extracted within several regions of interest are presented in Sect. \ref{sec:spectra_ionized},
and compared with state-of-the-art photoionization and shock models of the observed line ratios.
Newly identifed compact continuum sources that
may be Young Stellar Clusters (YSCs) are described in Sect. \ref{sec:continuum}.
We discuss our results in Sect. \ref{sec:discussion},
and give our conclusions in Sect. \ref{sec:conclusions}.
Subsequent papers will describe other aspects of the data including
the detection of warm molecular hydrogen, dust features, and the search for PAH emission (Hunt et al., Paper\,II, ApJ, submitted);
metallicity maps with \oiv\ \citep{rickards25};
emission-line maps (Hunt et al., in prep.);
and RF hardness inferred from line-ratio maps (Rickards Vaught et al., in prep.).

\section{MIRI/MRS data, analysis, and ancillary data \label{sec:data}}

\izw\ was observed on 8 March 2024 with the
\jwst\ MIRI MRS 
\citep[][]{argyriou23}
through the GO program \#3353 (PI Aloisi/coPI Hunt). 
Two pointings were executed, each of $\sim$8\,hrs duration, 
in order to sufficiently cover both the NW and SE OB associations; 
a separate background observation of the same duration was also acquired
in the context of an uninterruptible sequence.
We required a PA orientation range during the dedicated background observations 
to be able to acquire on-source MIRI imaging in
F560W (5.6\,\micron), F1130W (11.3\,\micron), and F2550W (25.5\,\micron)
to complement the filters acquired by GTO program \#1233 
\citep[PI Meixner, see][]{hirschauer24,bortolini24}. 
The MRS spectral cubes comprise all three gratings, SHORT, MEDIUM, and LONG, for spectral coverage from
$\sim 5-28$\,\micron\ with a spectral resolving power ranging from $\sim\,1550–3250$.
We adopted a 4-point dither pattern optimized for extended sources, and used the SLOWR1 readout pattern,
with 40 groups per integration and 2 integrations per exposure.


The MIRI data were reduced using the \jwst\ calibration pipeline version 1.15.1 \citep{bushouse24}, 
with the corresponding CRDS (Calibration Reference Data System) context 
\texttt{jwst\_1276.pmap} and CRDS file version 11.17.25. 
The data were processed through all three steps of the pipeline, with additional corrections applied to intermediate products outside the pipeline.
Details of the reduction procedures are given in Appendix \ref{sec:reduction}.
The final cubes were combined using the \texttt{drizzle} method \citep{law23}, 
for each of the four channels,
conserving the native wavelength coverage and pixel sizes
(e.g., 0\farcs13, 0\farcs17, 0\farcs20, 0\farcs35, for Channels 1, 2, 3, and 4, respectively).

\subsection{Ancillary data \label{sec:ancillary}}

We include in our analysis images/cubes of \izw\ at optical wavelengths.
The \hst/ACS F606W image of \izw\ is taken from \citet{aloisi07}, but here
realigned to the Gaia DR3 \citep{gaiadr3} astrometric system.

We also rely on the optical data cube from the Keck Cosmic Web Imager (KCWI) 
provided by \citet{rickards21}.
The angular resolution in the KWCI cube was derived from standard-star observations and is characterized
 by a full-width half-maximum (FWHM) of 0\farcs7, with a pixel-scale of $\sim$\,0\farcs15. 
The spectral resolving power is R$\,=\,3600$, with a wavelength coverage between 3700-5500\,\AA. 
For more details, see \citet{rickards21}.
After alignment with the astrometrically corrected \hst\ F606W image, 
the KWCI cube was convolved to the 27\,\micron\ MIRI PSF using the same procedure described 
below in Sect. \ref{sec:spectra}.
Optical spectra within the apertures described below have been extracted from this convolved cube.
In particular, \hb\ and \hd\ from KCWI will be used for the optical extinction estimate,
and \hb\ for line ratios relative to the MIRI \hi\ recombination lines.

\begin{figure*}[t!]
\includegraphics[height=0.409\textwidth]{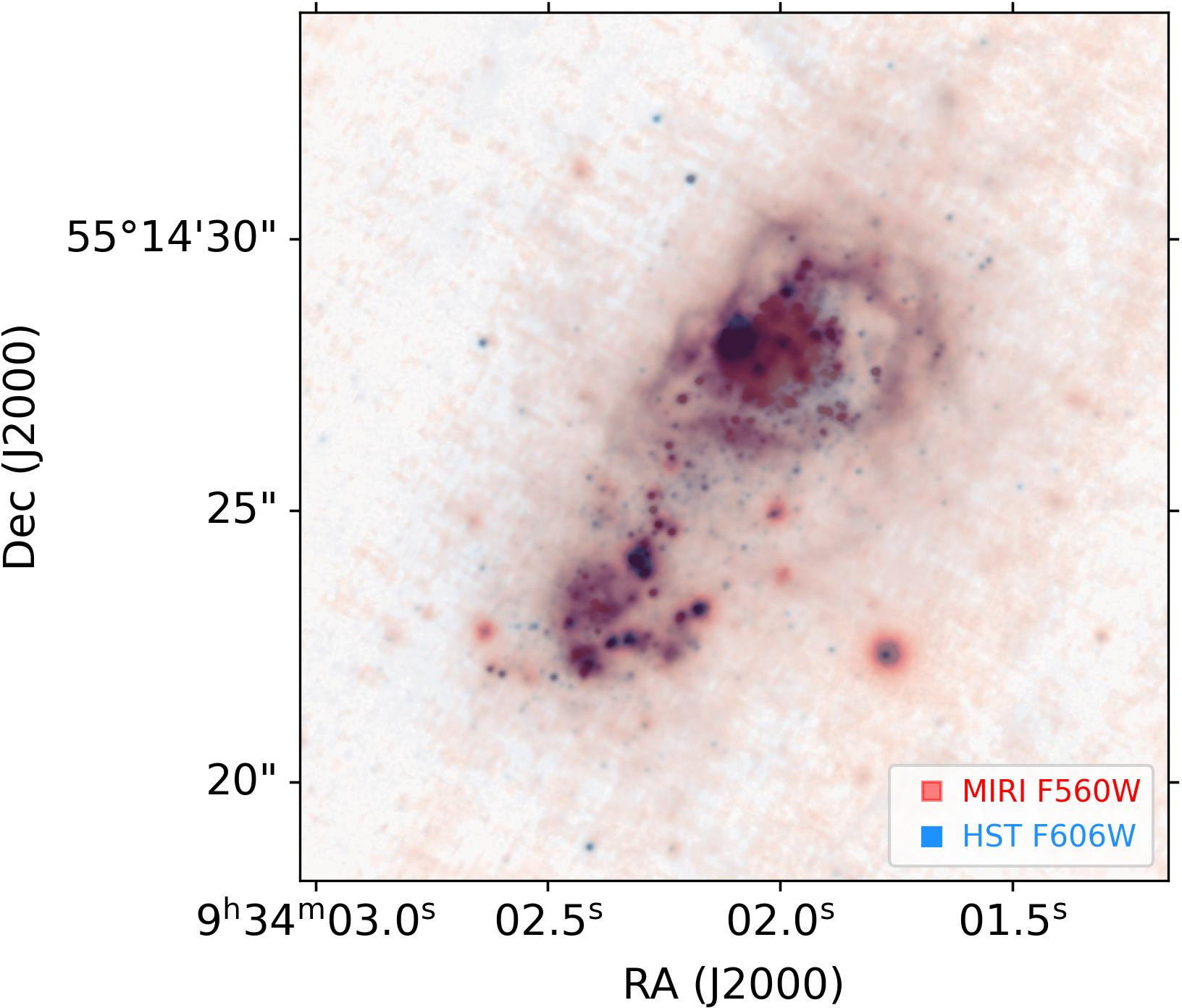}
\hspace{0.01\textwidth}
\includegraphics[height=0.409\textwidth]{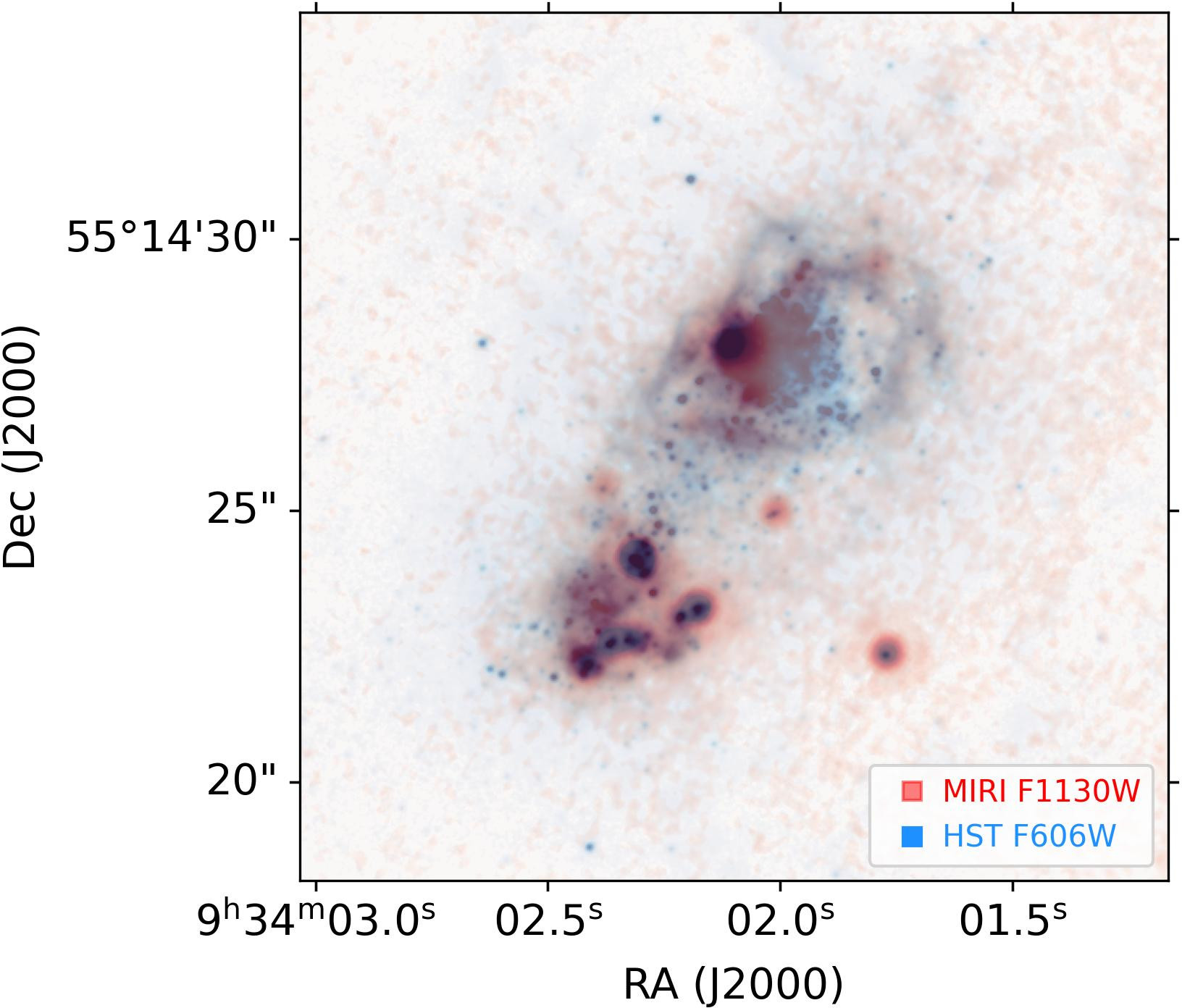} \\
\includegraphics[height=0.409\textwidth]{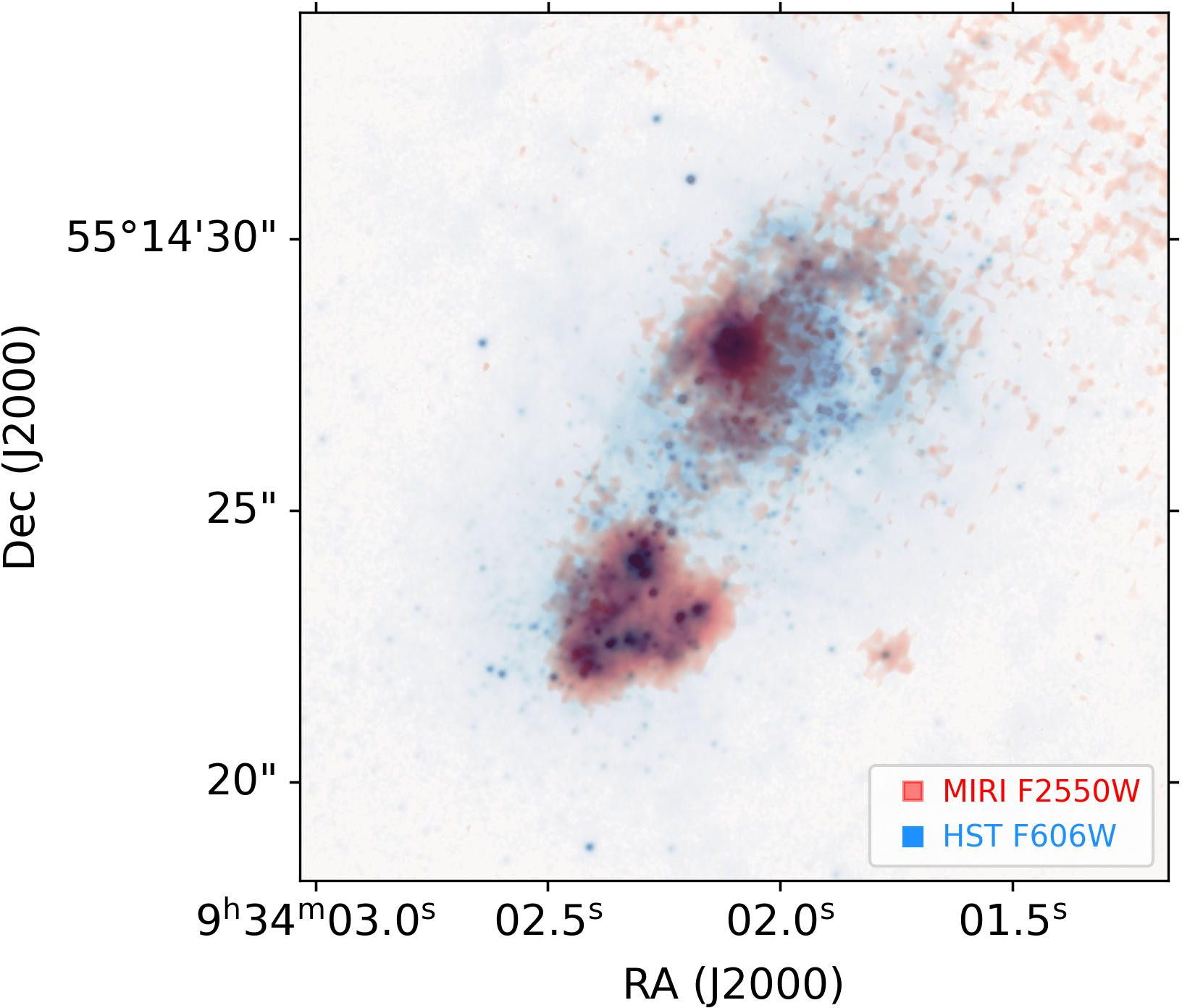}
\hspace{0.01\textwidth}
\includegraphics[height=0.409\textwidth]{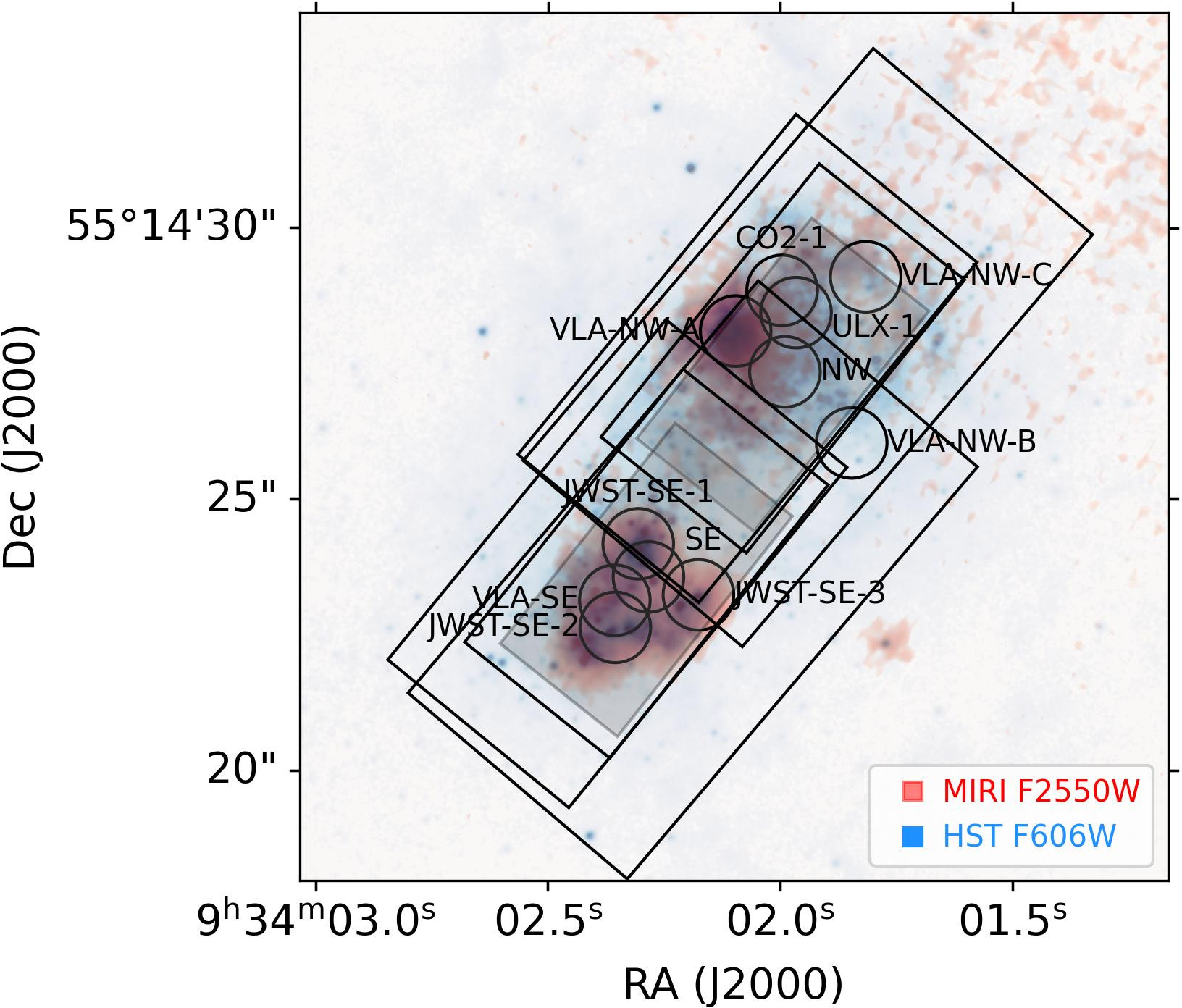} \\
\vspace{-\baselineskip}
\caption{16\arcsec$\times$16\arcsec\ overlays of MIRI images of \izw\ on the \hst\
F606W image astrometrically corrected to Gaia.
The MIRI F560W filter is shown in the top left panel, and F1130W in the top right.
The F2550W filter appears in the bottom panels, with the MIRI FoVs from both pointings
in the four overlapping channels overlaid in the right panel, 
together with the circles giving the apertures 
(0\farcs65 radius, $\sim 120$\,pc diameter) for spectral extraction.
The smallest FoV of Channel 1 is indicated by a slightly opaque background.
The underlying \hst\ image is illustrated in bluish tones, while the MIRI images in reddish ones.
}
\label{fig:apertures}
\end{figure*}

\subsection{Aperture spectra \label{sec:spectra}}

In order to ensure that the line ratios within each aperture are mutually consistent,
it was necessary to convolve the original data cubes to the 
point-spread function (PSF) of the longest wavelength of interest ($\sim 27$\,\micron).
We first assumed that the MIRI PSF is described by a 
Gaussian; then the MIRI PSF FWHM at each wavelength was estimated
according to \citet[][Eqn. (1)]{law23}. 
The final Gaussian kernel $\sigma_\lambda$ was calculated as $\mathrm{FWHM}_\lambda = \sqrt{8\ln2}\sigma_\lambda$.
The kernel width is defined as the quadrature difference for each $\lambda$ plane in the cube,
with a width $\sigma\,=\,\sqrt{\sigma_{27}^2  - \sigma_\lambda^2}$,
where $\sigma_\lambda$ is the intrinsic PSF $\sigma$ at a given wavelength. 
These kernels were then convolved with the original cube to achieve 27\,\micron\ spatial resolution
over each of the wavelength planes in the cube.

Spectra were extracted from the convolved cubes
from apertures with a radius of 0\farcs65 ($\sim 120$\,pc diameter),
centered on 11 regions of interest.
These regions of interest include the four \hii\ regions, dubbed VLA-NW-A, B, C, and VLA-SE, 
identified by the radio continuum maps 
with $\sim 2$\arcsec\ beam presented by \citet{cannon05};
they also include the peaks of the $^{12}$CO(2--1) detection near the NW OB association 
by \citet{zhou21},
and the ultraluminous X-ray source ULX-1 found by \citet{thuan04}.
In addition, we placed apertures on the approximate centers of the NW and SE 
OB complexes.

Finally, from visual inspection along the wavelength planes of the MIRI spectral cube, 
besides VLA-NW-A, which appears as a continuum source,
we noticed three additional continuum sources near the VLA SE \hii\ region, 
although not exactly coincident with it. 
Apertures for spectral extraction were also placed on the 14\,\micron\ peaks of this emission,
as measured from the cube; these are
denoted as JWST-SE-1, JWST-SE-2, and JWST-SE-3. 
These continuum sources are also apparent in the MIRI images presented by \citet{hirschauer24},
and will be discussed in Sect. \ref{sec:continuum}.
Table \ref{tab:aper} reports the aperture centers used here.

The aperture size is designed to encompass at least 65\% of the light 
in the PSF of the convolution kernel.
\citet{rigby23} find that an aperture of 0\farcs64 radius encloses 65\% of the energy in the 
F2550W MIRI filter,\footnote{See also 
\url{https://jwst-docs.stsci.edu/jwst-mid-infrared-instrument/miri-performance/miri-point-spread-functions\#gsc.tab=0}.}
similar in wavelength to the 27\,\micron\ kernel.
Although a larger aperture would be desirable, it 
results in incomplete spectra in several of the regions of interest,
because of the smaller field of view (FoV) at shorter wavelengths.
Thus, we adopt an aperture with radius 0\farcs65 as a compromise.
Moreover, in order to reduce sampling artefacts in the recovered spectra,
\citet{law23} recommend an extraction 
diameter at least equal to the FWHM of the PSF.
Since the \oiv\ wavelength in MIRI/MRS has a PSF FWHM $\sim\,$0\farcs96
\citep[Eq. (1) in][]{law23}, the adopted diameter of 1\farcs3 satisfies this recommendation
even for the longest wavelength of interest. 
Because we cannot know a priori whether or not the emission is unresolved (point-like)
or extended, an aperture of constant size across wavelength was deemed the best option.

\begin{table}
\caption{Aperture centers for spectral extraction}
\begin{centering}
\begin{tabular}{lcc}
\hline
\hline
\multicolumn{1}{c}{\rule{0pt}{3ex} Region} &
\multicolumn{1}{c}{RA } &
\multicolumn{1}{c}{Dec.} \\
&
\multicolumn{2}{c}{(J2000)} \\
\hline
\\ 
NW        & 9:34:01.99 & +55:14:27.3 \\
VLA-NW-A  & 9:34:02.10 & +55:14:28.1$^a$\\
VLA-NW-B  & 9:34:01.85 & +55:14:26.0$^a$\\
VLA-NW-C  & 9:34:01.82 & +55:14:29.1$^a$\\
ULX-1     & 9:34:01.97 & +55:14:28.4$^b$\\
CO2-1     & 9:34:02.00 & +55:14:28.8$^c$\\
SE        & 9:34:02.29 & +55:14:23.5\\
VLA-SE    & 9:34:02.36 & +55:14:23.1$^a$\\
JWST-SE-1 & 9:34:02.31 & +55:14:24.1\\
JWST-SE-2 & 9:34:02.36 & +55:14:22.6\\
JWST-SE-3 & 9:34:02.18 & +55:14:23.2\\
\\
\hline
\hline
\end{tabular}
\end{centering}
\\
\newline
$^{a}$~\citet{cannon05}; \\
$^{b}$~\citet{thuan04,ott05,rickards21}; \\
$^{c}$~\citet{zhou21}.
\\
\label{tab:aper}
\end{table}

The apertures are shown in the bottom right panel of Figure \ref{fig:apertures},
where we have overlaid in red the 
three MIRI images, F560W, F1130W, and F2550W, on the \hst\ F606W image, shown in blue.
Clearly, the apertures are not independent.
However, our objective is not to sample independent regions, but rather a variety of 
individual regions to assess differences in ionization state. 
The reddish/purple hue of the environment of the SE OB complex in Figure \ref{fig:apertures}
indicates that it has more warm dust
than the NW in general, with the exception of the VLA-NW-A \hii\ region.
The curved gaseous filaments around the NW OB complex are clearly visible, appearing also in the F560W image, 
resulting in a purple color.

\begin{figure*}[t!]
\includegraphics[width=\textwidth]{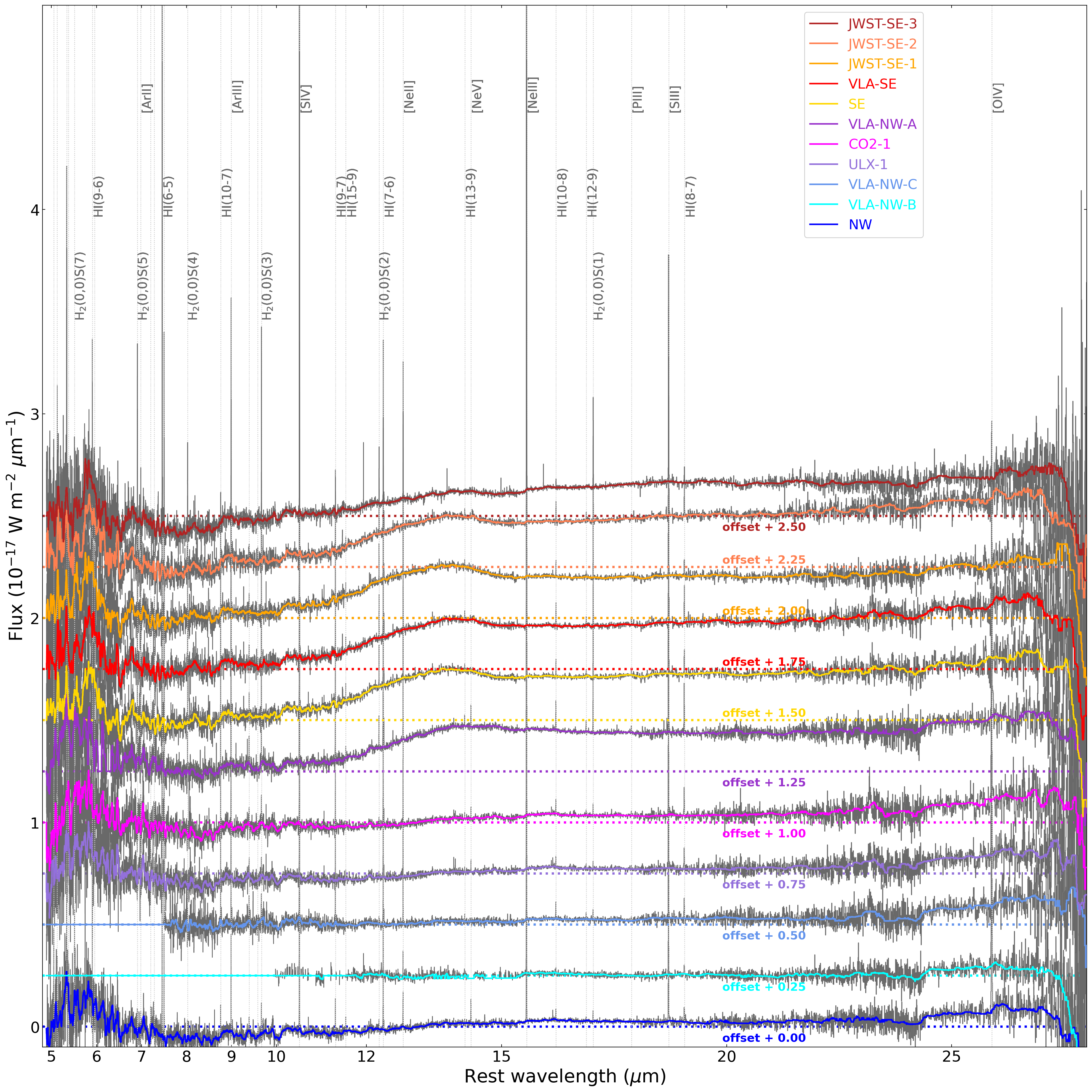}
\vspace{-\baselineskip}
\caption{Spectra extracted the 27\,\micron\ convolved cubes in the 0\farcs65-radius apertures shown in 
Figure \ref{fig:apertures}.
The vertical axis for flux density is in units of $10^{-17}$ W\,m$^{-2}$\,\micron$^{-1}$,
and the horizontal wavelength axis in \micron.
The gray curves show the spectra, while the heavy colored curves show the smoothed continuum;
the dotted horizontal colored lines indicate the zero level of the spectra.
For better visibility, the spectra are offset by small increments (in the vertical-axis
units of $10^{-17}$\,W\,m$^{-2}$\,\micron$^{-1}$) as denoted in the figure.
The vertical dotted lines correspond to the detected transitions given in Tables \ref{tab:nwflux}
and \ref{tab:seflux}; to avoid overcrowding, not all lines are labeled. 
The spectra for VLA-NW-B and VLA-NW-C are missing the short-wavelength MIRI channels because 
part of the aperture falls outside the MIRI FoV at those wavelengths.
}
\label{fig:spectra}
\end{figure*}

The spectra extracted from the convolved cubes are shown in Figure \ref{fig:spectra}, 
with offsets 
to enhance the visibility of the individual spectra. 
For each spectrum, the light gray curves show the original spectra, and the heavy
colored curves the smoothed continuum, as reported in the legend;
horizontal dotted colored lines show the zero level for each spectrum.

Emission lines were identified with an automatic algorithm based on 
\texttt{find\_lines\_threshold} within \texttt{Astropy/specutils} \citep{specutils}.
The fits to the piece-wise smoothed continuum were subtracted from the observed spectra, and 
the noise in the individual spectral regions was estimated for input to the line-finding algorithm,
by calculating the standard deviation in the piece-wise smoothed continuum
with \texttt{noise\_region\_uncertainty}.
Potential line detections were defined with an initial signal-to-noise (S/N) threshold of 3.
Each of these potential detections was then fit with the sum of a Gaussian and a linear continuum
in a window around the potential line. 
Lines were identified by comparing the rest-frame wavelength of the best-fit Gaussian
with known line lists
\citep{draine96,vanhoof18}.
The total line flux was calculated by integrating the Gaussian, and flux uncertainties by propagating
the errors on the fitted parameters given by the Hessian matrix associated with the fit.
Finally, we extracted the list of detected lines by requiring a S/N\,=\,3 on the fitted flux,
and the absolute value of the velocity offset to be $\leq\,250$\,\kms.
A subset of these is shown in Figure \ref{fig:spectra} as vertical dotted lines.

In \izw, the MIRI 1D spectra show (S/N$\geq$3) detections for a total of
10 fine-structure (FS) lines, 
15 \hi\ recombination lines, and 
7 \htwo\ transitions.
Table \ref{tab:fslines} gives the ionization potentials to create the corresponding ions (IPs) for the FS lines 
detected in the MIRI spectra of \izw.  
The detected ionized-line fluxes and their uncertainties are reported in Appendix \ref{sec:flux} in
Tables \ref{tab:nwflux} and \ref{tab:seflux}.
The Gaussian fits to the \oiv\ and \nev\ 
are also shown in Appendix \ref{sec:flux}
in Figs. \ref{fig:linefits_oiv} and \ref{fig:linefits_nev}.
In this paper, we restrict the discussion to the ionized gas; 
the warm \htwo\ and the dust emission in aperture spectra are described in a companion paper (Paper\,II).

\section{Ionized gas\label{sec:spectra_ionized}}

The ionized gas in \izw\ is in an extremely high ionization state. 
MIRI significantly detects \oiv\ within eight apertures, and 
the high-ionization line \nev\,14.32 in four 
(which can be mapped, Hunt et al., in prep.). 
Although \nev, with an IP of 97.1\,eV, is detected
at optical wavelengths (\nev\,$\lambda$3426) in several BCDs \citep[e.g.,][]{thuan05,izotov12}, 
it has not yet been 
detected in \izw\ in the optical \citep[e.g.,][]{izotov21}, 
perhaps because of insufficient spatial or wavelength coverage.

\begin{table}
\caption{Detected FS lines$^a$}
\begin{centering}
\begin{tabular}{lrr}
\hline
\hline
\multicolumn{1}{c}{\rule{0pt}{3ex} Line} &
\multicolumn{1}{c}{Rest wavelength} &
\multicolumn{1}{c}{Ionization} \\
&
\multicolumn{1}{c}{(\micron)} &
\multicolumn{1}{c}{Potential (eV)} \\ 
\hline
\\ 
\arii\  & 6.98527 & 15.76 \\
\ariii\ & 8.99138 & 27.63 \\
\feii\  & 5.34017 & 7.90 \\
\neii\  & 12.81355 & 21.56 \\
\neiii\ & 15.55510 & 40.96 \\
\nev\   & 14.32170 & 97.12 \\
\oiv\   & 25.89030 & 54.93 \\
\piii$^{b}$  & 17.88500 & 19.77 \\
\siii\  & 18.71300 & 23.34 \\
\siv\   & 10.51050 & 34.79 \\
\\
\hline
\hline
\end{tabular}
\end{centering}
\\
\newline
$^{a}$~Wavelengths and IPs taken from \citet{vanhoof18} and
\url{https://www.mpe.mpg.de/ir/ISO/linelists/FSlines.html}\\
$^{b}$~Possible weak detection only in aperture VLA-NW-A.
\label{tab:fslines}
\end{table}


The \nev\,24.32 line is not detected, but we can use the upper limits (ULs) on the line flux
to investigate the electron density $n_e$ in the four regions where \nev\,14.32 is detected:
CO2-1, NW, VLA-NW-A, ULX-1. 
The $3\sigma$ ULs for \nev\,24.32 are, respectively,
$2.4\times10^{-20}$\,\wmsq,
$8.5\times10^{-21}$\,\wmsq,
$1.6\times10^{-20}$\,\wmsq,
$3.1\times10^{-20}$\,\wmsq, 
corresponding to ULs on \nev\,24.32/\nev\,14.32 (see Table \ref{tab:nwflux})
of $\la 7$ (CO2-1), $\la 1$ (NW), $\la 2$ (VLA-NW-A), and $\la 8$ (ULX-1).
The line ratio is expected to vary from $\sim 1$
for $n_e < 10^3\,\mathrm{cm}^{-3}$ to $\sim 0.1$ for $n_e > 10^5\,\mathrm{cm}^{-3}$ \citep{draine11}. 
Thus, the present upper limits for [NeV]24.32/[NeV]14.32 do not constrain $n_e$.

\subsection{Extinction from HI recombination lines \label{sec:extinction1d}}

We have investigated the internal extinction in the aperture spectra using the \hi\ recombination lines given
in Table \ref{tab:hilines},
combined with those taken from the KCWI \citep[see Sect. \ref{sec:ancillary},][]{rickards21}.
A S/N of 10 was adopted for the individual detections to maximize the reliability of the measurements.
To calculate the intrinsic line ratios, 
we relied on Case B emissivities from \texttt{PyNeb} \citep{luridiana15},
fixing the electron temperature to $T_\mathrm{e}$\,=\,20\,000\,K and the electron density 
to $n_e\,=\,$ 100\,\cmthree\ \citep{izotov99,kehrig15}. 

\begin{table}
\caption{MIRI \hi\ lines used for extinction calculations$^{a}$}
\begin{centering}
\begin{tabular}{lr}
\hline
\hline
\multicolumn{1}{c}{\rule{0pt}{3ex} Line} &
\multicolumn{1}{c}{Rest wavelength} \\
&
\multicolumn{1}{c}{(\micron)} \\
\hline
\\ 
\hug\ \hi\ 9--6 & 5.90821 \\
\pfa\ \hi\ 6--5 & 7.45986 \\
\hub\ \hi\ 8--6 & 7.50249 \\
\hi\ 10--7 & 8.76006 \\
\hi\ 9--7 & 11.30870 \\
\hi\ 15--9 & 11.53949 \\
\hua\ \hi\ 7--6 & 12.37190 \\
\hi\ 10--8 & 16.20910 \\
\hi\ 12--9 & 16.88063 \\
\hi\ 8--7 & 19.90619 \\
\\
\hline
\hline
\end{tabular}
\end{centering}
\\
\newline
$^{a}$~These are the \hi\ detections with S/N$\,\geq\,10$ in the aperture spectra, as discussed in the text.
\label{tab:hilines}
\end{table}

\begin{figure*}[t!]
\includegraphics[width=\linewidth]{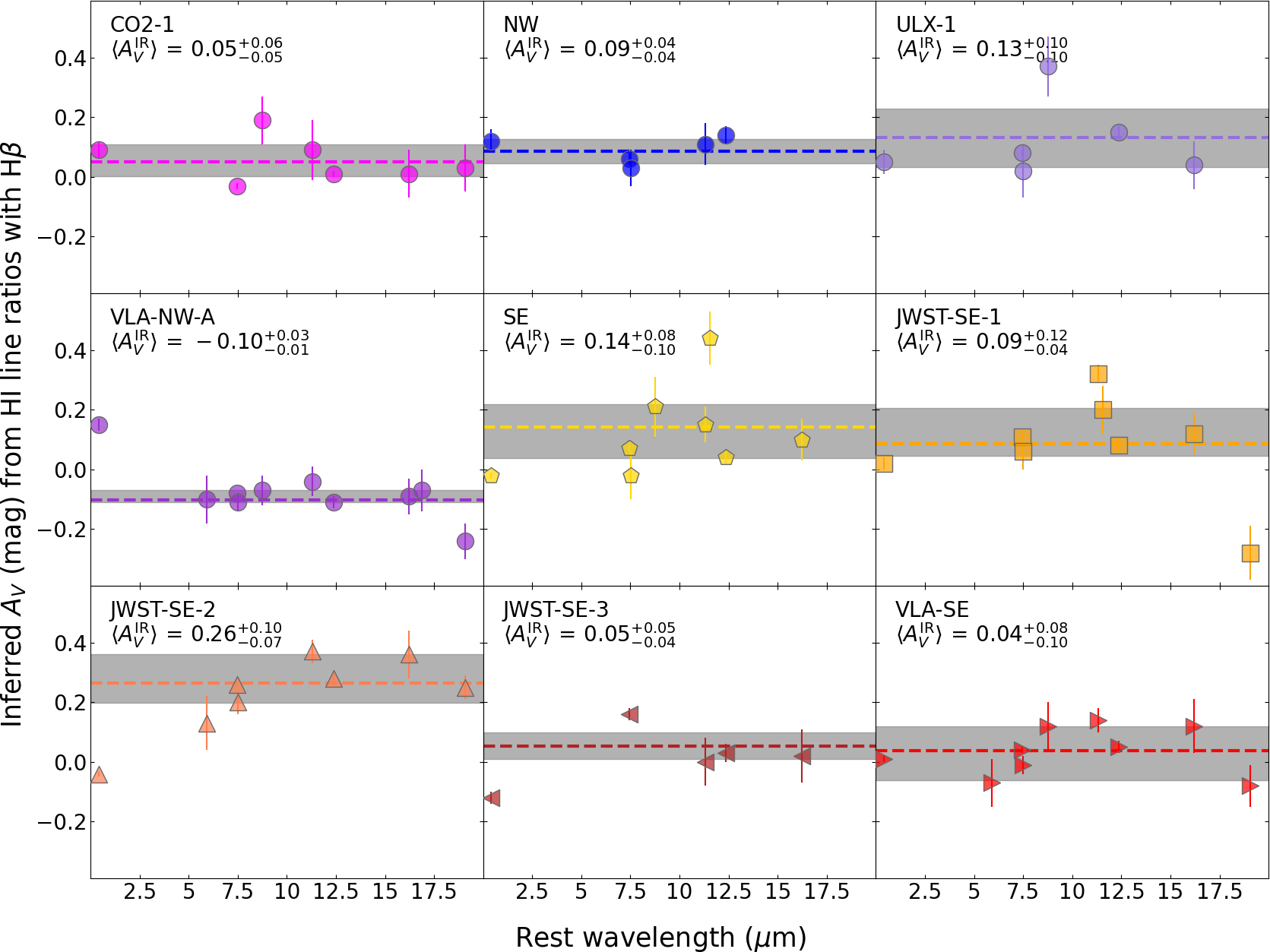}
\caption{Inferred extinction relative to \hb\ using \hd\ from KCWI, 
and the detected MIRI \hi\ recombination lines with a S/N\,$\geq$\,10;
G23 is assumed for the extinction curve, and no foreground extinction has been applied. 
The extinction values for each aperture are plotted against the wavelength of the MIRI \hi\
lines to illustrate the power of the MIRI wavelengths to probe deeper into embedded regions.
The horizontal shaded light-gray rectangles in each panel show the $1\sigma$ spread of the ensembles.
The \av\ values in the upper-left corner of each panel correspond to the MIRI average and spread,
not taking into account the optical \hd/\hb\ values.
VLA-NW-B and VLA-NW-C are not shown because they are not uniformly within the
FoV of the individual MIRI channels (see Figure \ref{fig:apertures}).
}
\label{fig:extinction1d}
\end{figure*}

We experimented with two extinction curves, \citet[][hereafter WD01]{wd01} 
and \citet[][hereafter G23]{gordon23},\footnote{The G23 extinction curve also incorporates 
results by \citet{gordon09}, \citet{fitzpatrick19}, \citet{gordon21}, and \citet{decleir22}.} 
as implemented in \texttt{dust-extinction},\footnote{Available at 
\url{https://dust-extinction.readthedocs.io/en/stable/} \citep{gordon24}.}
an affiliated package of \texttt{Astropy}. 
Several trials were run using only the MIRI \hi\ lines to infer extinction, but the extinction 
is so small that the uncertainties are too large to be reliable.
Thus, we have compared the MIRI lines to \hb, as shown in Figure \ref{fig:extinction1d}.
There are neglible differences between results from the WD01 and the G23 extinction curves,
so we illustrate only G23.

Figure \ref{fig:extinction1d} plots as a function of wavelength the G23 extinction inferred relative to \hb\ 
from KCWI 
for the \hi\ recombination lines detected with S/N$\,\geq\,$10
in the apertures defined in Sect. \ref{sec:spectra} (Table \ref{tab:aper}). 
VLA-NW-B and VLA-NW-C are not shown because they do not fall within all the
MIRI channels.
We have checked whether \heii\ lines could be potentially blended with the \hi\ lines
(e.g., \hua), and thus the flux overestimated;
however, no evidence was found for asymmetric line profiles with a blue shoulder
as would be expected if this were the case.
Thus, the mid-infrared (MIR) H recombination-line fluxes should not be significantly contaminated by \heii.

Figure \ref{fig:extinction1d} shows that the extinction in \izw\ is small (foreground extinction
has not been corrected for and will be discussed below).
The grey regions and mean \avir\ values and spreads given in the upper-left corner of
each panel are calculated using only MIRI lines.
We have done this to assess if the optical were unable to penetrate deeply embedded sources
\citep[e.g.,][]{calzetti94,hunt01};
however, there is no systematic dependence on MIR wavelength which might be expected if this were true.
Nevertheless,
\av\ inferred from the optical is $\sim 0.1 - 0.2$\,mag lower than the MIR estimates
in the three JWST-SE continuum and SE apertures (but not in VLA-SE). 
The extinction measured by MIRI is highest in JWST-SE-2, \av$\,\sim 0.3$, although
it is not similarly high in all of the SE apertures. 
In VLA-NW-A, the optically-inferred \av\ lies $\sim 0.2$ mag above the MIRI values, which are significantly negative.
After tests, we find that the inferred MIR extinction would be roughly zero, if, for that aperture,
we assume a higher $T_\mathrm{e}$ ($\sim 24\,000-25\,000$\,K) and a higher electron density 
to $n_e$ ($\sim 10^4$\,\cmthree), consistent with the higher temperatures and dense gas
located near VLA-NW-A found by Rickards Vaught et al. (Paper\,III).

The extinction measured in \izw\ is comparable with that estimated for the foreground,
due to dust in the Milky Way.
Foreground extinction for each aperture was determined from the \citet{schlegel98} 
dust maps recalibrated
to the scale of \citet{schlafly11}, as implemented in the publicly available Python package 
\texttt{dustmaps}\footnote{\texttt{dustmaps} is found at 
\url{https://dustmaps.readthedocs.io/en/latest/maps.html} and the dust maps themselves can be
accessed and downloaded in the context of this package.} \citep{green18}.
For a given location on the sky, the module returns the corresponding \ebv\ value derived by 
linearly interpolating the dust maps;  \rv\,=\,3.1 was used to convert \ebv\ to \av.
This process was reproduced for each of the 11 apertures considered, 
but there is very little variation (a standard deviation of $\,\pm\,0.0001$ over the 11 apertures,
presumably because of the low spatial resolution of the dust maps).
We find a mean foreground extinction for \izw, \av\,=\,0.091,
in agreement with the \av\ value from \citet{schlafly11} tabulated by NED.
Thus, it is probable that the extinction in \izw\ is so low as to be
extremely difficult to measure against the foreground, 
even leveraging the large wavelength coverage provided here.


The inferred \av\ also depends on the method for determining foreground extinction.
Using the Planck extinction maps \citep{planck16}, the estimated foreground extinction
\av\,=\,0.137, larger than the \av\,=\,0.091 that we measure,
so most of the extinction in \izw\ would be attributed to Milky Way dust.
Because of the small value and large scatter for the internal extinction in \izw,
none of the emission lines studied here have been corrected for dust.

Figure \ref{fig:spectra} suggests that there may be greater dust content in the SE.
Warm dust, evident in the 
emission bump around 14\,\micron, appears not only
in the brightest \hii\ region, VLA-NW-A,
but also in the SE, around the \hii\ region VLA-SE, the SE OB complex, 
and the JWST-SE continuum sources.
This is a qualitative confirmation of what is seen also in Figure \ref{fig:apertures}. 
From Figure \ref{fig:extinction1d}, we can compare the inferred extinction in the NW 
to that in the SE.
The extinction averaged over all the NW regions is \av\,=\,$0.07$ 
(where VLA-NW-A is taken to have \av\,=\,0), and in the SE, mean \av\,=\,$0.12$.
Such a difference is negligible, and within estimates of foreground extinction,
but may hint of more dust in the SE,
as also found by \citet{cannon02} using narrowband \ha\ and \hb\ \hst\ images.

\subsection{Fine-structure line ratios\label{sec:highion}}

The hardness\footnote{By hardness, we mean the slope of the 13.6$-$100\,eV RF spectrum.} 
of the RF is an important diagnostic to determine the
dominant energy source powering the ionizing radiation.
It is well established that the effective hardness of the stellar RF
increases with decreasing metallicity \citep[e.g.,][]{campbell86,vilchez98}.
Because of this, \oiv\ with an IP of 54.9\,eV just beyond the \heii\ edge,
has been frequently detected in BCDs \citep{wu07,hao09,hunt10}.
However, only with \jwst/MIRI is it possible to detect even higher
ionization MIR lines such as \nev.
Before \jwst, no BCD has been clearly detected in the \nev\,14.32\,\micron\ line.
Now, in addition to \izw, \nev\,14.32 has also been detected in another metal-poor BCD,
\sbs, at $\sim 5$\% Solar, slightly more metal rich than \izw\
\citep{mingozzi25}.
\jwst\ is also detecting such high-ionization lines in more massive star-forming galaxies
\citep[e.g., M\,83,][]{hernandez25}, and
is expected to revolutionize our perspective on high-ionization emission lines
in star-forming galaxies.

\begin{figure*}[t!]
\includegraphics[width=\linewidth]{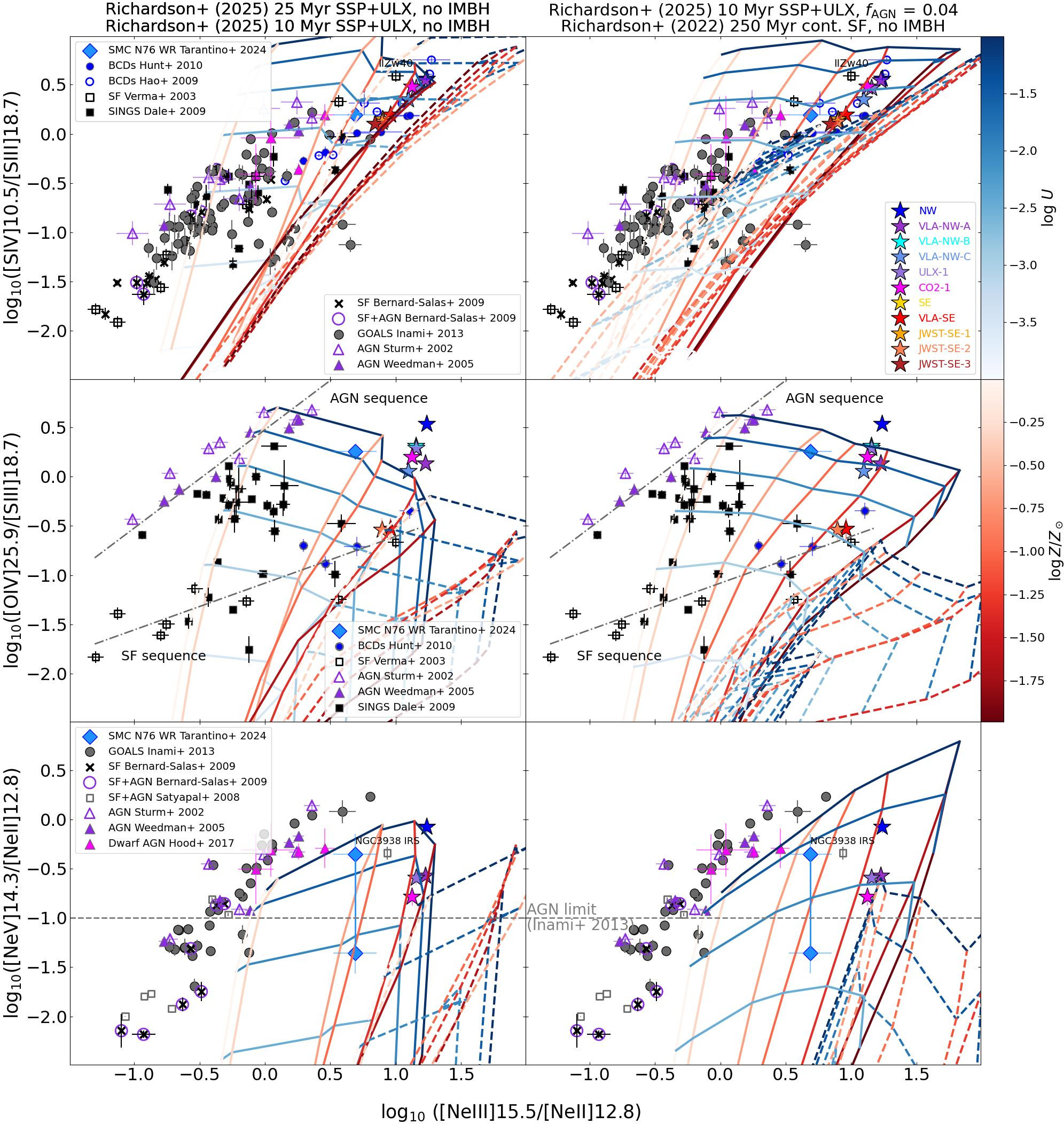}
\caption{Comparison of various RF diagnostic line flux ratios plotted versus \neiii/\neii.
Top panel: \siv/\siii; middle: \oiv/\siii; bottom: \nev/\neii.
Our MIRI points for \izw\ are given as stars, shown in the legend in the upper right panel.
Other samples of galaxies, AGN, BCDs, and dwarf galaxies with AGN are shown in
the legends according to the line ratio.
When the error bars are not evident, they are smaller than the symbol.
In the middle panel,
an ``AGN sequence'' and a ``SF sequence'' are illustrated by dot-dashed lines.
In the bottom panel, the horizontal dashed line 
corresponds to the AGN criterion of \nev/\neii\,$\geq 0.1$ defined by \citet{inami13}.
The same data are plotted in each row, but the 
left panels show the model grids by R25,
where the 10\,Myr SSPs are given with dashed contours, and the 25\,Myr SSPs
with solid ones.
The right panels show the 250\,Myr CSF population from R22 given as dashed curves, and,
as solid ones, the 
R25 model with \fagn\,=\,0.04 and a post-starburst age of 10\,Myr.
The colormaps for the model grids for log($U$) and log(\zzsun) are the same for all models. 
\label{fig:ion}
}
\end{figure*}

Because of their different IPs (see Table \ref{tab:fslines}),
 \neiii/\neii\ and \siv/\siii\ flux ratios trace the hardness of the RF
\citep[e.g.,][]{thornley00,sturm02,verma03}.
Combining lines with even larger differences in ionization potentials can 
distinguish between accreting massive black holes (or active galactic nuclei, AGN)
and stellar sources, although fast radiative shocks can sometimes mimic AGN,
particularly in low-metallicity environments \citep[e.g.,][]{izotov12,izotov21}.
Figure \ref{fig:ion} shows three commonly used diagnostics of the hardness of the RF
plotted against \neiii/\neii.
The line ratios from the MIRI apertures for \izw\ are shown as filled stars,
as described in the legend in the upper right panel.

To place \izw\ in the context of the broader galaxy population,
we have included for comparison other BCDs \citep{hao09,hunt10},
star-forming galaxies \citep{verma03,satyapal08,bernardsalas09,dale09},
luminous infrared galaxies (LIRGs) from the 
Great Observatories All-sky LIRG Survey \citep[GOALS,][]{inami13},
galaxies hosting AGN \citep{sturm02,weedman05}, and AGN in dwarf galaxies \citep{hood17}.
Indicated in the top panels are also previous MIR observations 
of \izw\ \citep{wu07}, and two other BCDs (roughly ten times more metal rich than \izw),
II\,Zw\,40 and NGC\,5253 \citep{verma03}.

Because \izw\ is known to harbor Wolf-Rayet (WR) stars, in particular in the NW
\citep{legrand97,izotov97,schaerer99,brown02}, 
we also include in the comparison sample the
WR nebula N76 \citep{naze03,tarantino24} in the Small Magellanic Cloud (SMC, $\sim 20$\% Solar metallicity)
The central ionizing source, AB7, is one of the most luminous WR stars in the SMC,
and provides a useful template for the ionization properties of an
extreme stellar source, able to create a highly ionized \hii\ region.
Models of hot, low-metallicity sources like AB7
suggest that they contain sufficient numbers of high-energy photons to
produce considerable \oiv\ and \nev\ emission \citep{schaerer98,schaererstasinska99}.
\citet{tarantino24} report detections from \spit/IRS observations of all the
ionized gas lines 
studied here.
However, they detect \nev\,24.32 rather than \nev\,14.32;
therefore, in the relevant plots, we divide the \nev\,24.32 flux by factors of 1 and 0.1,
corresponding to the low- and high-density limits, respectively, of the \nev\,24.32/\nev\,14.32
line ratio \citep[see above and][]{draine11}.

A clear correlation between \siv/\siii\ and \neiii/\neii\ is seen in
the top panels of Figure \ref{fig:ion}, similar to previous studies \citep[e.g.,][]{hao09};
our MIRI observations of the individual regions within \izw\ are consistent with these trends
at the extreme high-ionization end.
The similarity of these line ratios for galaxies with known AGN or composites with SF$+$AGN 
contributions implies that the IPs of these lines are insufficiently high to be diagnostic
of extremely hard ionizing radiation.
Higher ionization lines are needed, as we show in the middle and bottom panels.

The middle panels of Figure \ref{fig:ion} show \oiv/\siii\ versus \neiii/\neii.
Some of the observations shown in the top panels have not been included
\citep[e.g.,][]{wu07,bernardsalas09}, 
because the galaxies tend to fill the spectroscopic slits, and the different \spit/IRS LH (\oiv) 
and SH (\siii, \neiii, \neii) slit sizes have not been corrected for.
The NW regions in \izw\ are apparently in a higher ionization state than in the SE;
the SE regions fall on the extension of the trend of earlier observations, while in the NW,
the ratios are significantly above the \oiv/\siii\ trend and also have the highest \neiii/\neii\ values.
As also found by \citet{hao09} for \oiv\,25.89/\siii\,33.48, 
there appear to be two different sequences for
\oiv/\siii\ relative to \neiii/\neii, namely a SF sequence and an AGN one with higher \oiv/\siii\
for a given \neiii/\neii\ ratio.
The dot-dashed lines indicating the locus of the ``SF sequence'' and ``AGN sequence'' are merely to guide the eye;
no formal fitting procedure has been applied.

The bottom panels show the higher ionization line \nev\ ratio with \neii.
The galaxy indicated in the bottom panel, NGC\,3938, is a giant spiral
in which \nev\ was unexpectedly detected \citep{satyapal08}.
The high \neiii/\neii\ portion of the plot is occupied by this galaxy, 
the SMC N76 WR, and the four regions in the NW of \izw\ where \nev\ is detected.
These points clearly fall below and to the right of the trend shown by the AGN.
Here also, as for \oiv, there may be two kinds of behavior, reflecting different sources of
ionization for the emission.

While the SMC N76 WR from \citet{tarantino24} is indistinguishable from the 
general trends in the \siv/\siii\ and \neiii/\neii\ line ratios,
it stands out as one of the extreme sources in \oiv/\siii\ and \nev/\neii.
In these line ratios, it is similar to the NW apertures in \izw, although with
slightly lower \neiii/\neii.
The shape of N76, powered by the WR/O-star binary AB7, 
is an approximately spherical shell, of $\sim 40$\,pc in radius,
while the highly ionized gas, \oiv\ and \nev, also has a shell-like
morphology, but is smaller, with a projected radius of 10 to 20\,pc.
A WR binary like AB7 in the SMC could be powering some of the
photoionization in \izw; we will discuss this point further below.

\subsection{Comparison with models \label{sec:models}} 

Figure \ref{fig:ion} also illustrates a comparison with model grids.
The goal of this comparison is two-fold:
(1) to assess the success of the models to encompass the line ratios of a wide variety
of galaxies; and
(2) to compare the models in particular to \izw\ to constrain the excitation
mechanisms.
This broader comparison enables a sanity check for the general viability of the models,
and thus their applicability to \izw.
The left panels show selected Simple Stellar Population (SSP) models from \citet[][hereafter R25]{richardson25}, where
we have not included any contribution from the 
accreting intermediate-mass black holes (IMBHs)\footnote{IMBHs, by definition, have masses from $\sim 100$\,\msun\
to $\sim 10^5$\,\msun\  \citep[e.g.,][]{bhowmick24}.} also modeled by them
(for the left panels, we assume an AGN fraction $f_\mathrm{AGN}\,=\,0$).
The panels on the right show the stellar populations associated with a
continuous star-formation (CSF) history at an age of 250\,Myr by \citet[][hereafter R22]{richardson22}.
They also include R25 models with a small contribution from an IMBH as described below.

Both sets of models rely on the Binary Population and Spectral Synthesis (BPASS)
code \citep{stanway16,eldridge17} that considers stellar binary evolution, 
spanning a range of ages; the R22 models comprise metallicities from 0.05\,\zsun\ to \zsun,
while R25 metallicities from 0.01\,\zsun\ to \zsun. 
In R22 and R25, these stellar continua photoionize gas 
with ionization parameters \logu\ ranging from $-4$ to $-0.5$.

The R25 models also include self-consistent contributions
from XRBs in the form of ULXs as a function of age and stellar population metallicity,
as first implemented by \citet{garofali24}.
ULXs are a luminous subclass of XRBs that comprise a high-mass donor
star undergoing mass transfer and accretion onto a compact object,
assumed to be a stellar-mass BH in the models.
ULXs can therefore establish an upper limit to X-ray excitation in the case where
there is no IMBH.
This is important, in particular, for \izw\ which contains ULX-1 \citep{thuan04,kaaret13}.
The combination of the SSP$+$ULX SED models, 
which considers the timing of the onset of high-mass X-ray binaries, and the effects of metallicity on stellar
evolution and stellar wind properties, should ensure an overall realistic and self-consistent stellar population.
In these models, the shape of the ULX photoionizing SED does not vary with time,
but the normalization does; metallicity affects both the SED shape and normalization. 

\begin{figure}[t!]
\includegraphics[width=\linewidth]{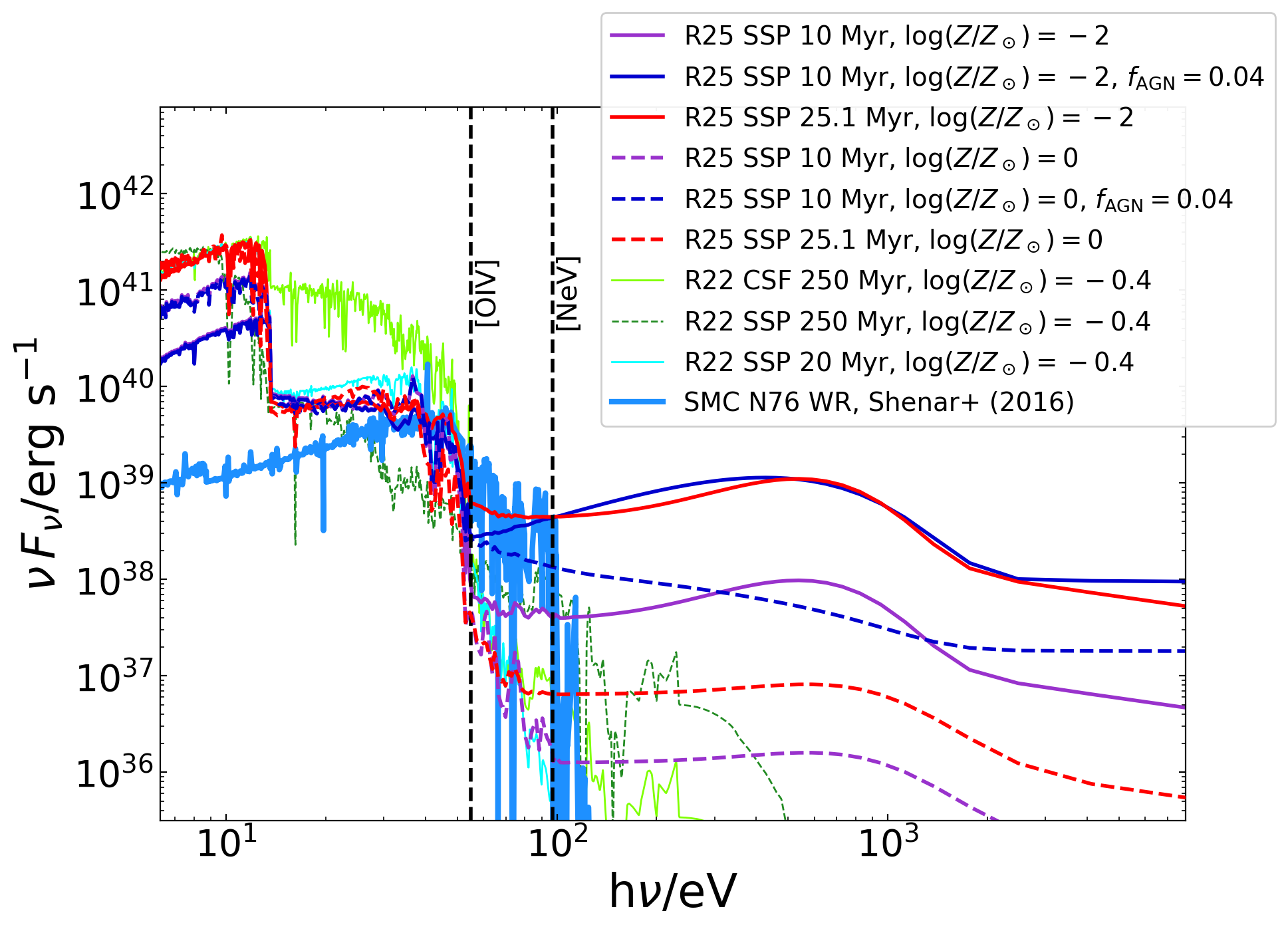}
\caption{SEDs of selected R22 and R25 SSP$+$ULXs plotted
against photon energy in eV. These R25 populations have ages of 10 to 25\,Myr,
and two values of metallicities: log(\zzsun)\,=\,$-2$ and log(\zzsun)\,=\,$0$.
The addition of \fagn\,=\,0.04 to the R25 10\,Myr SSP$+$ULX at the same two
metallicities is also shown, together with the SEDs of the stellar populations modeled by R22,
at two ages (SSPs with 20\,Myr, 250\,Myr; CSF 250\,Myr) with
a single, intermediate, metallicity, log(\zzsun)\,=\,$-0.4$.
The best-fit composite model SED of the WR/O binary star AB7 in the SMC is also shown \citep{shenar16}.
The IPs necessary to produce \oiv\ and \nev\ are illustrated as dashed vertical lines.
}
\label{fig:seds}
\end{figure}

As shown in Figure \ref{fig:seds}, where the ionizing SEDs are plotted as a function of energy,
age regulates the normalization relative to the remaining stellar populations.
Lower metallicity makes the ionizing SED harder (solid versus dashed curves),
and at 25\,Myr, the ULX contribution is maximized, relative to the stellar component (red curves).
Unlike the R22 models, the R25 models include a ULX population in addition to the SSPs \citep[e.g.,][]{garofali24}; 
this results in significant ionizing flux at energies $\ga$\,50\,eV.
This behavior is reflected in the right panels of Figure \ref{fig:ion}, where the 
R22 models without an IMBH are not able to reproduce the high \oiv/\siii, while
the R25 models without an IMBH in the left panel do a better job at reproducing this ratio.

Figure \ref{fig:seds} also compares the best-fit composite model SED of 
the WR binary AB7\footnote{We show the sum of the two separate components, WR$+$O6.} 
exciting SMC N76 \citep{shenar16}.
It provides an ionizing flux similar to
the most extreme SSP$+$ULX at $\sim 50$\,eV, consistently
with the observed high-ionization emission lines, \nev\ and \oiv, observed
by \citet{tarantino24} in this object and shown in Figure \ref{fig:ion}.

For completeness, we have included in the model comparison the R25 model of an SSP+ULX
at 10\,Myr where the ULX contribution is $\sim 10$ times lower relative to the SSP
than the maximum at 25\,Myr (see Figure \ref{fig:seds}), but with an AGN fraction \fagn\ of 4\%;
the ``AGN fraction'' \fagn\ corresponds to the fraction of
ionized photons attributed to the AGN. 
We have assumed the ``light seeding'' scenario; see R25 for more 
details.\footnote{The seeding mechanisms in R25 have a metallicity dependence;
for light seeding the IMBH masses range from 100\,\msun\ at the lowest metallicity
to $10^{5.89}$\,\msun\ at the highest.} 
This is a relatively arbitrary choice, but the aim here is to illustrate
the trends of the models with low \fagn.
The right panels of Figure \ref{fig:ion} shows these model grids,
illustrating that they are very similar to the maximum
ULX contribution at 25\,Myr (shown in the left panel).
This is due to the resemblance of the photoionizing SEDs in Figure \ref{fig:seds}. 
Such ambiguity makes it difficult to make decisive claims about
the need for an IMBH (or not) in \izw.

\begin{figure*}[t!]
\includegraphics[width=\linewidth]{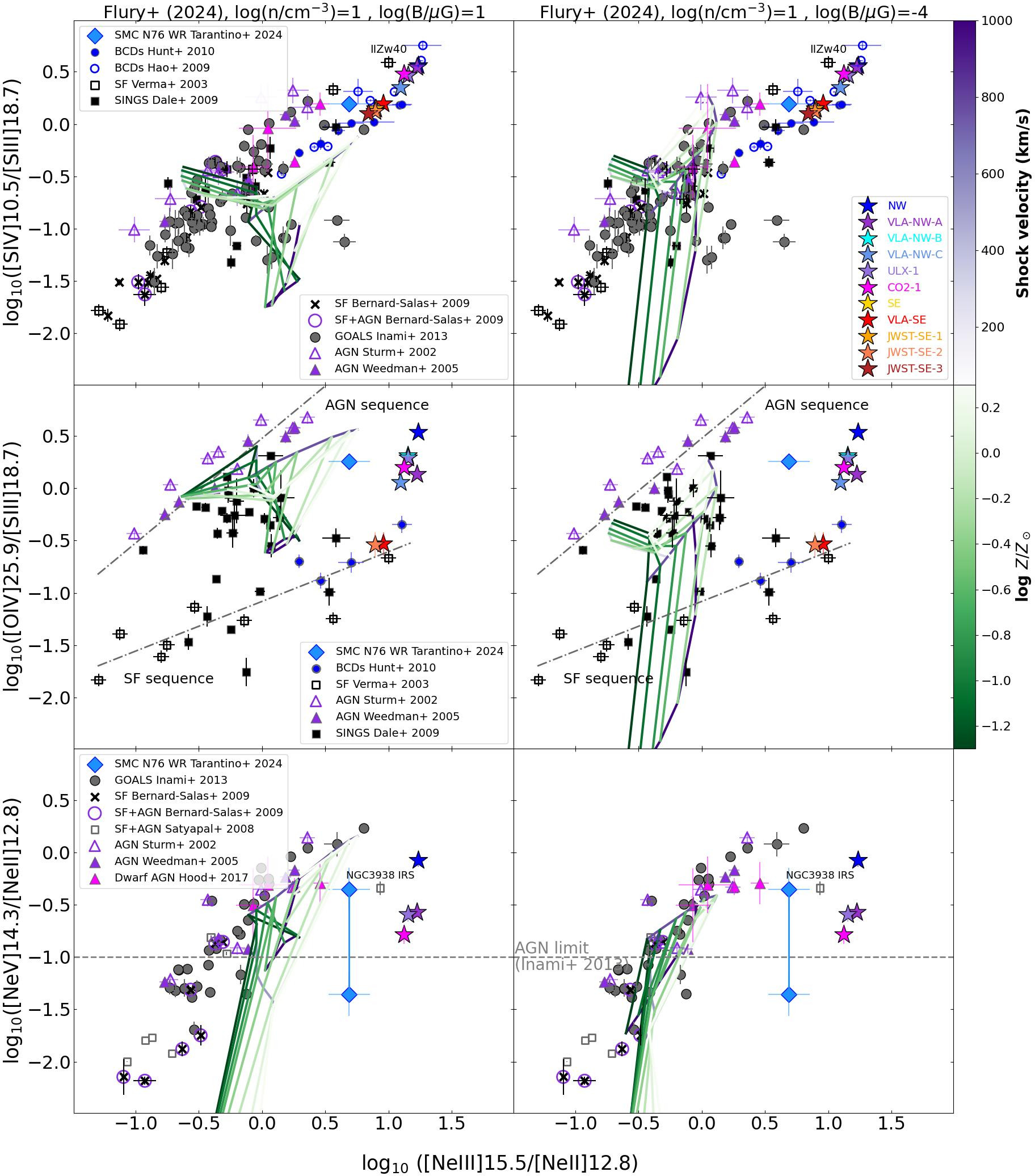}
\caption{Line ratios as in Figure \ref{fig:ion}, but here with
the shock models from \citet{flury24} overlaid on the data points.
As in Figure \ref{fig:ion}, when the error bars are not evident, they are smaller than the symbol.
The colormaps for the model grids for shock velocity and log(\zzsun) are the same for all panels.
Two of the F24 models are shown, both of which having a pre-shock density of 10\,\cmthree;
the left panel gives the highest magnetic field B modeled by them, with 10\,$\mu$G,
and the right the lowest value, $10^{-4}$\,$\mu$G.
In the bottom panel, the horizontal dashed line 
corresponds to the AGN criterion of \nev/\neii\,$\geq 0.1$ defined by \citet{inami13}.
}
\label{fig:shocks}
\vspace{3\baselineskip}
\end{figure*}

\begin{figure*}[t!]
\includegraphics[width=\linewidth]{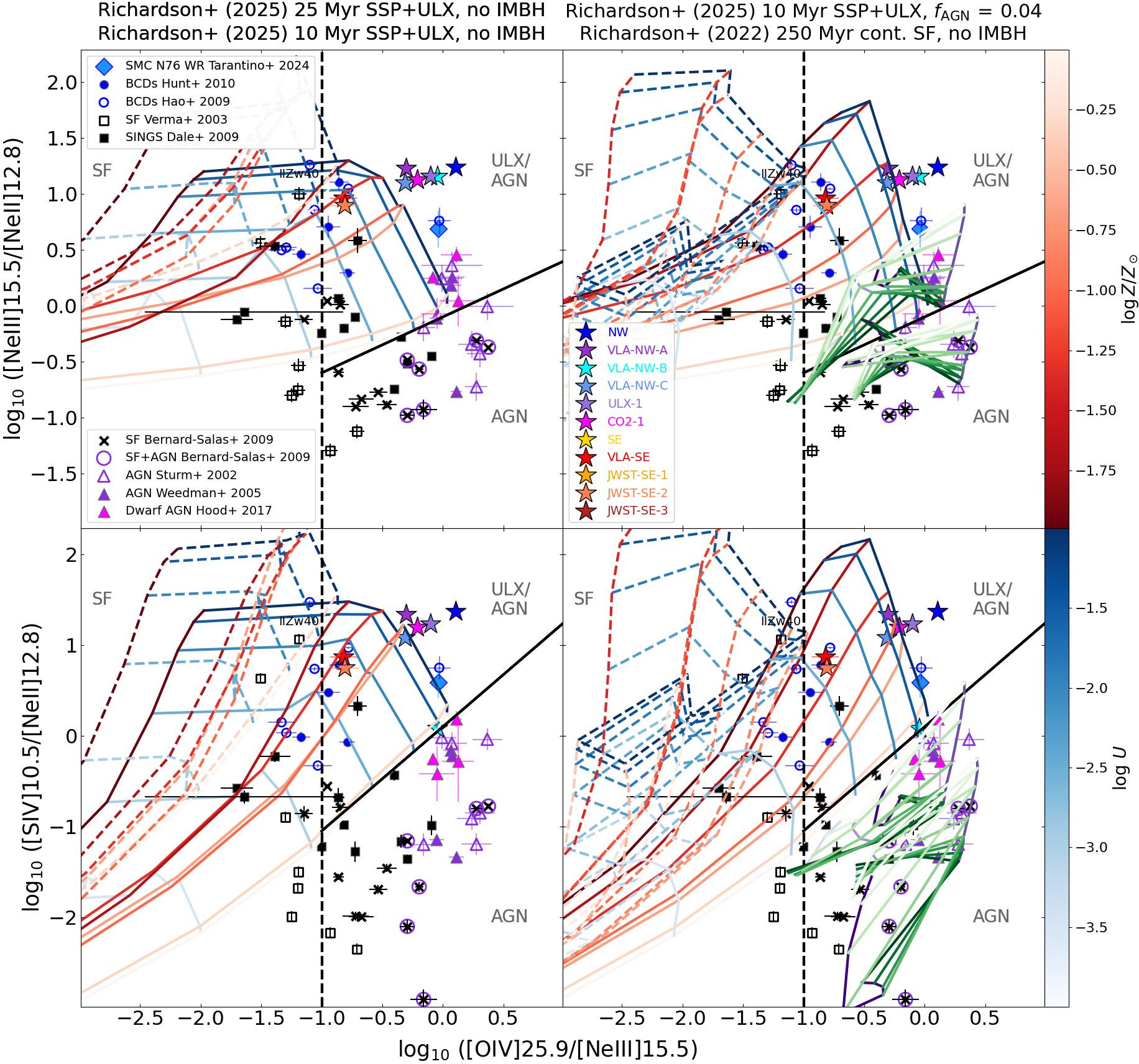}
\caption{The observed line ratios considered to be the strongest diagnostics by R25.
As in Figure \ref{fig:ion}, when the error bars are not evident, they are smaller than the symbol.
The demarcation lines shown from Eqs. (9) and (10) in R25 
separate stars$+$ULX excitation from that requiring an AGN,
and also clearly delineate shocks from photoionization from stars.
Left panel: the R25 models from Figure \ref{fig:ion} are shown as in Figure \ref{fig:ion}, 
with the 25 (10)\,Myr SSP+ULX given as solid (dashed) curves. 
Right panel: the F24 shock models are as in Figure \ref{fig:shocks} (with the same color coding, not shown here).
As in Figure \ref{fig:ion}, the R25 models with \fagn\,=\,0.04 are given as solid contours,
while the R22 models are dashed ones.
}
\label{fig:newratios}
\vspace{3\baselineskip}
\end{figure*}

In general, the R25 models are fairly successful
at reproducing the line ratios of the broader galaxy population.
The model grids show that increasing \siv/\siii\ is associated with increasing \logu, and
increasing \neiii/\neii\ with decreasing $Z$
(see color maps for the models in the right panels, the same for all panels in the plot).
\oiv/\siii\ also increases with decreasing $Z$ as expected, and with increasing \logu.
Moreover, the 25\,Myr models, with the maximum ULX contribution 
and the highest ionization parameter $U$ come very close to matching
the line ratios of \izw\ (left panels of Figure \ref{fig:ion}).
This is particularly true for the \nev/\neii\ ratio of \izw\
(bottom panel of Figure \ref{fig:ion}), where the most metal-poor
model curves with the highest $U$ are coincident with what is observed in \izw.
However, without an IMBH, the R22 CSF models lie somewhat far from the \nev/\neii\ ratios of observed AGN and some of the LIRGs,
mainly because the \neiii/\neii\ ratios tend to be somewhat over predicted.

The measured line ratios may reflect additional physical phenomena (e.g., shocks) that
are not described by photoionization models.
Thus, in Figure \ref{fig:shocks},
we examine the shock models by \citet[][hereafter F24]{flury24}.
These are based on \texttt{MAPPINGS V} \citep{sutherland17,sutherland18},
treating shocks as a time-dependent flow in the frame of the shock.
The precursor gas, the gas upstream of the shock,
has a given ionization balance, density, temperature, and composition.
The shock front is characterized by a velocity and magnetic field strength, B,
and gas beyond the shock front is collisionally excited to very high temperatures
($T\sim10^6$\,K) \citep[e.g.,][]{binette85,draine93,sutherland17}.
As this gas cools, high-ionization species such as \nev\
can persist in the gas even as it approaches lower temperatures ($T\sim10^4$\,K).
For more details see F24.
Metallicities in the models range from log(\zzsun) $-1.3$ to 0.3, so
may not be strictly appropriate for \izw;
nevertheless, our goal is to assess the differentiation of shocks
from photoionization with data spanning the broader galaxy population. 

In Figure \ref{fig:shocks},
the color maps indicate metallicity and shock velocity, and we have
limited the plots to pre-shock densities of 10\,\cmthree, but checked that this parameter
is not crucial for defining the range in predicted line ratios
(see also R25).
In the left panel, the magnetic field B has a value of 10\,$\mu$G,
the highest value modeled by F24, and in the right, $10^{-4}$\,$\mu$G, the lowest. 
The highest B field tends to be more successful than the lowest in reproducing the overall trends
in the line ratios, in particular for \oiv/\siii\ (middle panel of Figure \ref{fig:shocks}).
Shocks are also able to reproduce the observed range of \nev/\neii\ vs. \neiii/\neii\
for the overall galaxy populations (lower panel), but not for \izw\
which seems to be better approximated by the SSP$+$ULX populations of R25
(or the light-seed \fagn\,=\,0.04).

Finally, in Figure \ref{fig:newratios}, we compare a different set of line ratios 
with models.
These line ratios, \oiv/\neiii\ combined with \neiii/\neii\ and \siv/\neii, are proposed
by R25 to be the strongest diagnostics for separating stars$+$ULX and AGN contributions.
The R25 models shown in Figure \ref{fig:ion} are also given in the left panels
(no IMBH), and in the right, the R22 (no IMBH) and the R25 10\,Myr light-seed model with \fagn\,=\,0.04.
The two F24 shock models from Figure \ref{fig:shocks} are also given in the right
with the same color coding as in Figure \ref{fig:shocks}.
The demarcation lines correspond to Eqs. (9) and (10) in R25, as well
as the vertical line corresponding to an \oiv/\neiii\ ratio of 10\% to separate AGN/ULXs and SSPs.

Comparison of the models with the data in Figure \ref{fig:newratios}
shows that the demarcation proposed by R25
is reasonably accurate;
the observed line ratios of AGN in all panels fall where expected.
In the lower left panel the dwarf AGN \citep{hood17} fall cleanly in the AGN region,
while they lie at the border of AGN and ULX/AGN in the top left panel.
Interestingly, in the upper right panel of Figure \ref{fig:newratios}, 
the shock models span the ULX/AGN and pure AGN regimes,
but in the lower right, they fall cleanly in the AGN category.
The line ratios for \izw\ are clearly well away from the AGN regime,
and lie, as expected from Figure \ref{fig:ion}, in the region attributable to ULX/AGN.

Overall, this comparison of the models with broad galaxy populations and \izw\ suggests that:
(1) as stressed by R25, no single line ratio is a foolproof diagnostic of excitation mechanisms; 
(2) more than one physical mechanism, such as photoionization and shocks, 
can shape the properties of the ISM in galaxies, at least judging from our 
emission-line ratio analysis. 
Furthermore, for \izw, comparison of the models with the observed line ratios
clearly indicates that the high excitation of the gas, particularly in the NW,
is compatible with the hard RF produced by low-metallicity ULXs.
Nevertheless, the ambiguity with these and the IMBH contribution with light seeds 
and \fagn\,=\,0.04, does not allow ruling out a small contribution from a low-mass IMBH
\citep[see also][]{mingozzi25}.

\begin{figure*}[t!]
\hbox{
\includegraphics[width=0.25\linewidth]{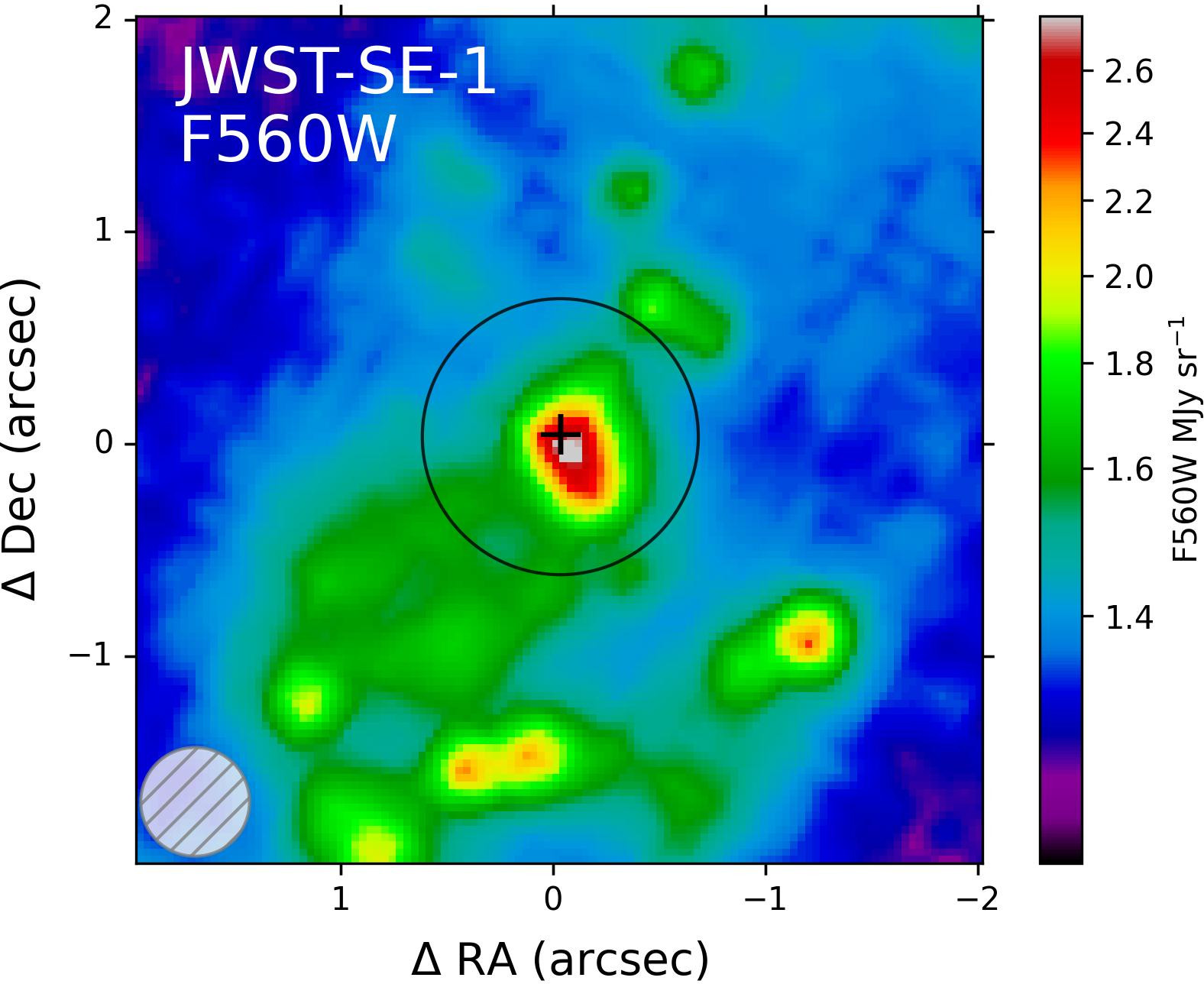}
\includegraphics[width=0.25\linewidth]{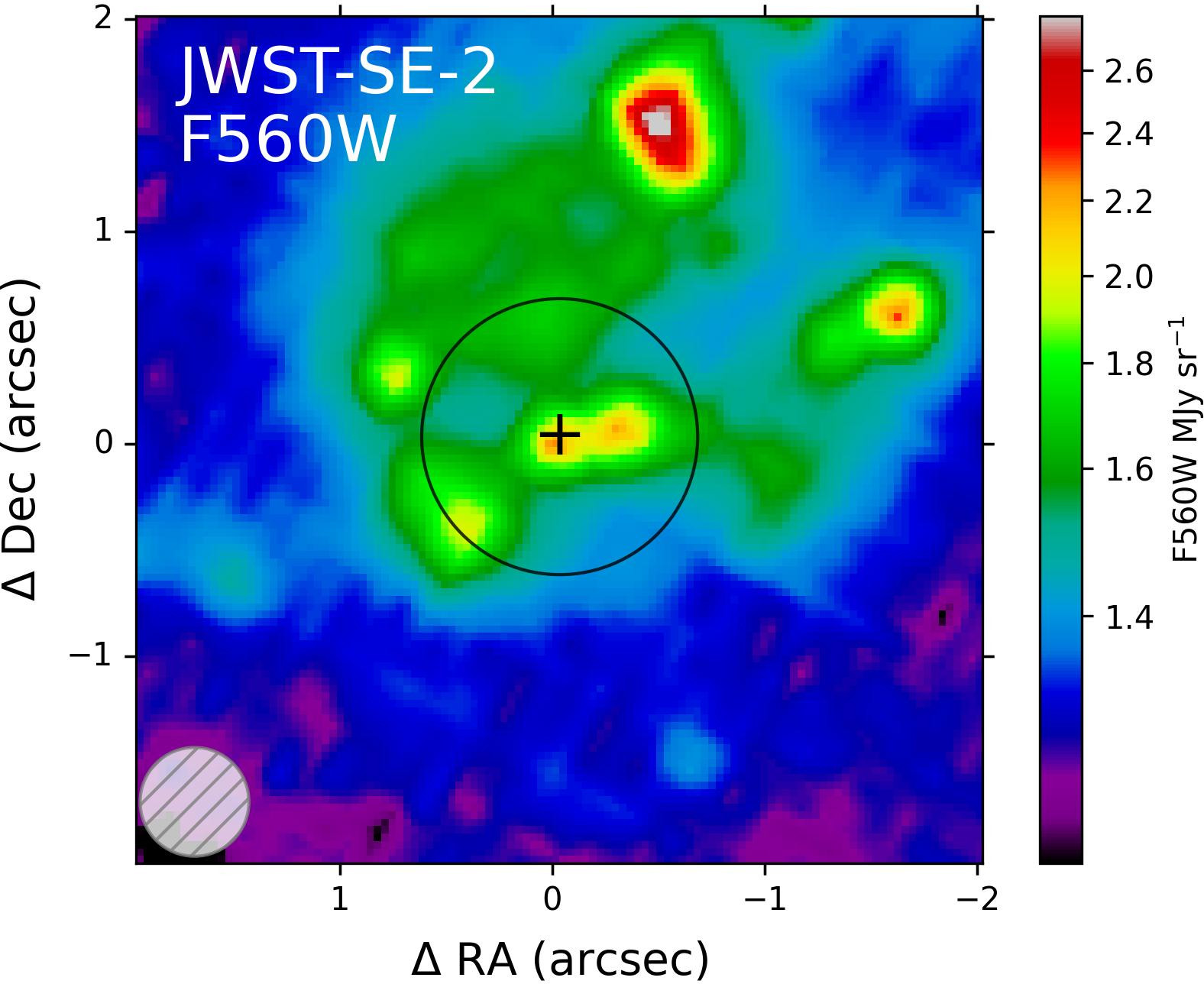}
\includegraphics[width=0.25\linewidth]{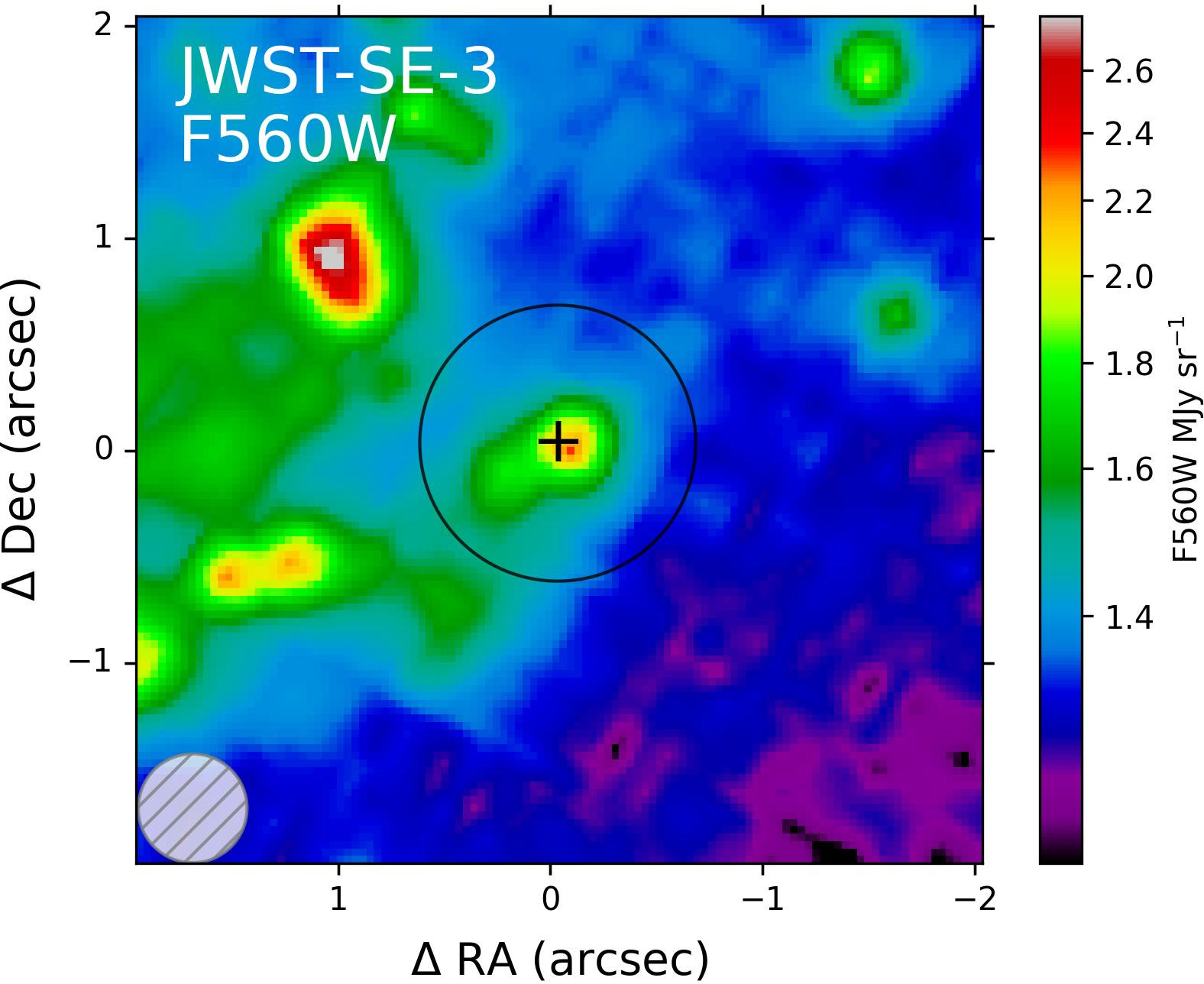}
\includegraphics[width=0.25\linewidth]{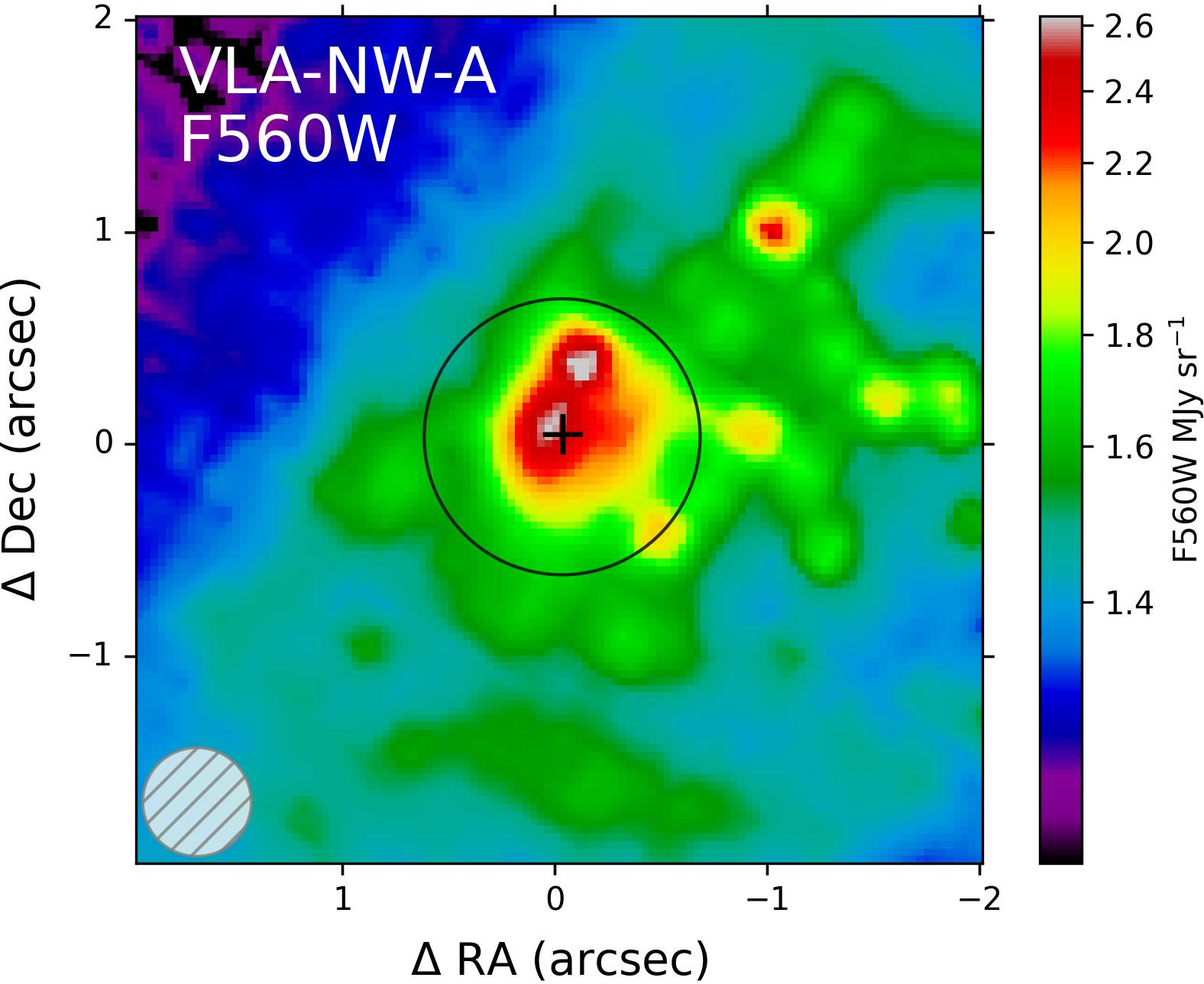}
}
\hbox{
\includegraphics[width=0.25\linewidth]{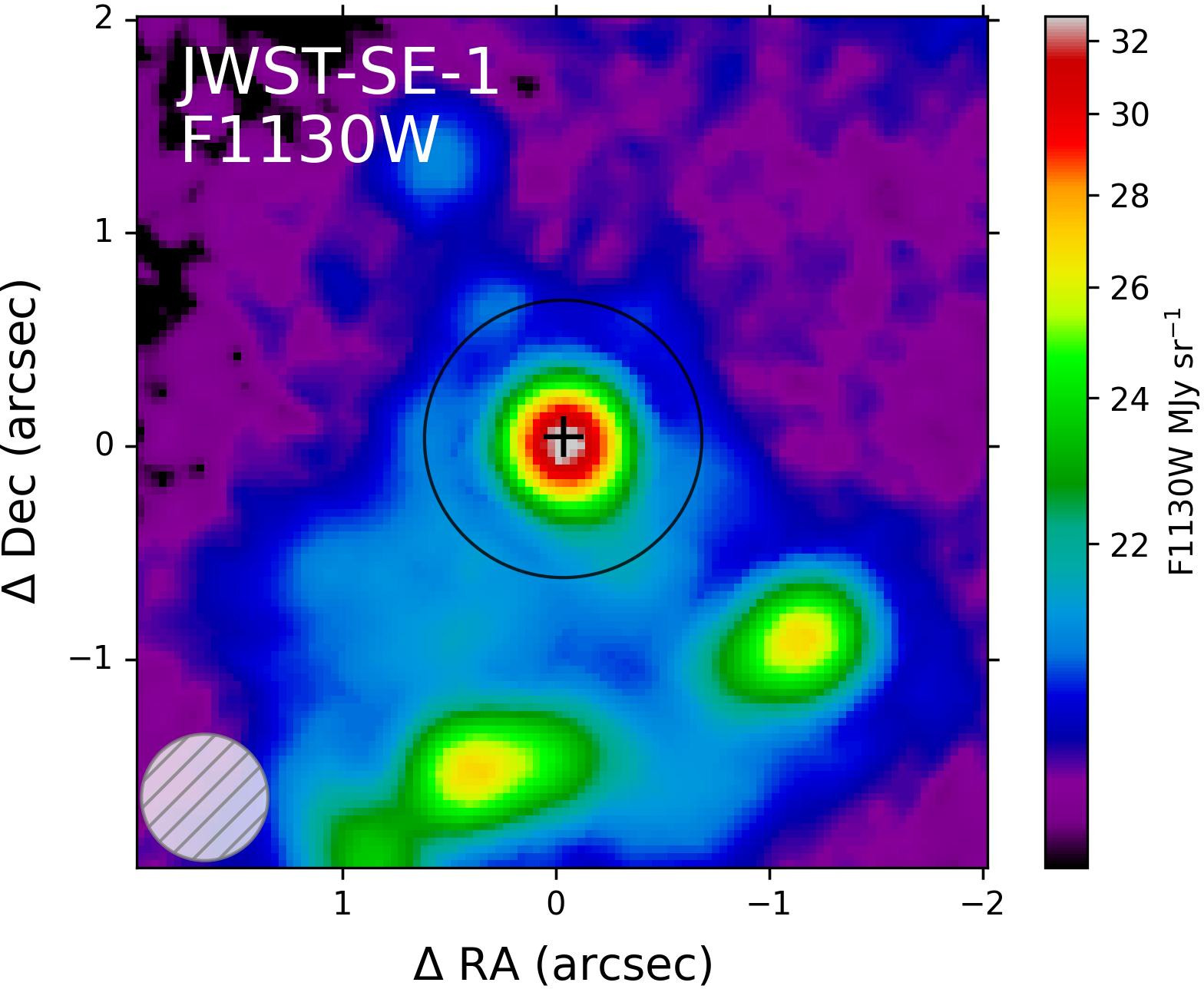}
\includegraphics[width=0.25\linewidth]{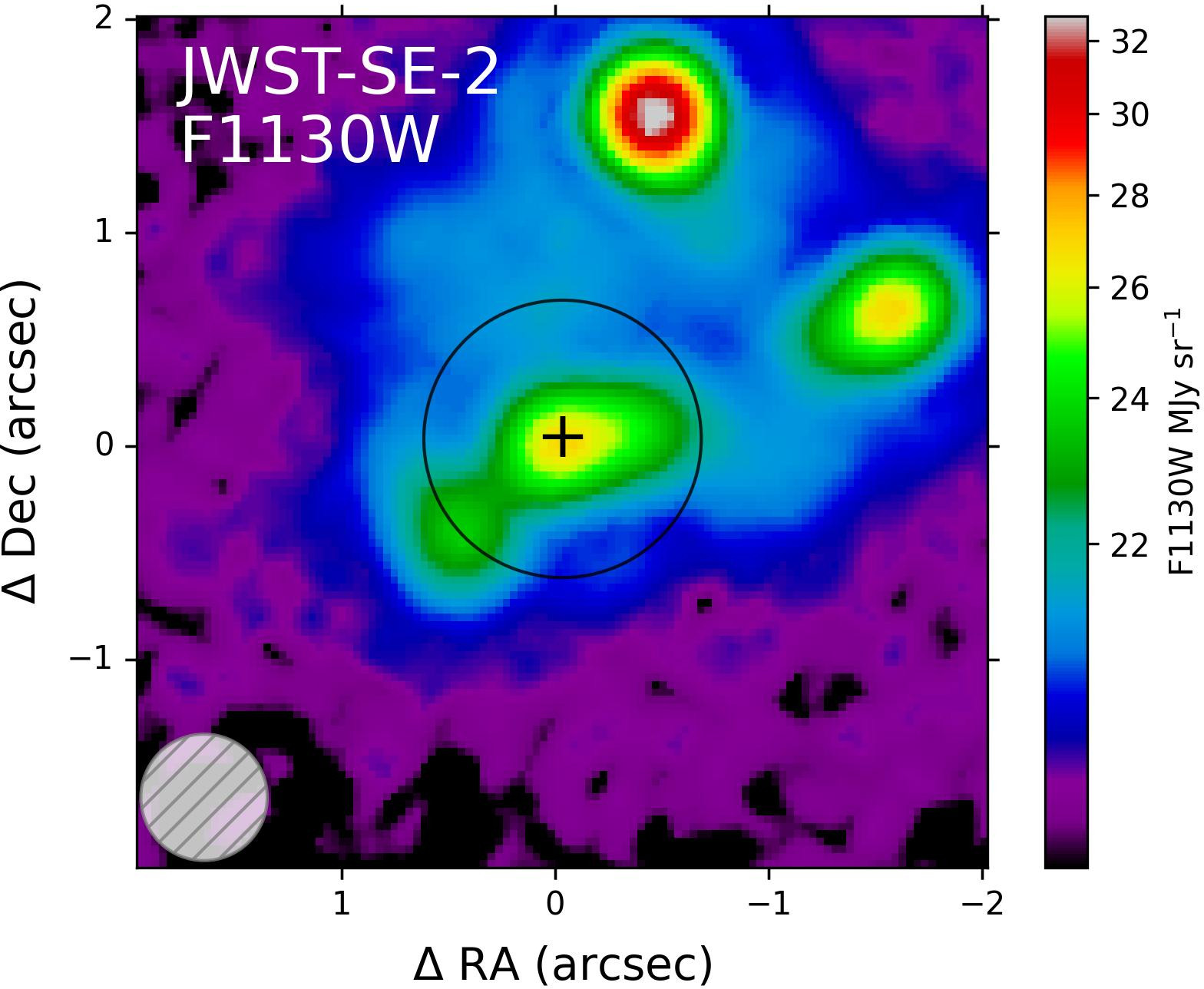}
\includegraphics[width=0.25\linewidth]{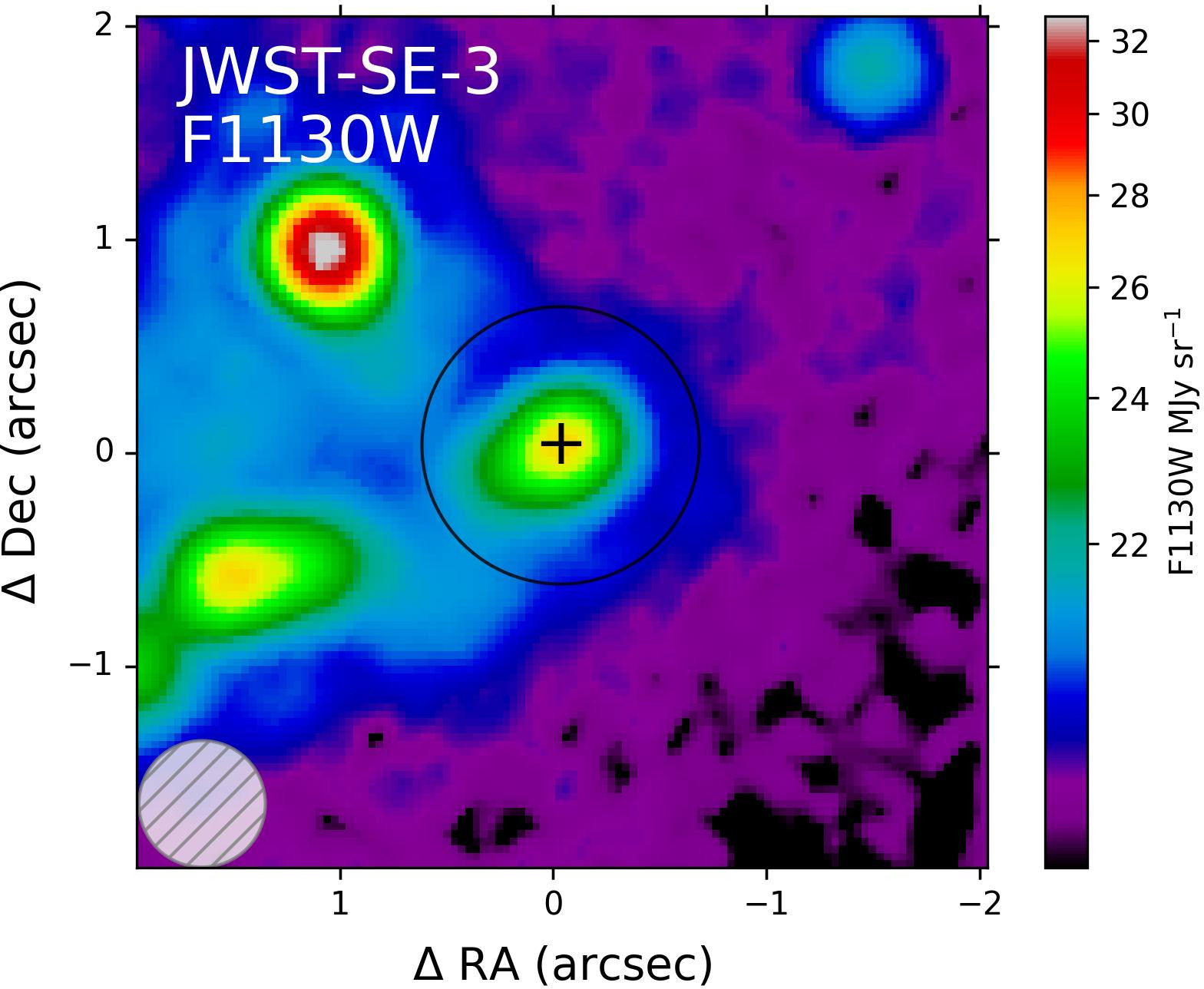}
\includegraphics[width=0.25\linewidth]{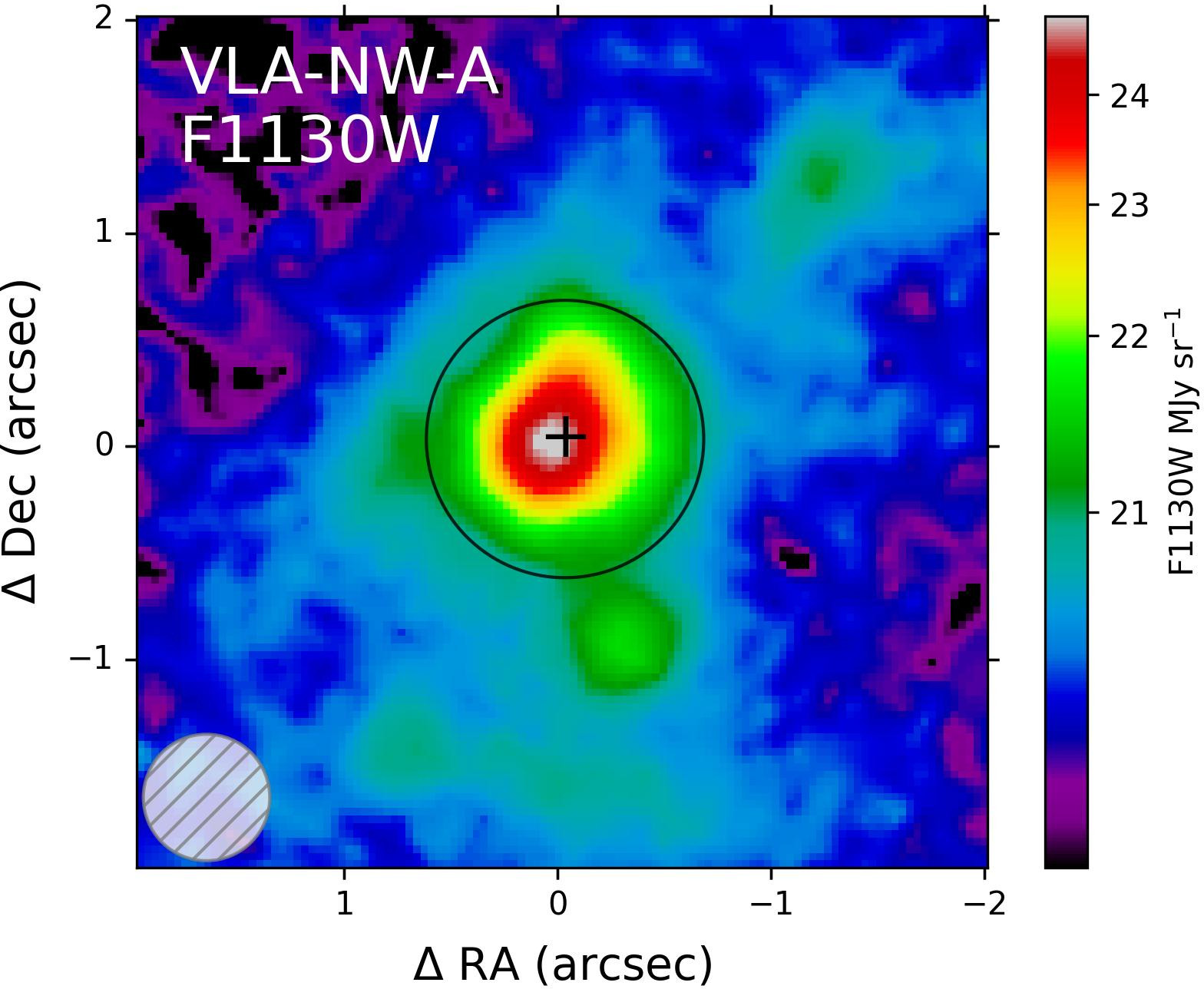}
}
\hbox{
\includegraphics[width=0.25\linewidth]{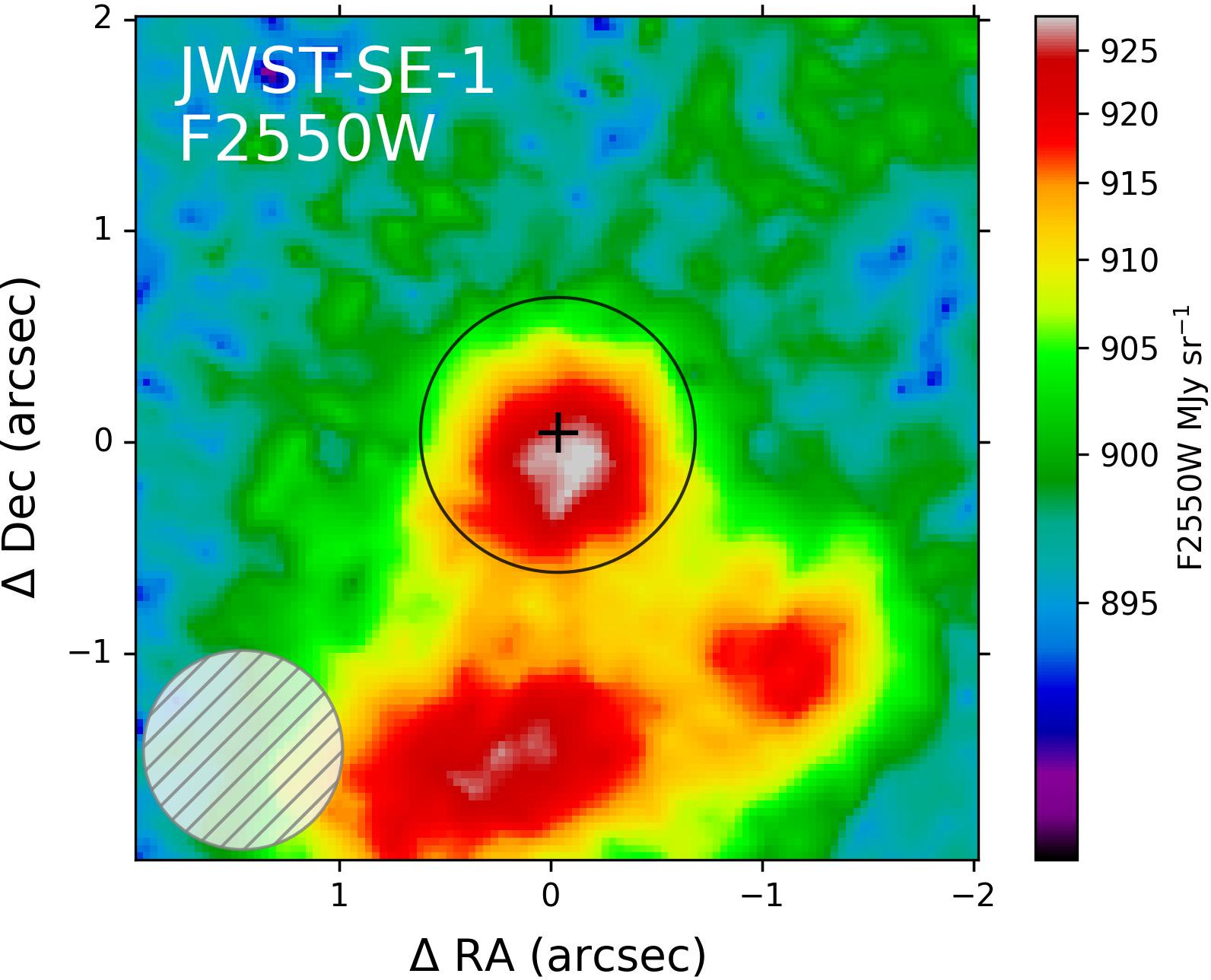}
\includegraphics[width=0.25\linewidth]{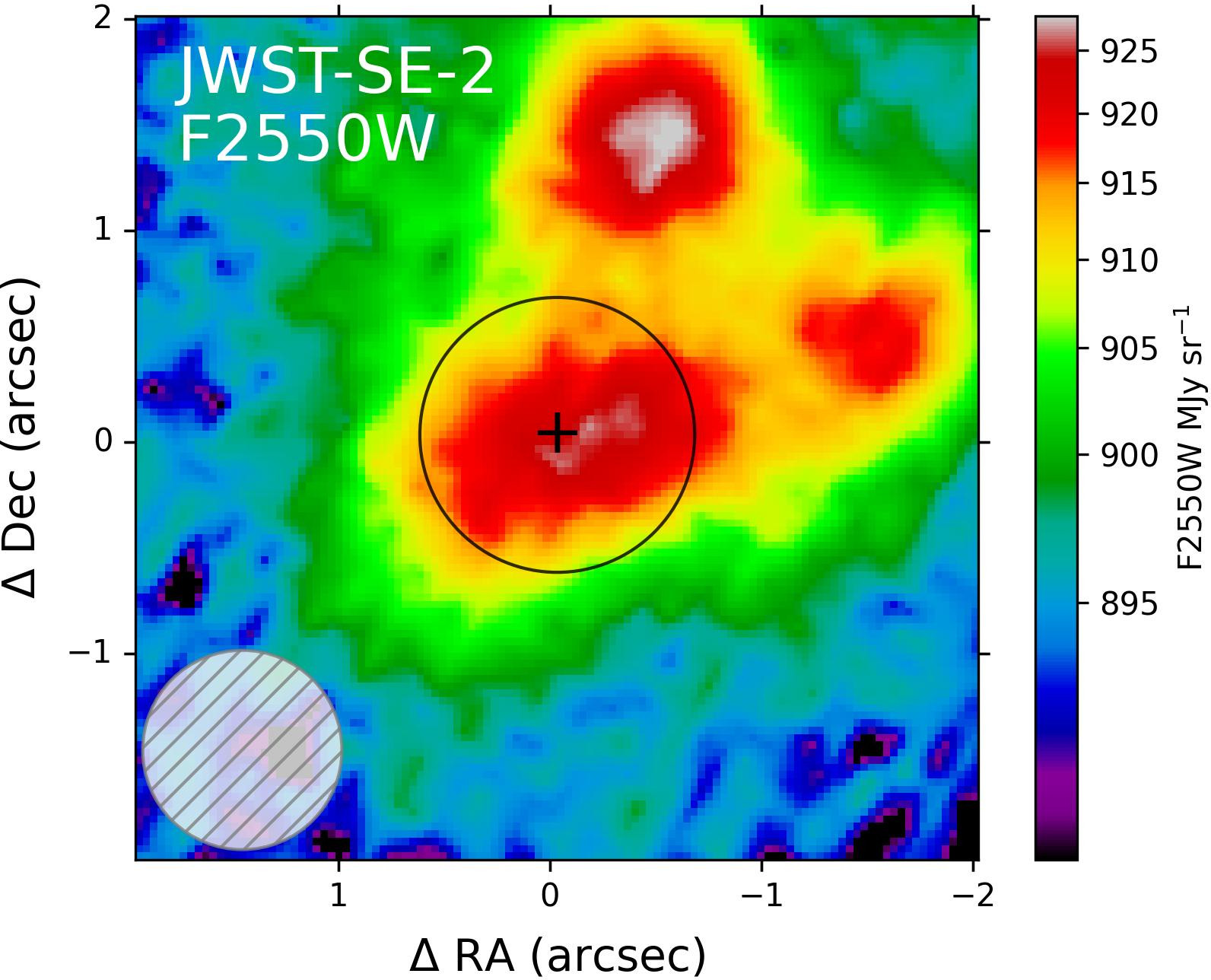}
\includegraphics[width=0.25\linewidth]{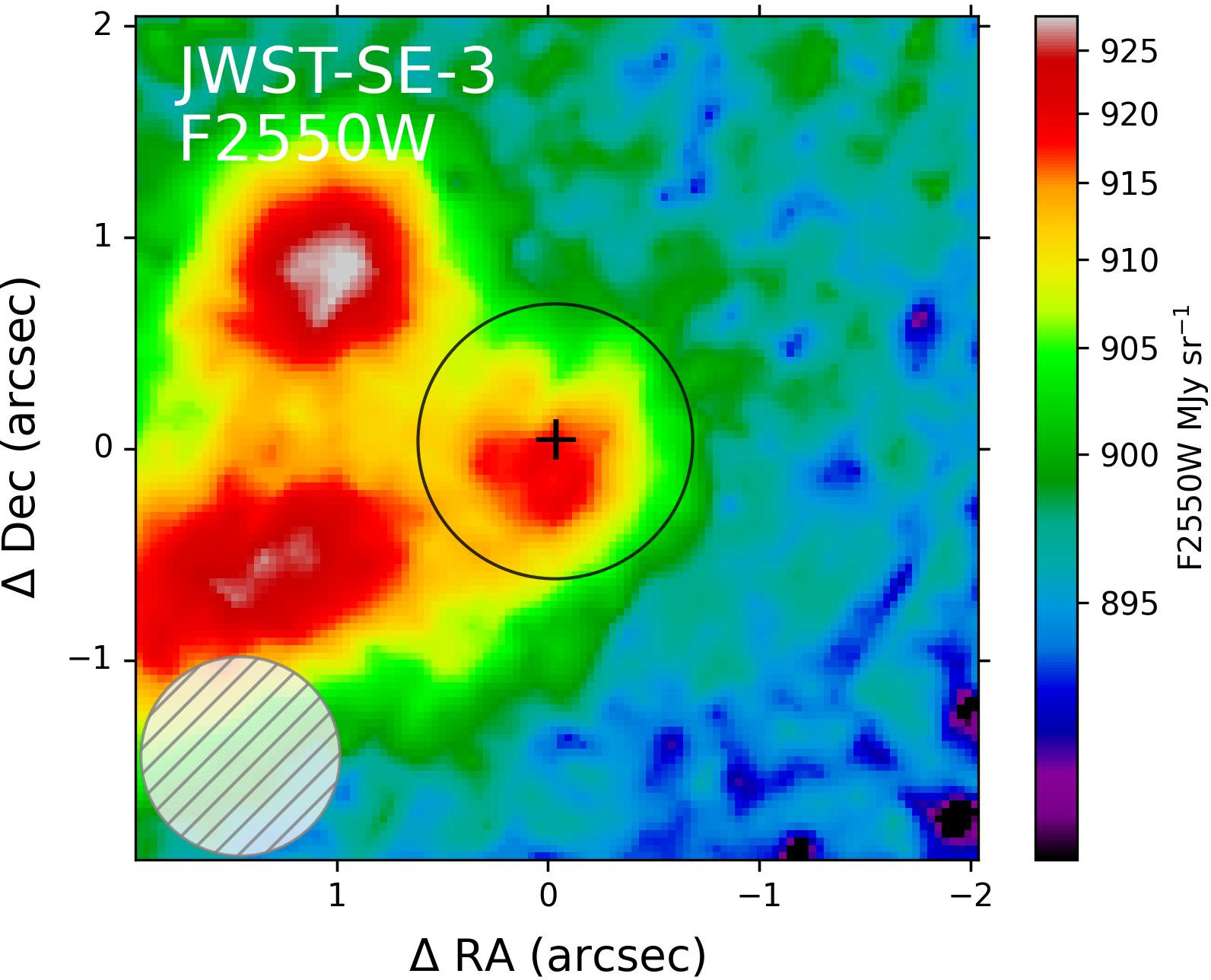}
\includegraphics[width=0.25\linewidth]{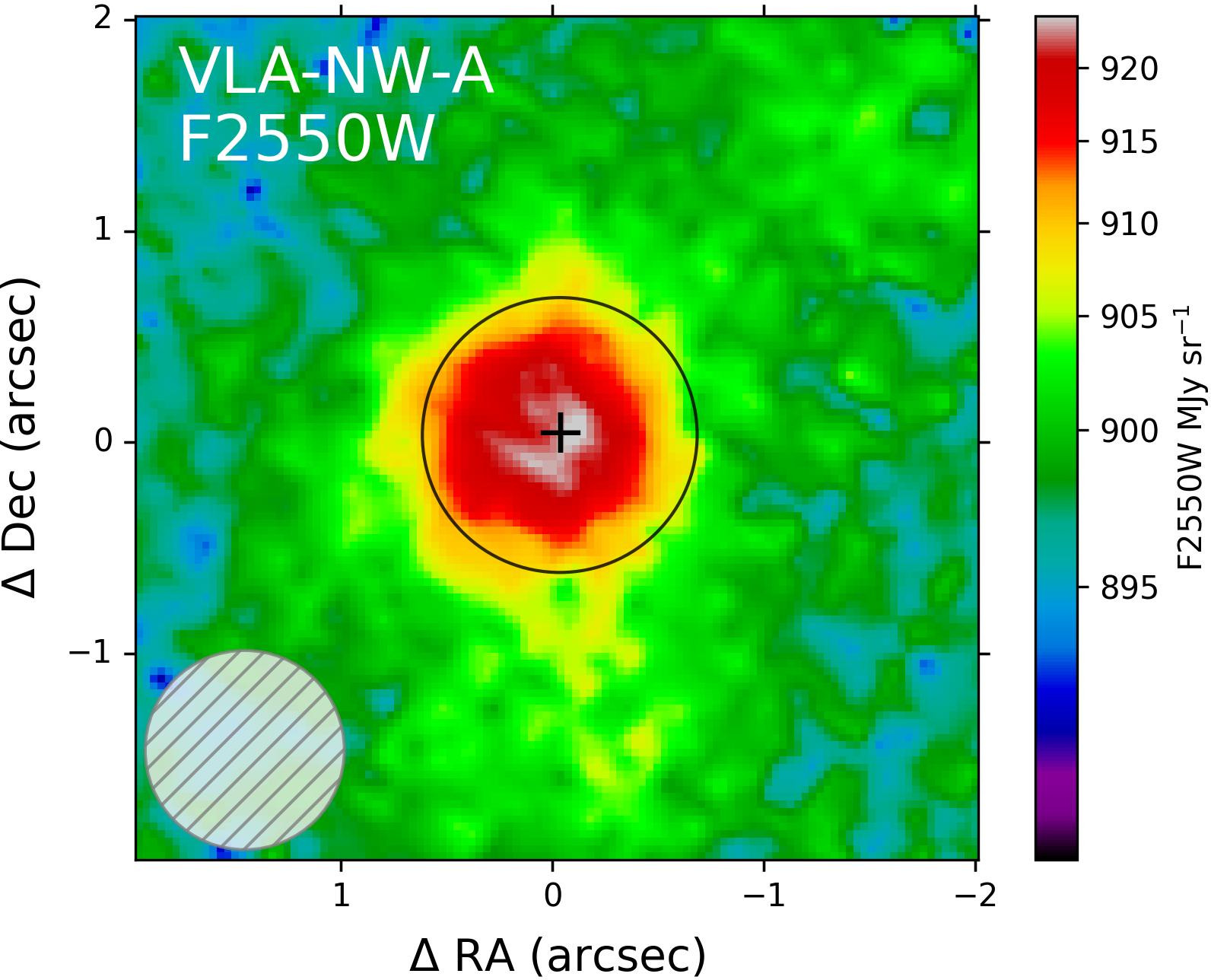}
}
\hbox{
\includegraphics[width=0.25\linewidth]{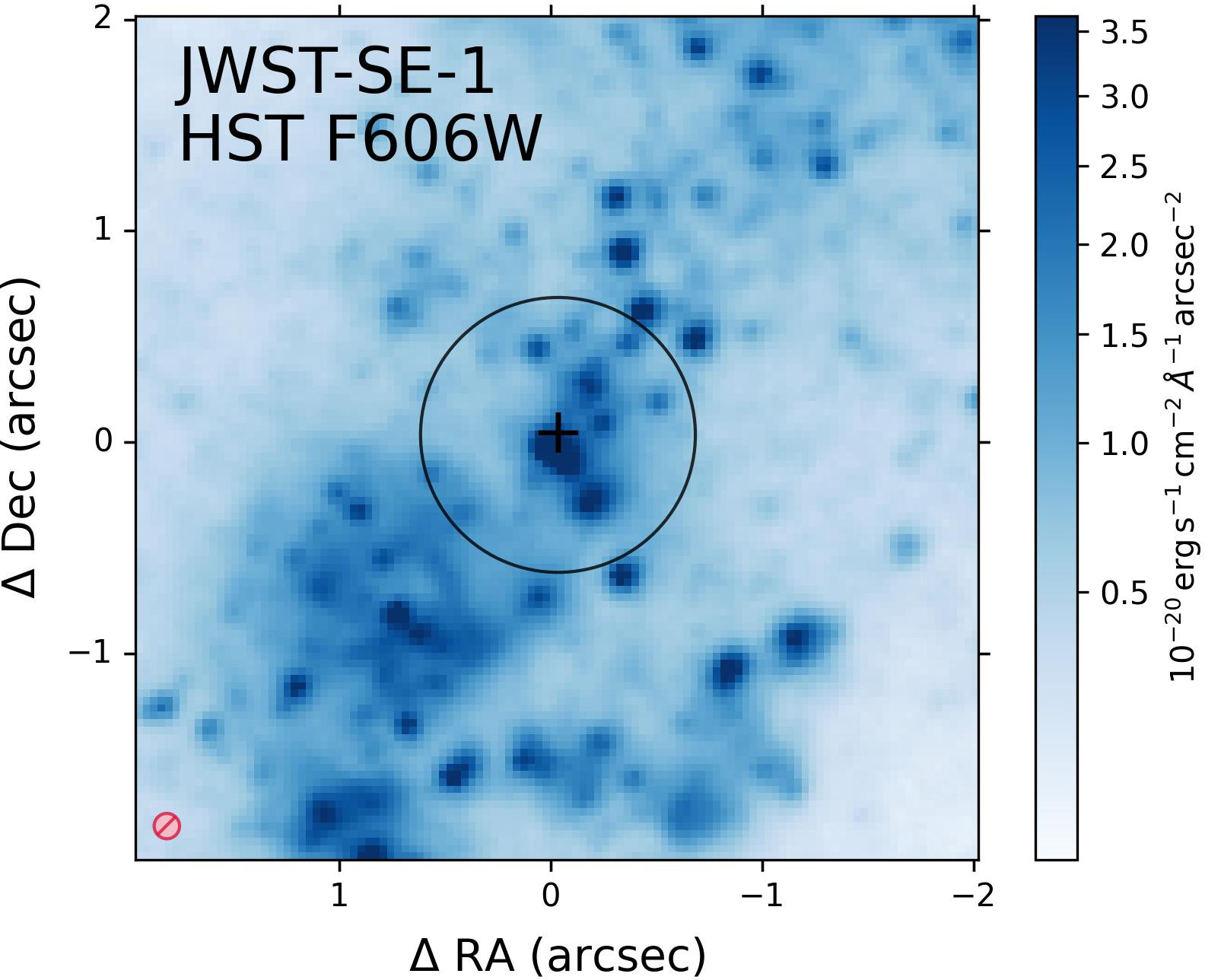}
\includegraphics[width=0.25\linewidth]{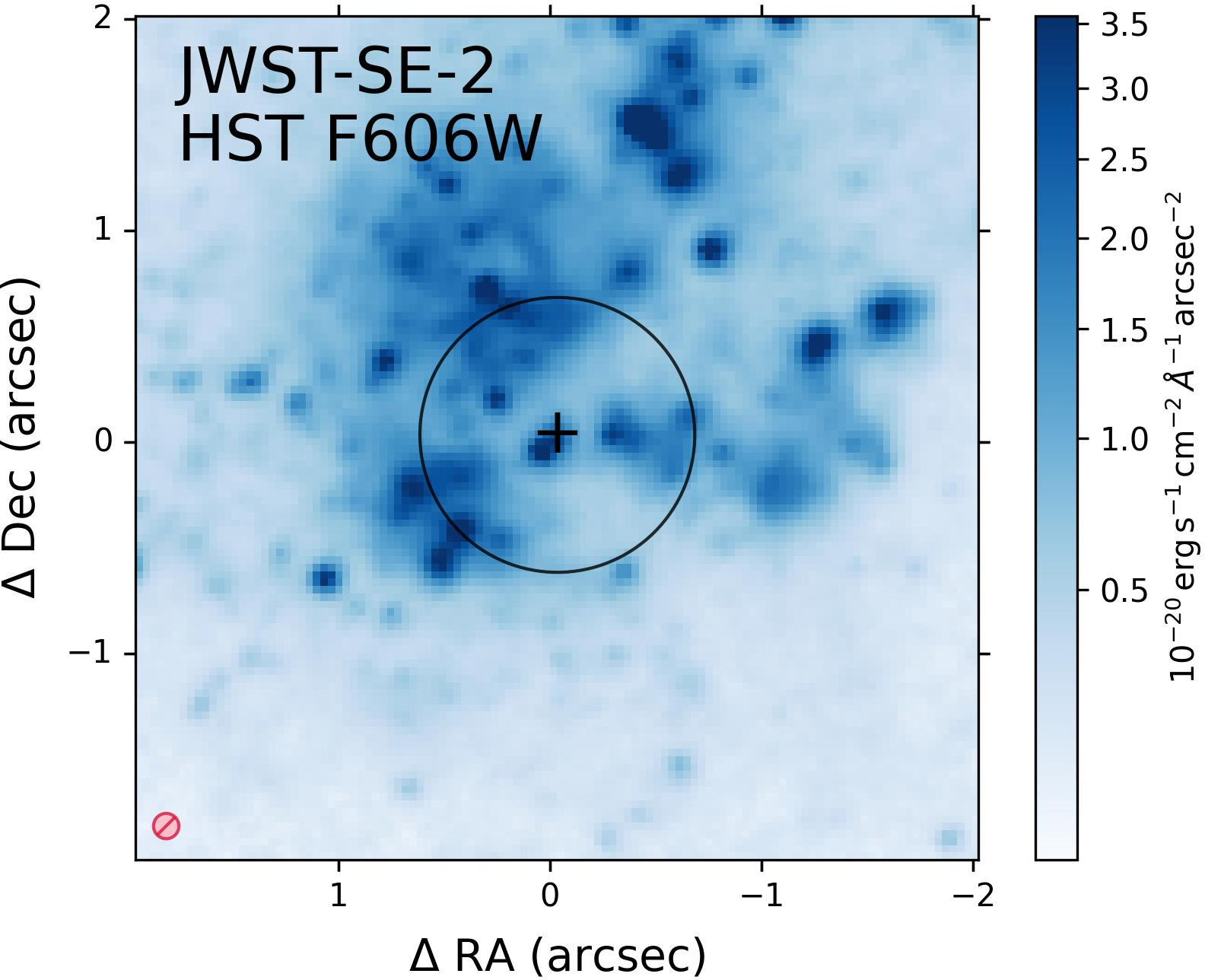}
\includegraphics[width=0.25\linewidth]{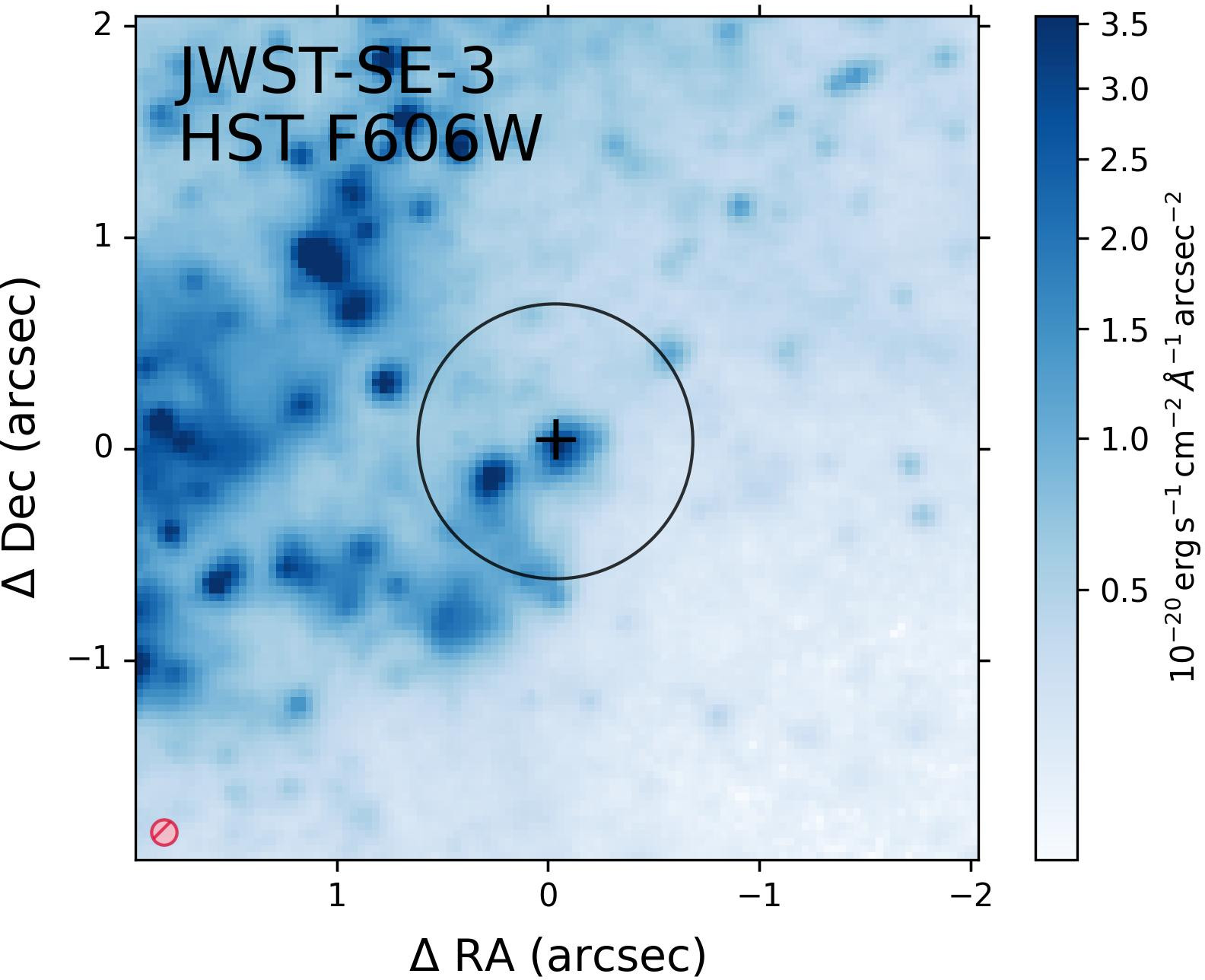}
\includegraphics[width=0.25\linewidth]{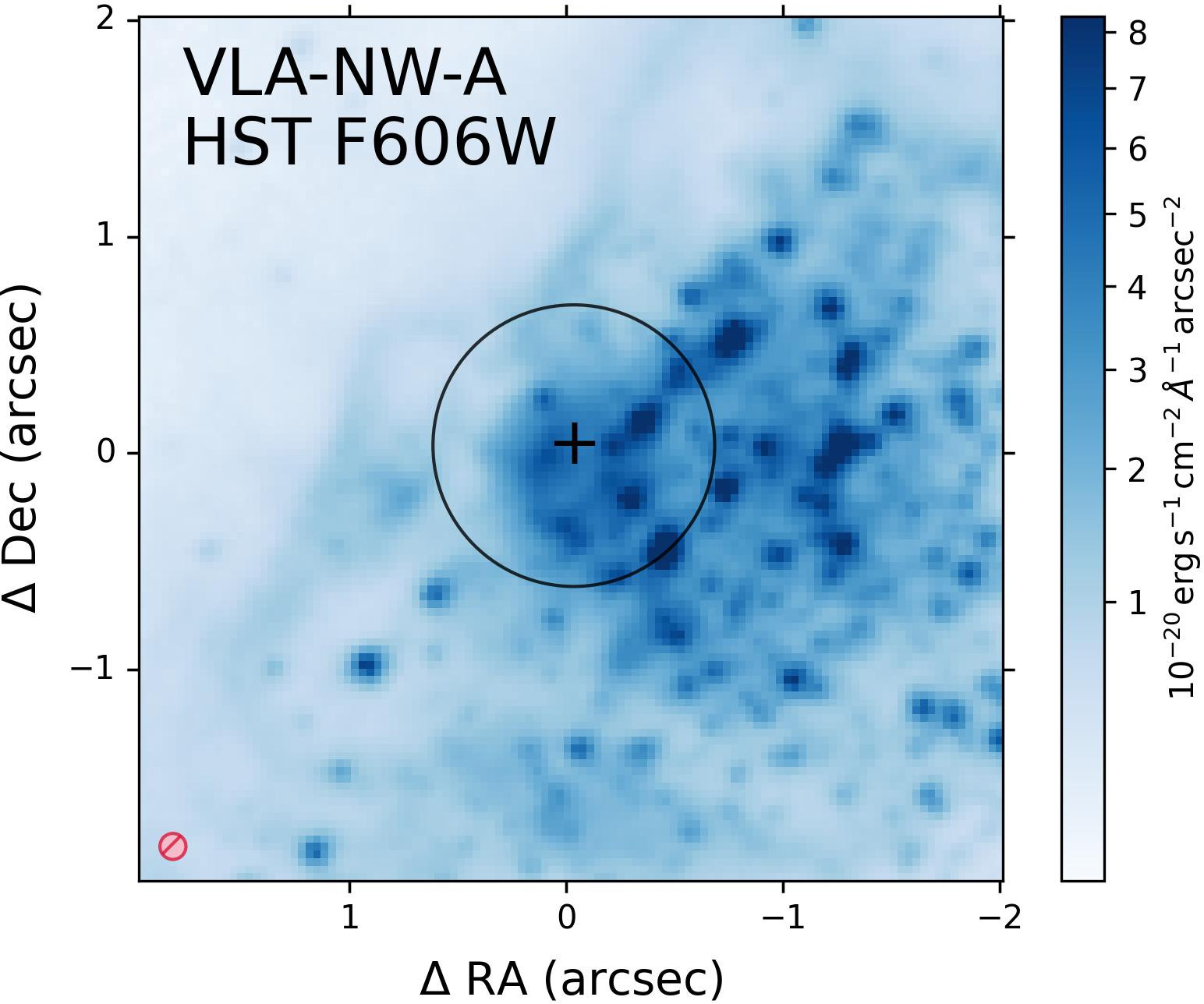}
}
\caption{4\arcsec$\times$4\arcsec\ cutouts of the MIRI 14\,\micron\ continuum sources: 
from left to right JWST-SE-1, JWST-SE-2, JWST-SE-3, and VLA-NW-A, and from top to bottom
F560W, F1130W, F2550W, and \hst/F606W.
The images are centered on the respective 14\,\micron\ emission peaks (Table \ref{tab:aper}),
shown by a $+$, and surrounded by the 0\farcs65-radius aperture used for spectroscopic extraction
shown by a concentric circle.
The FWHM of the PSF is shown as a hatched circle in the lower left corner of each panel.
}
\label{fig:continuum}
\end{figure*}

\section{MIRI continuum sources \label{sec:continuum}}

As described in Sect. \ref{sec:spectra}, 
four sources, only one of which was previously known, 
were identified as compact continuum emitters by visual inspection
of the MIRI spectral cubes.
Here we describe their properties in terms of size and estimated age.
Close-ups of these four objects from our MIRI imaging are shown in Figure \ref{fig:continuum}, with
F560W, F1130W, F2550W from the top row to the penultimate;
the bottom row gives a comparison with \hst\ ACS/F606W.
The spatial resolution of the images is shown as a hatched
circle in the lower left corner of each panel corresponding to
the FWHM of the PSF.\footnote{The HST PSF has been taken from 
\url{https://hst-docs.stsci.edu/acsihb/chapter-5-imaging/5-6-acs-point-spread-functions.}}
When Figure \ref{fig:continuum} is compared to Figure \ref{fig:apertures}, it is
easily appreciated that neither VLA-SE nor the central position of the SE star complex
correspond to peaks in the continuum maps.

\begin{figure*}[t!]
\includegraphics[width=\linewidth]{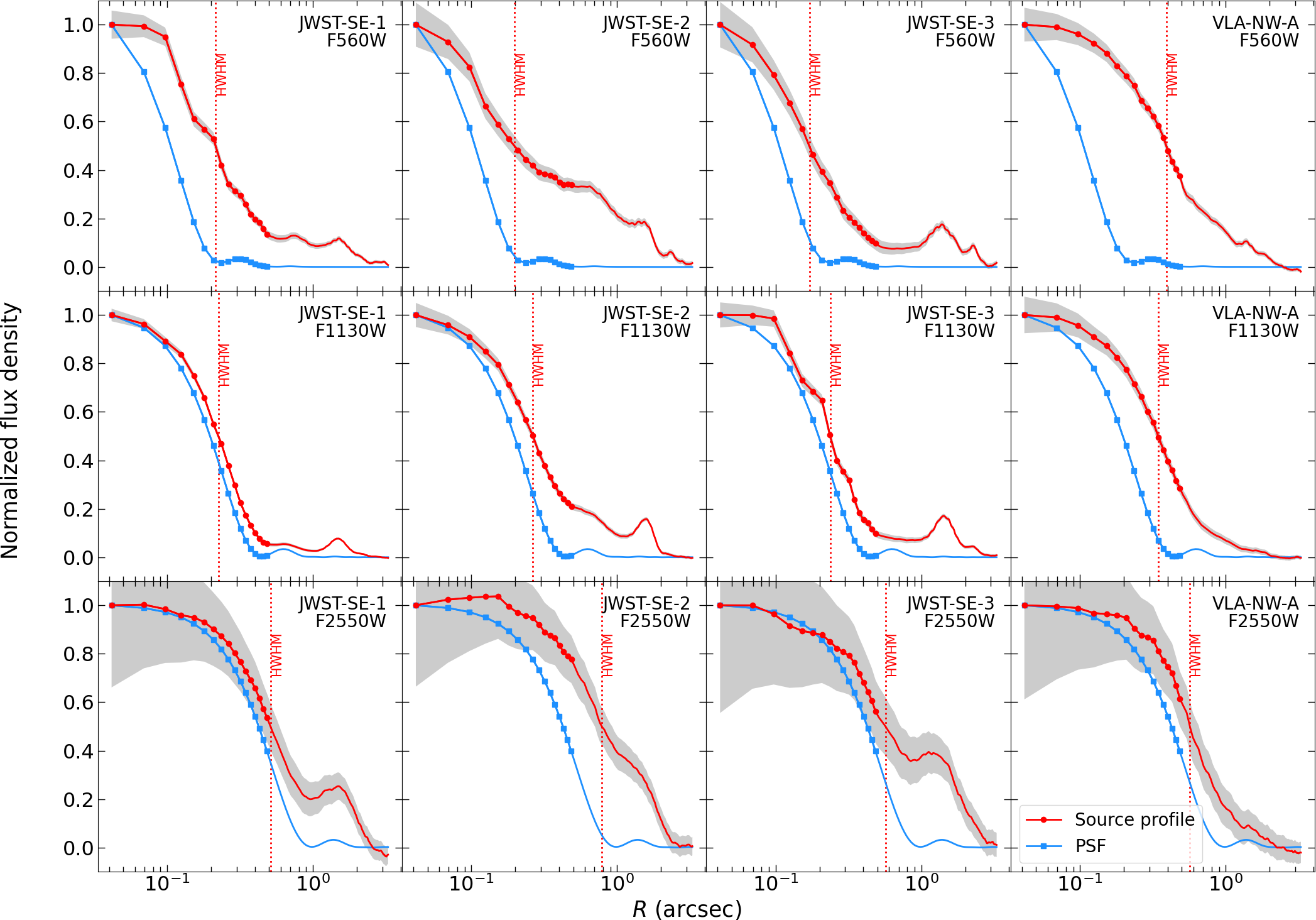}
\caption{Radial profiles centered on the four continuum sources in the three MIRI imaging filters.
The source profiles are shown as red curves, 
and the PSF is with blue curves, as shown in the legend in the lower right panel.
The source profiles have been resampled to the same pixel scale
as the PSFs, taken from the MIRI website;
see text for more details.
All profiles are normalized to the peak amplitude, and
individual radial sampling points are shown up to $R\,=\,$0\farcs5.
The HWHM of the source profile is given by a vertical dotted line.
The gray shaded regions illustrate the 1$\sigma$ uncertainties, 
calculated from the uncertainties in the source profile and the global background.
}
\label{fig:profiles}
\end{figure*}

\begin{table*}
\caption{Aperture photometry and sizes of continuum sources \label{tab:continuum}}
\begin{center}
\begin{tabular}{lcccccccccccc}
\hline
\hline
\multicolumn{1}{c}{\rule{0pt}{3ex} Region} &
\multicolumn{3}{c}{Flux (mJy)$^a$} &
\multicolumn{3}{c}{HWHM (arcsec)$^b$} & 
\multicolumn{3}{c}{Deconvolved FWHM$^c$} \\
&
\multicolumn{1}{c}{F560W} &
\multicolumn{1}{c}{F1130W} &
\multicolumn{1}{c}{F2550W} &
\multicolumn{1}{c}{F560W} &
\multicolumn{1}{c}{F1130W} &
\multicolumn{1}{c}{F2550W} &
\multicolumn{1}{c}{F560W} &
\multicolumn{1}{c}{F1130W} &
\multicolumn{1}{c}{F2550W} \\
\hline
\\ 
JWST-SE-1 & 0.0101   & 0.0754 & 0.5615     & 0.215   & 0.225   & 0.512 & 0\farcs375 (33.1\,pc) & 0\farcs216 (19.1\,pc) & 0\farcs571 (50.3\,pc) \\
          & (0.0009) & (0.0023) & (0.0239) & (0.007) & (0.004) & (0.08)\\
JWST-SE-2 & 0.0103   & 0.0677 & 0.7229     & 0.198   & 0.265   & 0.786 & 0\farcs333 (29.4\,pc) & 0\farcs352 (31.1\,pc) & 1\farcs322 (116.6\,pc) \\
          & (0.0009) & (0.0022) & (0.0241) & (0.026) & (0.009)  & (0.14)\\
JWST-SE-3 & 0.0048    & 0.0466 & 0.4528    & 0.171   & 0.237   & 0.568 & 0\farcs266 (23.5\,pc) & 0\farcs261 (23.0\,pc) & 0\farcs752 (66.4\,pc)\\
          & (0.0006) & (0.0019) & (0.0237) & (0.013) & (0.006) & (0.16)\\
VLA-NW-A  & 0.0163    & 0.0551 & 0.5461    & 0.392   & 0.344   & 0.569 & 0\farcs753 (66.5\,pc) & 0\farcs564 (49.8\,pc) & 0\farcs755 (66.7\,pc) \\
          & (0.0011) & (0.0020) & (0.0239) & (0.011) & (0.013) & (0.087)\\
\\
\hline
\hline
\end{tabular}
\end{center}
$^a$~Circular-aperture photometry with radii of 0\farcs65, with local background subtracted. 
The formal photometric uncertainties are given in the subsequent rows enclosed by parentheses.
More details are given in the text.\\
$^b$~The uncertainties in the HWHM determinations are given in the subsequent rows enclosed by parentheses.\\
$^c$~Deconvolved FWHM are calculated by subtracting in quadrature the PSF FWHM,
taking the square root, and converting to parsecs at the distance of \izw, 18.2\,Mpc. 
Measured HWHMs for the PSFs are 0\farcs2562, 0\farcs2962, 0\farcs4712, for F560W, F1130W, and F2550W, respectively.\\
\\
\end{table*}

We estimated the sizes of the continuum sources in the MIRI images by examining radial profiles
and comparing them with the profile of the PSF at that wavelength.\footnote{We have adopted the MIRI imaging PSFs from
\url{https://stsci.app.box.com/v/jwst-simulated-psf-library/folder/174727061405}.
In our analysis, we use PSFs with the oversampling, $\times 4$ the detector pixel scale.}
Figure \ref{fig:profiles} shows these profiles within
concentric circular apertures centered on the positions in Table \ref{tab:aper}.
The extraction of radial profiles and the photometry described
below was performed using \texttt{Astropy/Photutils} \citep{larry_bradley_2024_13989456}.
We have estimated empirically the size of these sources in the three MIRI images by 
calculating the half-width half-maximum (HWHM) from the profiles, and comparing
it to the HWHM of the PSF.
Because tests showed that results were more reliable with the oversampled PSF
($\times 4$ of the nominal pixel scale), we also extracted the profiles
at the same oversampled pixel scale.
The profiles have been normalized to have a maximum of unity;
the vertical dotted line shows the HWHM corresponding to the 50\% level,
relative to this normalized maximum.
The procedure was repeated in the same way for the PSF profiles;
the PSF FWHMs determined from our procedures are, on average,
5\% larger than those given in the \jwst\ documentation.\footnote{See
\url{https://jwst-docs.stsci.edu/jwst-mid-infrared-instrument/miri-performance/miri-point-spread-functions\#gsc.tab=0}}
This is reasonable agreement, so that the application of our algorithms 
to both the PSF and source profiles should ensure consistent results.
The intrinsic source sizes have been calculated by subtracting in quadrature the PSF FWHM from
the source FWHM, taking the square root, then converting to parsecs.
The results from our fits are reported in Table \ref{tab:continuum},
where the uncertainties are given in parentheses in the following row.

Deconvolved source diameters (FWHM) of the continuum sources
in \izw\ range from $\sim 20$\,pc to $\ga 70$\,pc at F560W,
and from $\sim 20$\,pc to $\sim 50$\,pc at F1130W;
they are slightly larger at F2550W.
JWST-SE-1 and JWST-SE-3 are only slightly resolved at F1130W, with deconvolved diameters $\sim 20$\,pc,
but more resolved and slightly broader at F560W.
The complexity of the region around JWST-SE-2 at F2550W enlarges its deconvolved size
to $\ga 100$\,pc, but, interestingly, at F1130W, its emission is more concentrated and slightly resolved.
The deconvolved dimensions $\sim 50-70$\,pc of VLA-NW-A in all filters tend to be 
slightly larger (except for JWST-SE-2, F2550W) than the other three continuum sources.

To further investigate the nature of these continuum sources, we have compared
photometry of the sources in the three MIRI filters 
with SED templates of star clusters as a function of age by \citet{whitmore25}.
Although such a comparison is not ideal, mainly because the templates are derived from
NGC\,628, a spiral galaxy with roughly Solar metallicity,
it can place the properties of the sources in \izw\ in a broader context.
To this end, we performed aperture photometry from the MIRI images on each of the continuum sources,
and compared it with the integrated flux from the profile;
the total fluxes agree to within 10\%.
However, because the profiles extend to radii of $\ga$\,2\arcsec\
(see Figure \ref{fig:continuum}),
we preferred to analyze the photometry within a circular aperture of radius 0\farcs65,
the same as the spectral extraction aperture. 
Local background in an annulus with radius of 2\arcsec\ and width of 0\farcs2
was subtracted before calculating the flux within the aperture.
Table \ref{tab:continuum} reports the photometry and the associated uncertainties.
Although the small aperture used here suffers from losses at the longest wavelength
(F2550W), it is a reasonable comparison with the \citet{whitmore25} templates,
since they use small-aperture photometry for the derivation of the templates, 
with MIRI apertures corresponding to 50\% encircled energy for point sources 
\citep[radius of 0\farcs42 at F2100W, see][]{rodriguez25}.
We have also checked the SED comparison with photometry for our continuum sources
measured in larger apertures, and the age classification is the same.

\begin{figure}[t!]
\hbox{
\includegraphics[width=\linewidth]{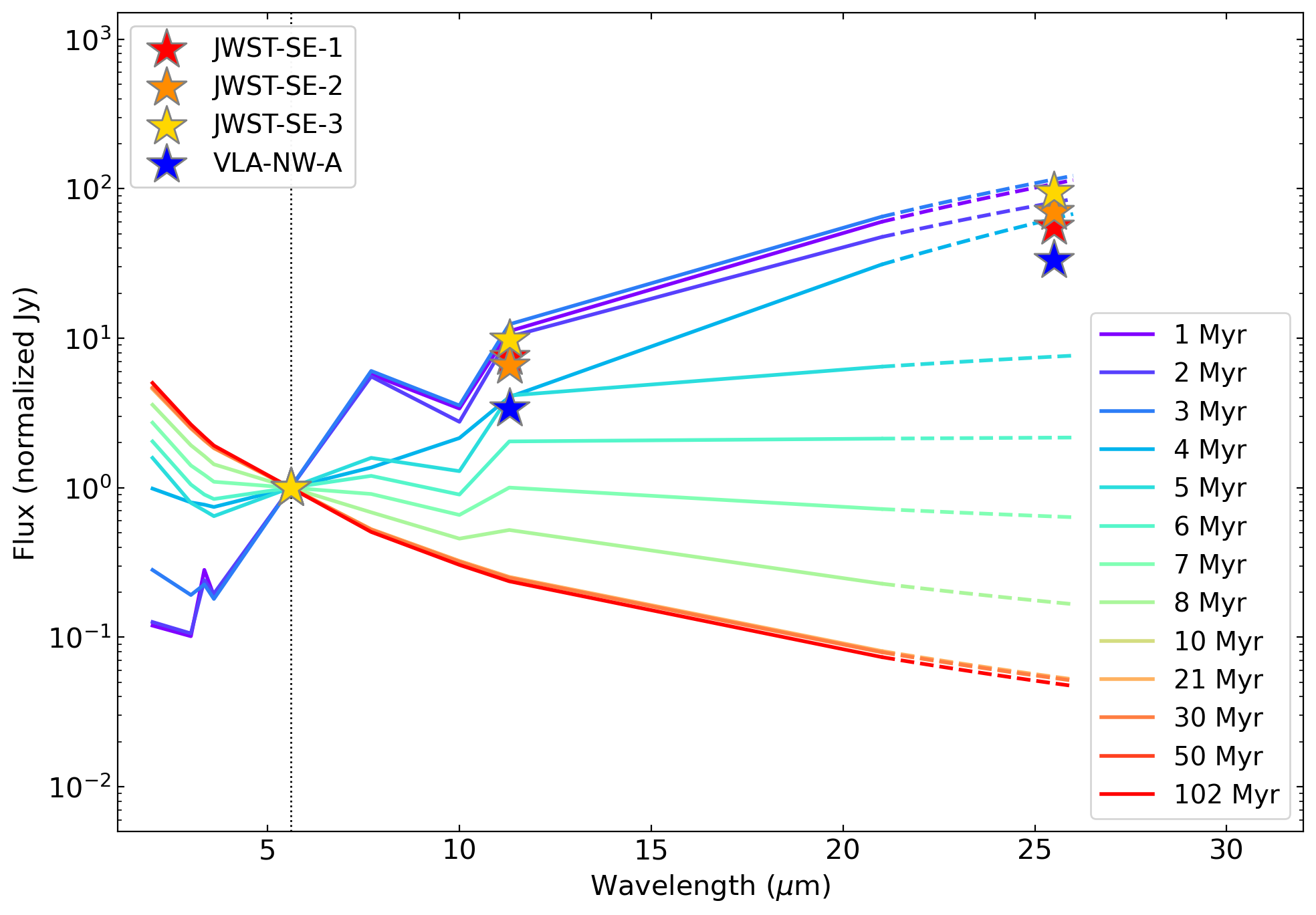}
}
\caption{Comparison of MIRI imaging aperture photometry of the four continuum sources
with photometric templates of star clusters as a function of age from \citet{whitmore25}.
All SEDs are normalized to 5.6\,\micron.
The continuum sources are shown as filled stars, and
the different ages of the templates correspond to the colors reported in the legend.
The dashed lines are the extrapolation of the templates to the F2550W filter observed here.
}
\label{fig:seds_clusters}
\end{figure}

Figure \ref{fig:seds_clusters} 
shows this comparison, where
the four \izw\ continuum sources are given as filled stars, and the SEDs have been
normalized to our shortest-wavelength MIRI filter, F560W.
Various ages from the templates\footnote{Obtained from \url{https://archive.stsci.edu/hlsp/phangs/phangs-cat}.}
are shown by colored curves as reported in the legend.
The templates do not extend to 25\,\micron, so the
F2550W photometry can be viewed as an extrapolation (shown as dashed lines in Figure \ref{fig:seds_clusters}).
When compared with the templates from \citet{whitmore25}, 
the ages of the \jwst-identified continuum sources appear to be quite young, $\la 3-4$\,Myr.
The radio-identified \hii\ region VLA-NW-A seems slightly older, $\sim 5$\,Myr.
Despite the silicate absorption feature at $\sim 10$\,\micron\
and the prominent PAH feature at 7.7\,\micron\ in the templates, not seen in \izw\
(see Fig. \ref{fig:spectra}),
all four \izw\ continuum sources would fall in the category of ``nearly embedded cluster candidates''
according to the SED templates provided by \citet{whitmore25}.

These JWST continuum sources seem to be metal-poor YSCs.
The low dust content and low extinction in \izw\ suggests that they cannot be 
partially embedded, like the YSCS in NGC\,628 and other nearby galaxies.
This is yet another limitation of the templates of \citet{whitmore25}
when applied to \izw.
Indeed, the cluster candidates are clearly visible in the \hst\ images (bottom panel of Figure \ref{fig:continuum}).
Typical young massive clusters have diameters of $\la 10$\,pc as found
in other nearby galaxies and starbursts such as M\,82, M\,83, NGC\,253, NGC\,3256, NGC\,3351, 
NGC\,4449, and NGC\,5253
\citep[e.g.,][]{levy24,deshmukh24,leroy18,sun24,linden24,mcquaid24,smith20}.
Hence, we cannot be seeing the star clusters directly with MIRI, 
since the sizes of the MIRI sources are larger than this.
Instead, with MIRI, we may be viewing the 
dust emission from the surrounding \hii\ regions and neutral (partially molecular) clouds.
In the Milky Way, emerging star clusters are very compact like those mentioned above,
but the MIR emission is more extended since it comes from the surrounding cloud
\citep[e.g.,][]{churchwell06}.
Moreover, given the crowding and compact size of the sources in the \hst\ image,
the larger MIR sizes may also reflect the possibility that the continuum sources comprise more than one star cluster.

From a statistical perspective, we would expect to see at least a few YSCs in \izw.
With an average SFR$\,\sim\,0.2-1$\,\msunyr\ over the last 10\,Myr \citep{hunt05,annibali13}, 
\izw\ would have formed at least $\sim 2\,\times\,10^6$\,\msun\ in that time.
Assuming a conservative cluster formation rate of $\sim 10$\%
\citep[meaning that essentially 10\% of the mass formed in the last 10\,Myr should be in star clusters,][]{krumholz19},
we would expect statistically to detect around 2 YSCs with mass $>\,10^4$\,\msun, and about
20 clusters between $10^3$ and $10^4$\,\msun.
This last estimate is based on a power-law cluster mass function with a power-law index of $-2$ \citep{krumholz19}.
Thus, it is possible that the \jwst\ continuum sources are YSCs, with masses between $10^3$ and $10^5$\,\msun,
identified by the dust continuum associated with the surrounding \hii\ regions. 

The \jwst\ 14\,\micron-continuum sources resemble the class of 21\,\micron\ compact ISM sources 
found by \citet{hassani23} in several nearby star-forming galaxies.
92\% of the sources identified by \citet{hassani23} sources are spatially associated with \hii\ regions and 74\% are coincident
with a stellar complex identified by \hst.
Less then 10\% of them are completely embedded in dust with no visibile counterpart.
Those 21\,\micron-selected sources also have \ha\ or CO emission, implying that 
they are associated with \hii\ regions or molecular clouds.
In Paper\,II, the rotational \htwo\ transitions of \izw\ in aperture spectra are presented; 
most of the warm \htwo\ in \izw\ is found in the SE, where all five apertures
in the SE have five or more significant detections in the rotational lines from 
$J_\mathrm{upper}$ from 3 to 10,
while in the NW, only VLA-NW-A, shows a similar number of detections.
The presence of warm \htwo, presumably associated with molecular clouds,
further corroborates that the continuum sources identified here
are of a similar class to the sources found by \citet{hassani23}.

The diameters of the \jwst\ 14\,\micron\ continuum sources including VLA-NW-A
are similar to those of the spatially resolved 24\,\micron-selected sources in M\,33 by \citet{sharma11},
and the \hii\ regions in M\,83 \citep{dellabruna22}.
Moreover, as evident in the bottom panel of Figure \ref{fig:continuum},
they are visible at 0.6\,\micron\ with \hst, so cannot be 
enshrouded by dust.
The extinction \av\ measured within the JWST-SE apertures 
is not significantly greater than the foreground extinction
(see Figure \ref{fig:extinction1d}), another indication that they are not 
embedded in their natal dust cloud.
Like the 21\,\micron\ compact sources identified by \citet{hassani23},
these MIRI continuum sources behave like \hii\ regions, emitting
\hi\ recombination lines,
as well as various FS lines (\siii, \siv, \neii, \neiii, \ariii) discussed in Sect. 
\ref{sec:highion}.
The \hst\ images show multiple sources, possibly associated with the YSCs exciting the \hii\ regions.





\section{The ionization sources in \izw \label{sec:discussion}}

In Section \ref{sec:spectra_ionized}, we showed that eight of the eleven 120\,pc diameter regions
selected in \izw\ harbor \oiv\,25.89 emission with an IP to create the ion of 54.9\,eV; 
four of the regions also emit \nev\,14.32 with an IP of 97.1\,eV 
(see Figures \ref{fig:linefits_oiv} and \ref{fig:linefits_nev} in Appendix \ref{sec:flux}).
The highest ionization gas is found in the NW, 
within the NW OB complex, the VLA-NW-A \hii\ region, 
the CO(2--1) aperture, and also, possibly most importantly, around the site of the ULX.
In the SE, 
one of the three continuum sources, together with VLA-SE,
contains \oiv\ emission, and none contain significant \nev.
The high ionization in the NW, relative to the SE, is seen not only in these highly ionized lines,
but also in elevated 
\neiii/\neii\ and \siv/\siii\ ratios 
(see Figure \ref{fig:ion}).
These ratios for \izw\ fall on the extreme end of the range of previous observations
across a wide range of galaxy types.

These results pose the question of the physical process or processes exciting this gas.
In the following, we assess the various theoretical models analyzed in Sect. \ref{sec:models},
as well as the empirical comparison with SMC N76 WR, and other models for \izw\ in the literature.

\subsection{Shocks}

Fast radiative shocks have been frequently proposed 
as the mechanism powering the high-ionization lines in BCDs
\citep[e.g.,][]{thuan05,izotov12,izotov21}.
However, we find that although they reproduce quite well some of the line ratios
from known AGN, the F24 shock models are unable
to reproduce the MIRI line ratios of \izw.
Other shock models such as \citet{alarie19}, tested by \citet{mingozzi25},
give similar results.
Thus, we conclude that shocks could help regulate the line ratios of massive galaxies
with or without AGN,
but they are not responsible for
the MIR high-ionization emission lines in \izw.

\subsection{IMBHs}

As shown in Figures \ref{fig:ion} and \ref{fig:newratios},
even a small \fagn\ of 4\% can reproduce fairly well the line ratios observed in \izw.
Because of the similarity in ionizing SEDs (see Figure \ref{fig:seds})
and the modeled line ratios for the 25\,Myr SSP$+$ULX populations
and those including IMBH \fagn\,=\,0.04,
we cannot rule out a small IMBH in \izw\ on the basis of line ratios alone.
Indeed, \citet{mingozzi25} find that a 4--16\% contribution from a $\sim 10^5$\,\msun\ 
IMBH is needed to reproduce the strong \nev/\neii\ emission seen in \sbs,
a BCD with \zzsun\,$\sim 5$\%.
This will be further discussed in 
a forthcoming paper (Rickards Vaught et al., in prep.).

\subsection{Feedback from star clusters}

Shocked stellar winds from star clusters and supernovae can produce
hot diffuse gas that emits X-rays \citep[e.g.,][]{chevalier85,cervino02}.
Such gas has been modeled extensively in the literature, but recent
models focus on explaining the \heii\ emission in dwarf galaxies via X-ray heating.
Because of slower stellar winds at low metallicity,
\citet{oskinova22} find that plasma temperatures are lower,
and the production rate of \heii\ ionizing photons is higher.
Thus, they argue that their model is consistent with the \heii\ observed in \izw\
\citep[e.g.,][]{kehrig21,rickards21}.

Alternatively, the models by \citet{franeck22} use a different approach to estimate
the soft X-ray emission of hot gas from star-cluster winds. 
The X-ray emission that would be observed in the
scenario of hot diffuse gas produced by star clusters and supernovae
is expected to be faint.
Thus, it falls below the observed X-ray emission of the comparison sample of \citet{franeck22};
\citet{oskinova22} argue that the diffuse hot gas in \izw\ is also too faint to be detected.
Thus, with current observations, it is difficult to test this hypothesis,
without radiative transfer models that predict line ratios under such
conditions.

\subsection{WR stars}

Because of the similarity of the line ratios observed in the SMC N76 WR nebula 
and \izw,
stars like AB7 \citep{naze03,shenar16,tarantino24} could be responsible for ionizing the gas in \izw.
As mentioned above, \izw\ hosts WR stars found with long-slit spectroscopy \citep{legrand97,izotov97}, 
and a WR blue bump has been identified in the NW by \citet{kehrig16}.
Moreover, the SED of WR N76 shown in Figure \ref{fig:seds} from \citet{shenar16}
is very similar below $\sim 100$\,eV to the R25 25\,Myr 1\% Solar metallicity ionizing source,
which comes close to reproducing the observed line ratios in \izw.

Although WR stars in \izw\ have been previously dismissed as unable to power
the highly ionized gas in \izw\ \citep{kehrig15}, 
the resemblance of the MIR high-ionization line ratios in \izw\ with SMC WR N76
suggests that one or a few stars like AB7 in \izw\ could be responsible.
Such a mechanism would be particularly important in the SE where
two apertures show \oiv\ emission, relatively far away from ULX-1 in the NW.
\citet{rickards21} also suggested that the \heii\ emission seen in the SE
could be powered by a single star like AB7.

It is probable that the YMCs in the MIRI-identified
continuum sources host at least one WR star or WR binary such as AB7;
\citet{leitherer14} predict that at sub-Solar metallicity,
the ratio between WR and O stars peaks at $\sim 5$\% between 3 and 4\,Myr; 
with $\ga 500$ O stars in the SE alone, we would thus expect $\ga 25$ WR stars in the SE.
The inferred young age of 3-4\,Myr of the SE continuum sources from Figure \ref{fig:seds_clusters} 
would be consistent with this, and with the ionizing sources comprising WR stars/binaries. 

\subsection{ULXs}

The ULX in \izw\ is located in the NW, and the ULX-1 120-pc diameter
aperture overlaps slightly with VLA-NW-A, CO2-1, and NW (see Figure \ref{fig:apertures}).
All six NW apertures show \oiv\,25.89 emission, and the four apertures near the ULX
also show \nev\,14.32.
The latest measurements of the 0.3-10 keV X-ray luminosity of ULX-1 in \izw\
give $L_X\,\ga\,1.1\times10^{40}$\,erg\,s$^{-1}$, 
suggesting a possible supercritical accretion state \citep{kaaret13,yoshimoto24}. 
It is well known that ULXs are capable of generating large highly ionized nebulae
a few hundreds of pc in diameter, observable both in the radio continuum \citep[e.g.,][]{berghea20,soria21}
and the optical \citep[e.g.,][]{moon11,kaaret17,gurpide22}.
Such nebulae could be what we are observing in \izw;
indeed, \citet{lebout17} attribute the heating source in the NW region of \izw\
to photoionization by radiation from a bright X-ray binary.

The centers of VLA-NW-A and NW are at a projected distance of $\sim 150$\,pc from ULX-1,
while CO2-1 is slightly closer in projection.
If ULX-1 were to explain the \heii, \oiv, and \nev\ emission seen in those regions
\citep[for \heii, see][]{rickards21,kehrig21},
the ULX would have to maintain a large \heiii\ 
region, with radius $R_{\rm He\,III}\approx 200$\,pc, with \heiii\ recombining to \heii\ at a rate
{\small
\begin{eqnarray}
\dot{N}_{\rm He\,II}
&~\approx~& 
\frac{4\pi}{3} R_{\rm He\,III}^3 0.08 n_e^2\alpha_{\rm He\,III} \\
&\approx& 
10^{49.9}
\left(\frac{R_{\rm He\,III}}{200\mathrm{pc}}\right)^3
\left(\frac{n_e}{\mathrm{cm}^{-3}}\right)^2
\left(\frac{\alpha_{\rm He\,III}}{10^{-12}\mathrm{cm}^3\mathrm{s}^{-1}}\right)\mathrm{s}^{-1}\ , \nonumber
\end{eqnarray}
}
\noindent
where $\alpha_{\rm He\,III}$ is the rate coefficient for
radiative recombination \heiii\,$\rightarrow$\,\heii. 
If these recombinations are balanced by photoionizations, the required
ULX luminosity is
{\small
\begin{eqnarray}
L_X &~\approx~& 100{\rm \,eV} \times \dot{N}_{\rm He\,III} \\
&\approx& 10^{40.1}\left(\frac{R_{\rm He\,III}}{200\mathrm{pc}}\right)^3
\left(\frac{n_e}{\mathrm{cm}^{-3}}\right)^2
\left(\frac{\alpha_{\rm He\,III}}{10^{-12}\mathrm{cm}^3\mathrm{s}^{-1}}\right)\mathrm{erg}\,\mathrm{s}^{-1} \nonumber
\ .
\end{eqnarray}
}
\noindent
Thus, if the rms electron density is $\sim 1$\,\cmthree,
the ULX in \izw\ could plausibly sustain the highly ionized gas, \nev\ and \oiv\ (and \heii), seen
in these NW apertures.

\subsection{\nev\ and AGN}

\nev\ is widely regarded as an indicator of an accreting massive BH
\citep[e.g.,][]{inami13,feltre16,chisholm24}, so if we were to 
take the \nev\ detections in the four NW apertures at face value,
we would conclude that there is an accreting massive BH or AGN in \izw.
However, the location of \izw\ in the line-ratio diagnostics proposed by R25 
shown in Figure \ref{fig:newratios} suggests that either a $\sim 25$\,Myr SSP$+$ULX or
a $\sim 10$\,Myr SSP$+$ULX$+$AGN with a very small \fagn\ of 4\%
could be responsible. 
The difference between a pure SSP$+$ULX, and 
one including a very small AGN fraction from IMBH emission, 
is not easy to distinguish only from line ratios.
Given the similarity of the gas properties of \izw\ with high-$z$ metal-poor dwarf galaxies,
our observations furnish a cautionary tale
for automatically associating highly ionized gas detections such as \nev\ with an AGN.
The extreme low-metallicity environment in \izw, more pronounced in the NW,
can mimic the harsh RF environment surrounding an accreting massive BH.

\section{Summary \label{sec:conclusions}}

In order to characterize the ISM in an extreme, metal-poor dwarf galaxy,
we obtained MIRI/MRS observations of the main body of \izw.
We have selected 11 apertures, $\sim 120$\,pc in diameter,
centered on regions of interest across the galaxy, and extracted 1D spectra within them
(Figure \ref{fig:spectra}).
Our findings can be summarized as follows:
\begin{itemize}
\renewcommand\labelitemi{$-$}
\item By comparing the MIRI \hi\ recombination lines with the \hb\ and \hd\ lines 
from KCWI IFU maps of \izw\ \citep{rickards21}, 
we have verified the very low extinction of \av\,$\sim 0.1-0.2$\,mag (Figure \ref{fig:extinction1d})
previously found in the optical \citep[e.g.,][]{izotov99,kehrig15}.
Because of the small extinction in the MIR, it was not possible to perform
the measurements using MIRI alone.
\item 
In addition to numerous \hi\ recombination lines,
we have detected a plethora of MIR fine-structure lines in \izw, including 
\arii, \ariii, \feii, \neii, \neiii, \siii, \siv,
as well as \piii\ (weak detection in only one aperture,
see Table \ref{tab:fslines}).
Also detected are extremely high-ionization emission lines including 
\oiv\,25.89 with an IP of 54.9\,eV and \nev\,14.32 (IP\,=\,97.1\,eV).
\nev\,14.32 has never been detected in a dwarf galaxy before \jwst\
\citep[see][for the detection of \nev\,14.32 in \sbs]{mingozzi25}. 
\item
The line ratios of the lower ionization lines in \izw\ (\neii, \neiii, \siii, \siv) are 
consistent, at the extreme end, with previous observations of BCDs and other galaxies. 
However, the higher ionization lines (\oiv, \nev), found mainly in the NW, fall in regions of the
line-ratio diagrams that do not coincide with massive star-forming galaxies or known AGN.
Comparison with state-of-the-art photoionization and shock models shows
that the line ratios in \izw\ cannot be well reproduced by shocks.
However, they are fairly well approximated with low-metallicity stellar populations
containing a self-consistent substantial ULX component.
Wolf-Rayet stars like AB7 in the SMC or a small 4\% contribution from an IMBH 
could also be responsible (see Figures \ref{fig:ion}, \ref{fig:shocks}, \ref{fig:newratios}). 
Ultimately, the confluence of more than one of the physical processes discussed here
could be powering the ionization state of \izw.
\item 
From the MIRI cube, we identified four \jwst/MIRI continuum sources at 14\,\micron,
one of which is associated with the radio \hii\ region VLA-NW-A, and
three of which were not previously characterized (see Figure \ref{fig:continuum}). 
Their deconvolved diameters of $\sim 30-100$\,pc are larger than the $\sim 10$\,pc diameter of typical young massive star clusters,
and they have apparently emerged from their natal dust cloud because of the low
measured extinction, and the existence of an \hst\ counterpart at F606W.
Comparison of the SEDs with templates of young stellar clusters by \citet{whitmore25} suggests
that they are associated with YSCs, and that we are seeing the dust continuum 
from the surrounding \hii\ regions and neutral (partially molecular) clouds.
Such objects would be expected statistically, given the cluster nature of star formation.
Their sizes are consistent with 21$-$24\,\micron-selected sources in M\,33 and in other nearby galaxies
\citep{sharma11, hassani23}.
\end{itemize}

It is becoming more and more evident that \jwst/MIRI is a uniquely powerful facility for
characterizing the ISM in nearby galaxies, even in faint, extremely metal-poor dwarfs such as \izw.
The extreme conditions in the ISM of \izw\ can serve as a local benchmark
to compare to the metal-poor dwarf-galaxy population now being identified 
by \jwst\ at high redshift.

\section*{Acknowledgments}
We thank the referee for a timely and insightful report, and for the attentive comments that helped improve the paper.
We are also grateful to Tomer Shenar who kindly made available the SED of AB7 in the SMC \citep{shenar16}.
This work is based on observations made with the 
NASA/ESA/CSA James Webb Space Telescope. The
data were obtained from the Mikulski Archive for Space Telescopes at the Space Telescope Science Institute, 
which is operated by the Association of Universities for Research in Astronomy, Inc., under NASA contract NAS 5-03127 for \jwst. 
These observations are associated with program JWST-3533. 
RRV is grateful for the support of this program, provided by NASA through a grant from the Space Telescope Science Institute.
AB-Z acknowledges support by NASA under award number 80GSFC24M0006.
BLJ and M. Mingozzi are thankful for support from the European Space Agency (ESA).
M. Meixner acknowledges that a portion of her research
was carried out at the Jet Propulsion Laboratory, California
Institute of Technology, under a contract with the National Aeronautics and Space Administration 
(80NM0018D0004).
LKH thanks Martha Haynes for very interesting discussions, and for
sharing her physical insight into low-metallicity galaxies in general, and \izw\ in particular.
This research made use of \texttt{Photutils}, an \texttt{Astropy} package for
detection and photometry of astronomical sources \citep{larry_bradley_2024_13989456}. 

%

\vspace{5mm}
\facilities{JWST(MIRI), HST(ACS), Keck(KCWI)}


\software{Astropy \citep{2013A&A...558A..33A,2018AJ....156..123A}, Numpy \citep{harris2020array}, SciPy \citep{2020SciPy-NMeth}, 
Dustmaps \citep{green18} }


The JWST data presented in this article were obtained from the Mikulski Archive for Space Telescopes (MAST) at the Space Telescope Science Institute. 
The specific observations analyzed can be accessed via \dataset[doi: 10.17909/n80x-b534 ]{https://doi.org/10.17909/n80x-b534}.



\appendix

\section{Detailed description of MRS data reduction \label{sec:reduction}}

Background subtraction was performed separately from the pipeline at the exposure level, using the rate files after the \texttt{calwebb\_detector1} step. 
A master background was created from the background observation rate files for each detector and subtracted accordingly. 
The presence of warm pixels in the final products was mitigated using \texttt{LaCosmic}, 
a well-known Python package for cosmic-ray removal based on \citet{vandokkum01}. 
\texttt{LaCosmic} was applied to the rate files with a contrast setting of 2 and a cosmic-ray threshold (\texttt{cr\_threshold}) of 25. 
The \texttt{neighbor\_threshold} was set to 5, 
which effectively flagged cosmic rays that had not been detected in earlier pipeline steps. 
Additionally, a warm pixel mask was created for each detector and added to the 
DQ extension of the calibrated files after the \texttt{calwebb\_spec2} step.

The \texttt{calwebb\_spec2} step was executed with the \texttt{residual\_fringe} correction enabled. 
This step addresses residual fringes that may persist after the initial fringe flat correction. 
Even small variations in the position observed can result in significantly different fringe patterns. 
The \texttt{residual\_fringe} step models these residual fringes as a summation of 
sinusoidal components. 
Additional fringe corrections were applied to the final extracted spectra. 
At this stage, the warm pixel mask was incorporated into the DQ extension of the calibration 
files before the cube-building process.
The astrometric and distortion corrections were also applied in this step according
to \citet{patapis24}.

Finally, the \texttt{calwebb\_spec3} step was executed. 
During the \texttt{outlier\_detection} step, any remaining unflagged warm pixels 
were identified by setting \texttt{threshold\_percent} to 99.8,
and \texttt{kernel\_size} to [11,11]. 
In the \texttt{cube\_build} step, 
the \texttt{output\_type} was set to band, allowing a third fringe correction to be
applied on a band-by-band basis.

\section{Flux measurements of detected lines and Gaussian fits for detected \oiv\ and \nev\ \label{sec:flux}}

\begin{figure}[h!]
\centering
\includegraphics[width=0.2\linewidth]{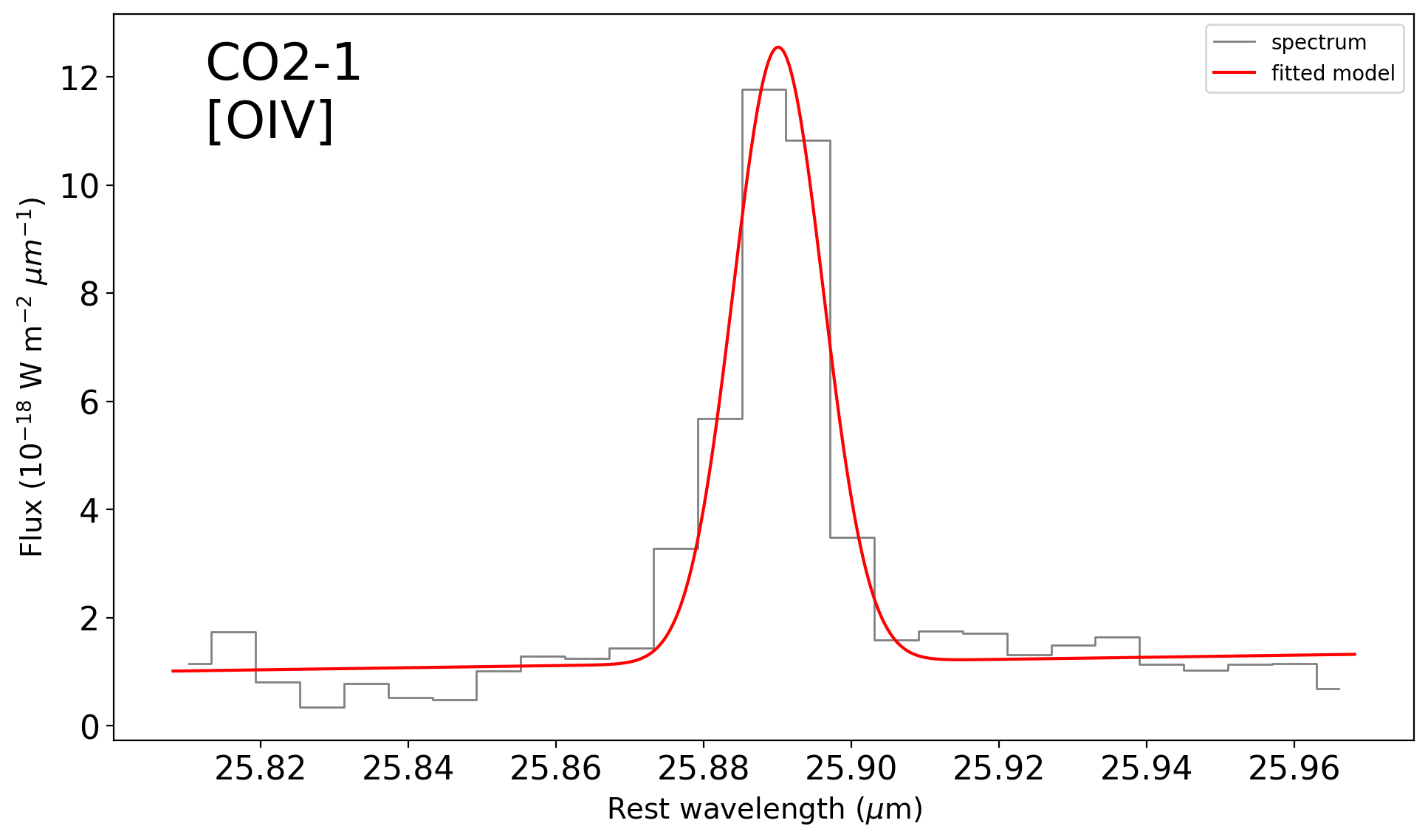} 
\includegraphics[width=0.2\linewidth]{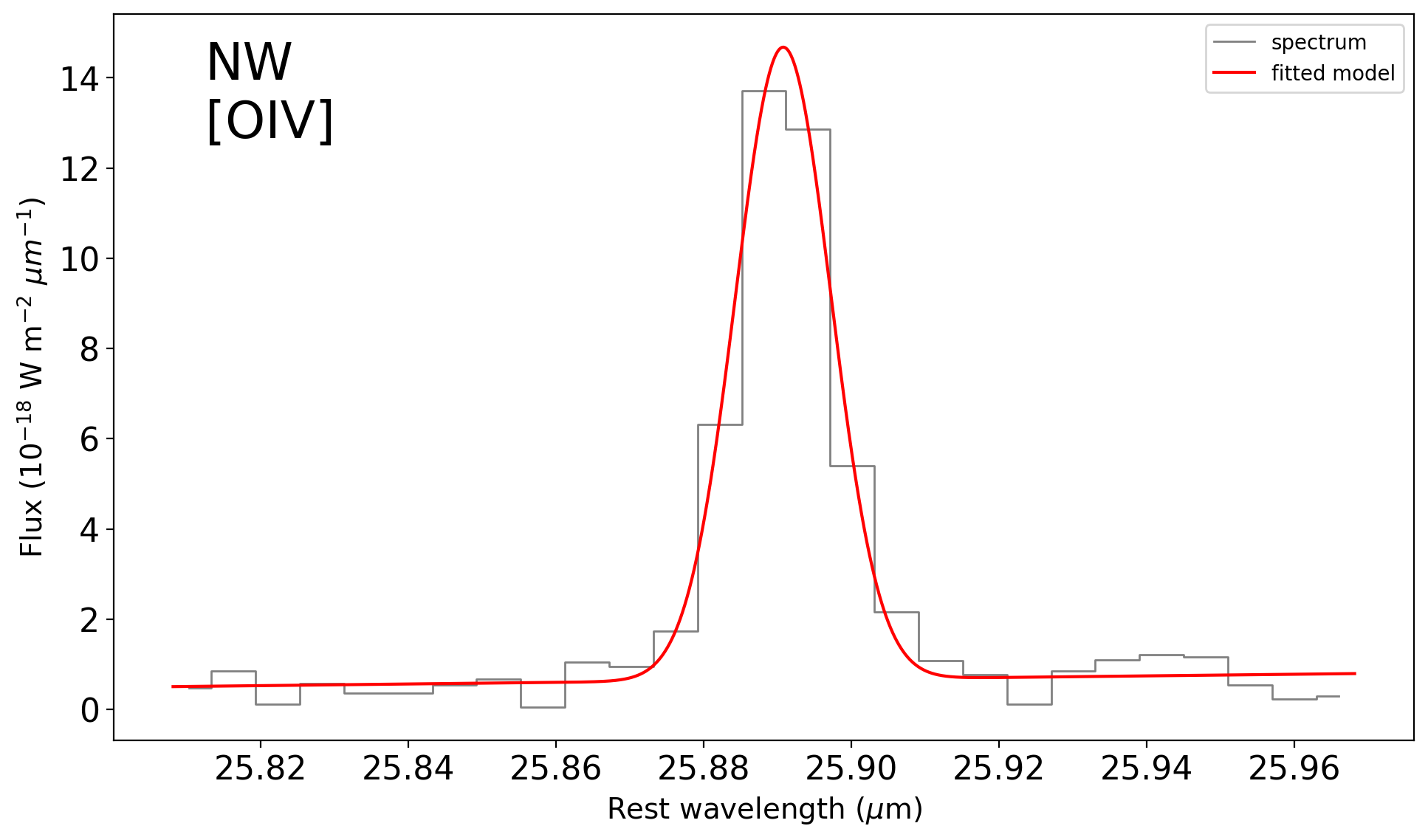}
\includegraphics[width=0.2\linewidth]{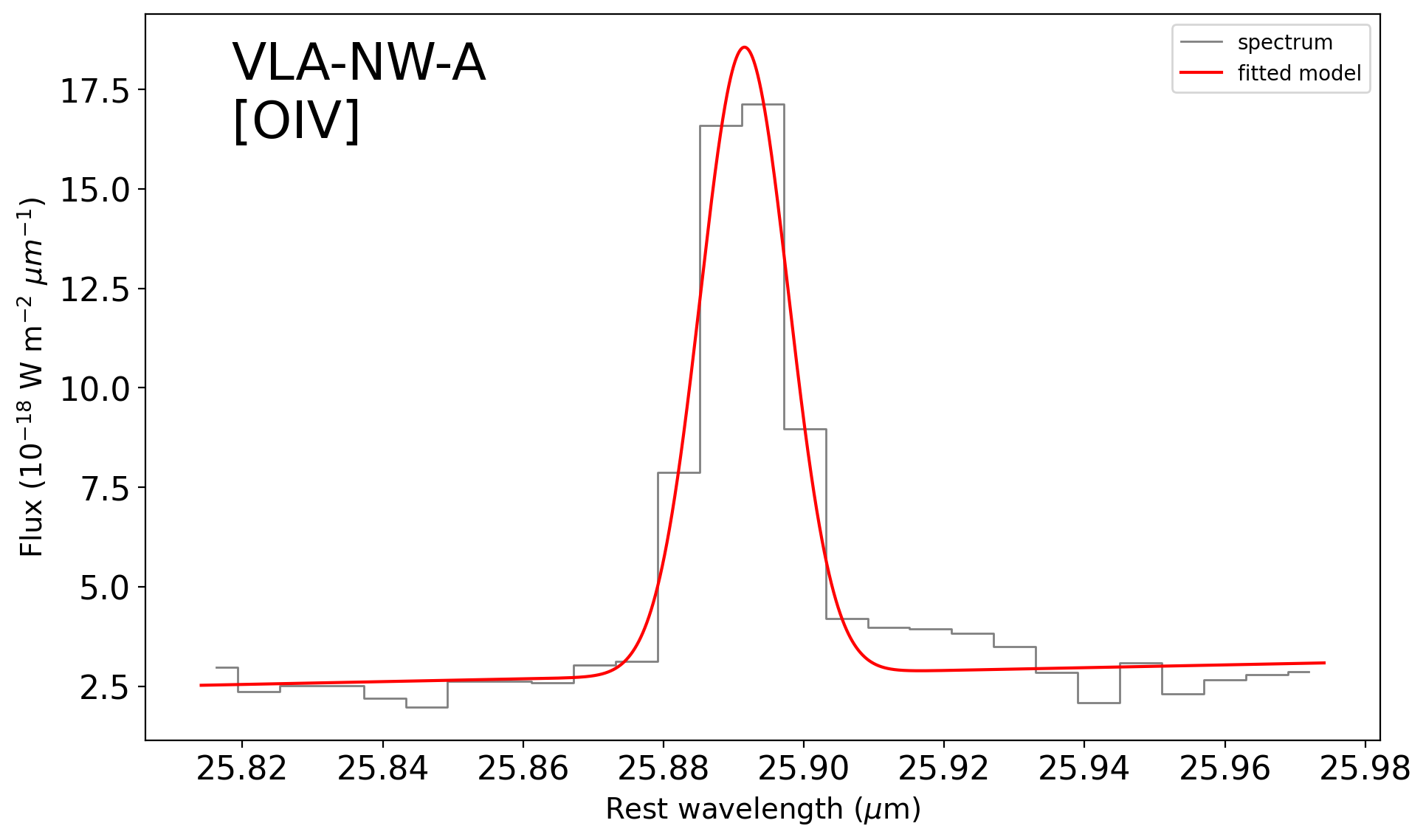}
\includegraphics[width=0.2\linewidth]{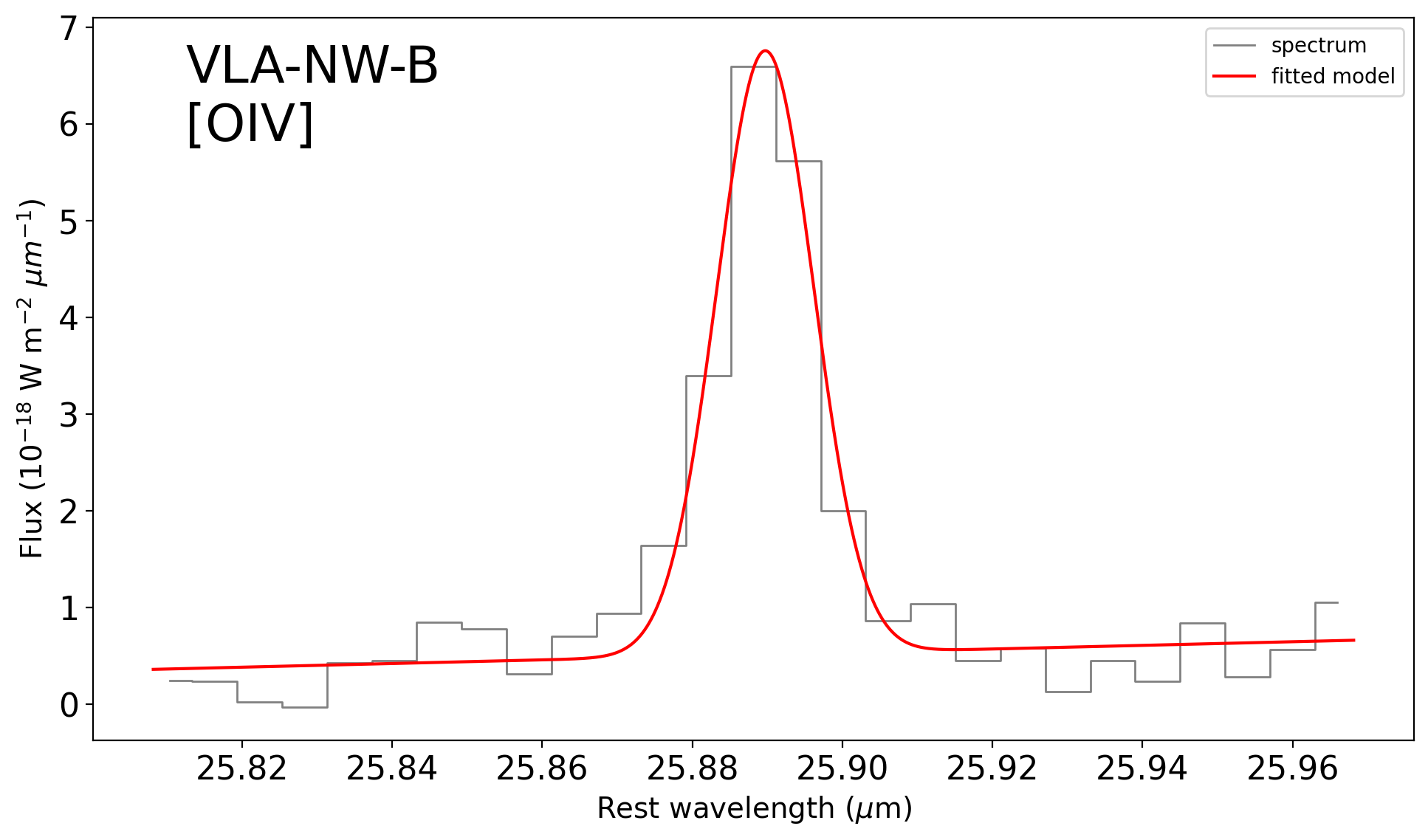} \\
\centering
\includegraphics[width=0.2\linewidth]{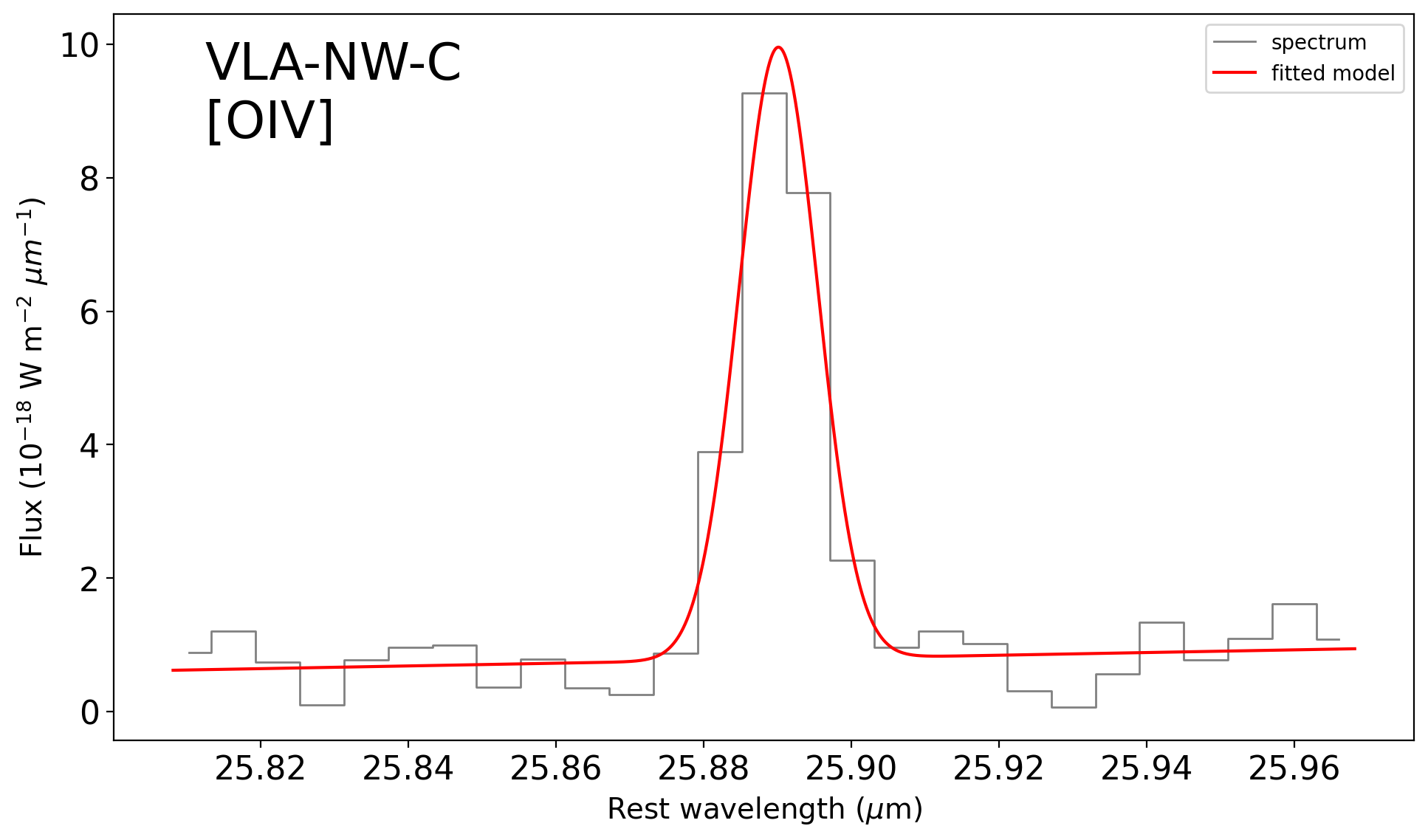}
\includegraphics[width=0.2\linewidth]{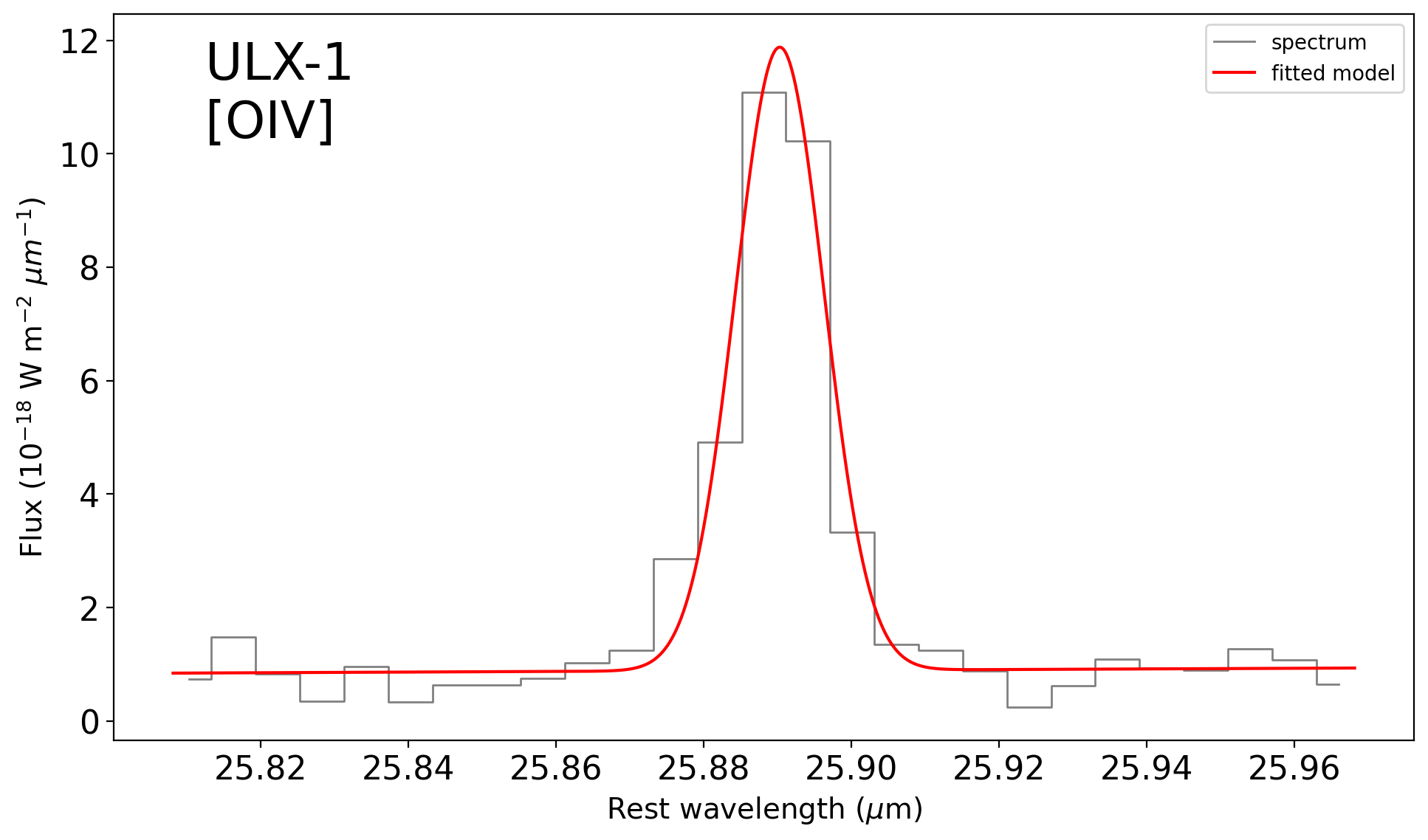}
\includegraphics[width=0.2\linewidth]{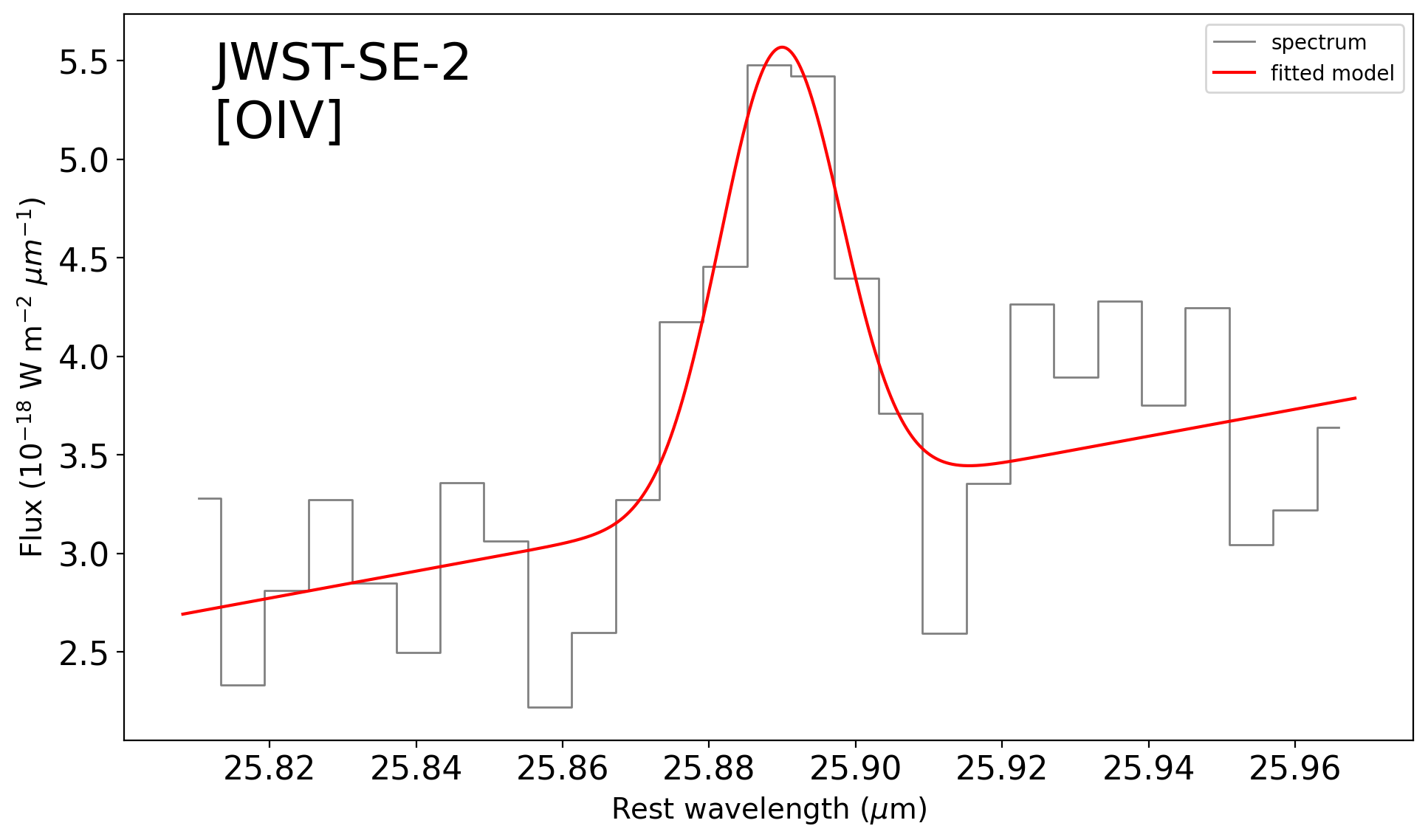}
\includegraphics[width=0.2\linewidth]{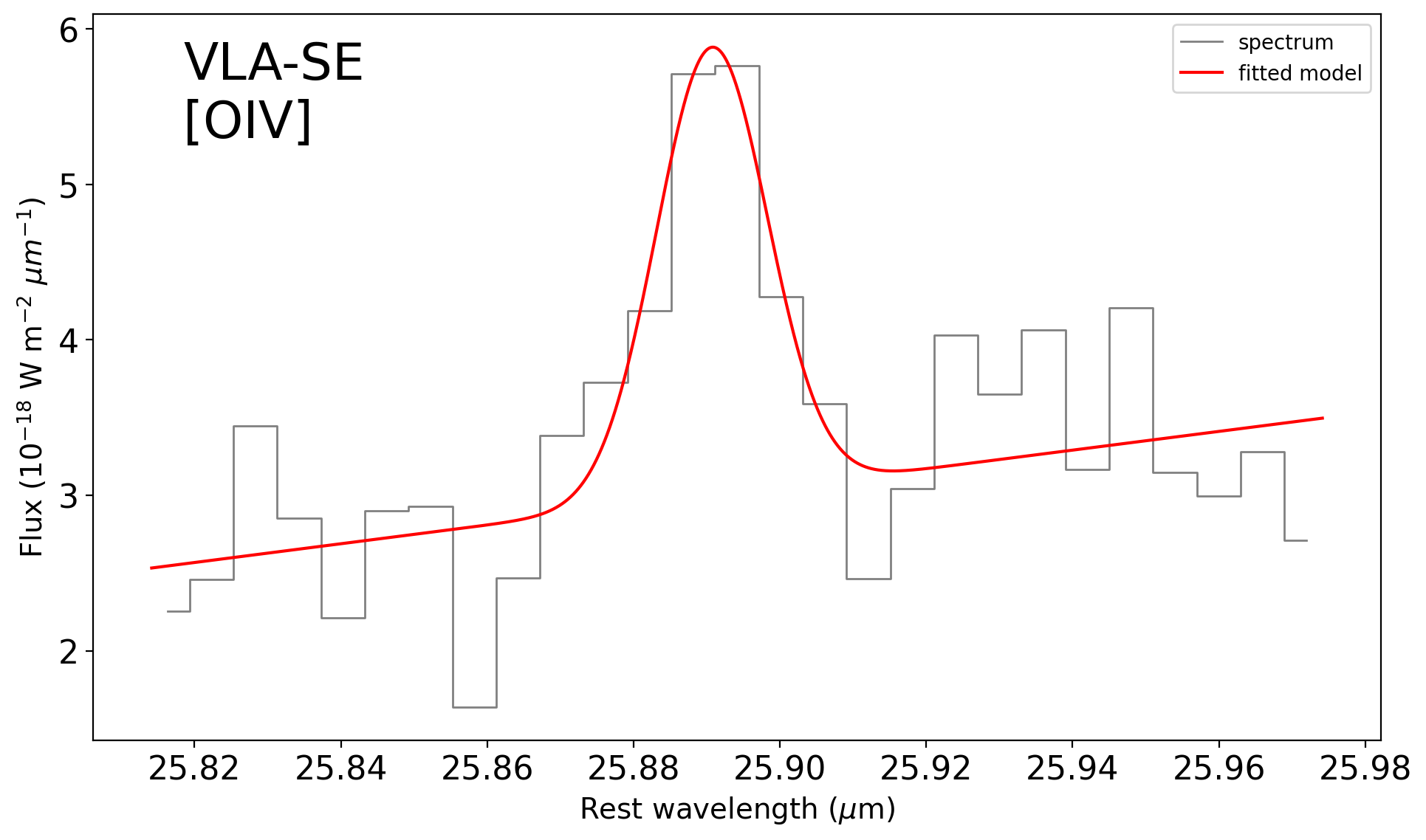}
\caption{Gaussian line fits to the detected \oiv; 
the vertical flux density scale is in units of $10^{-18}$ W\,m\,$^{-2}$\,\micron. 
}
\label{fig:linefits_oiv}
\end{figure}

\begin{figure}[h!]
\centering
\includegraphics[width=0.2\linewidth]{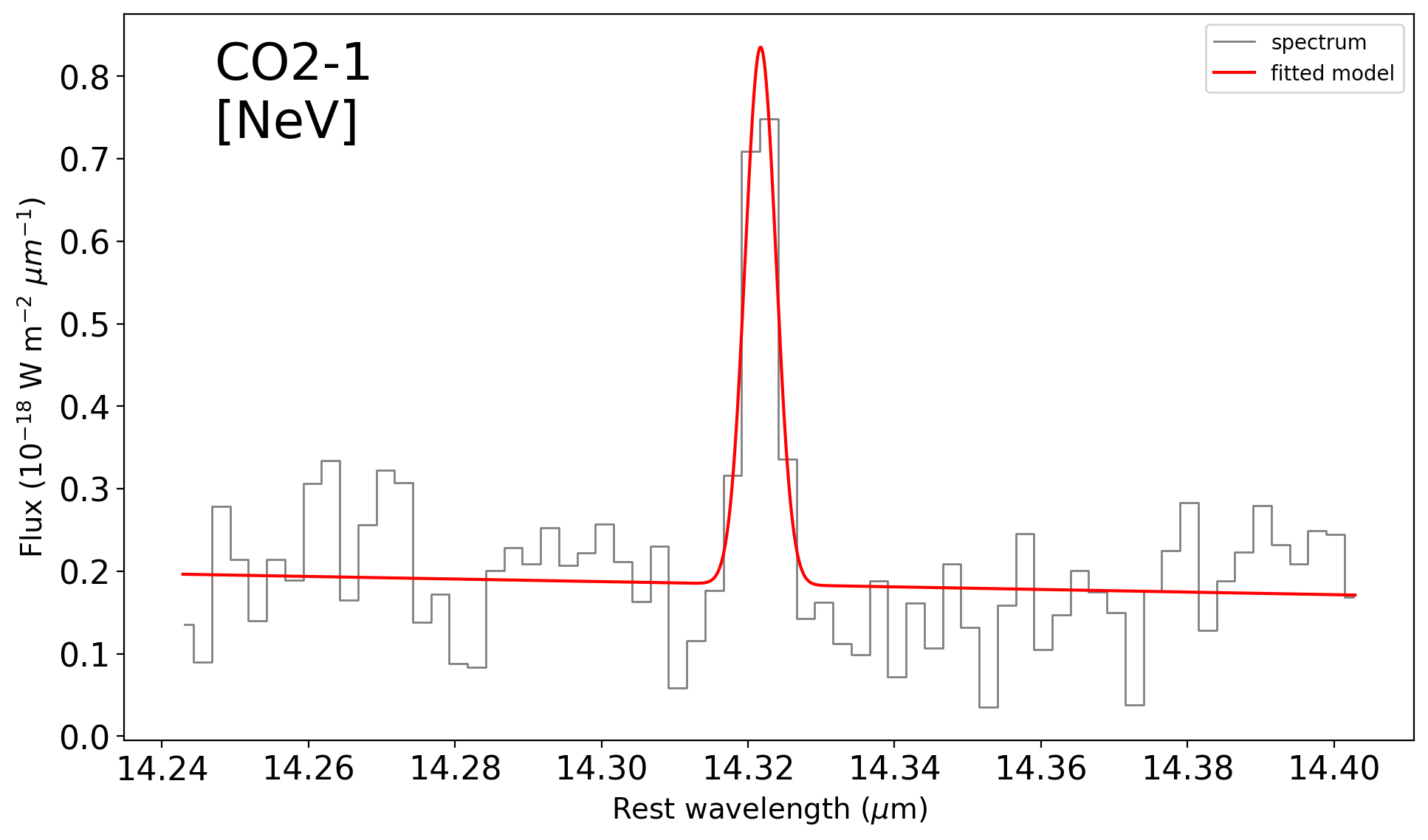}
\includegraphics[width=0.2\linewidth]{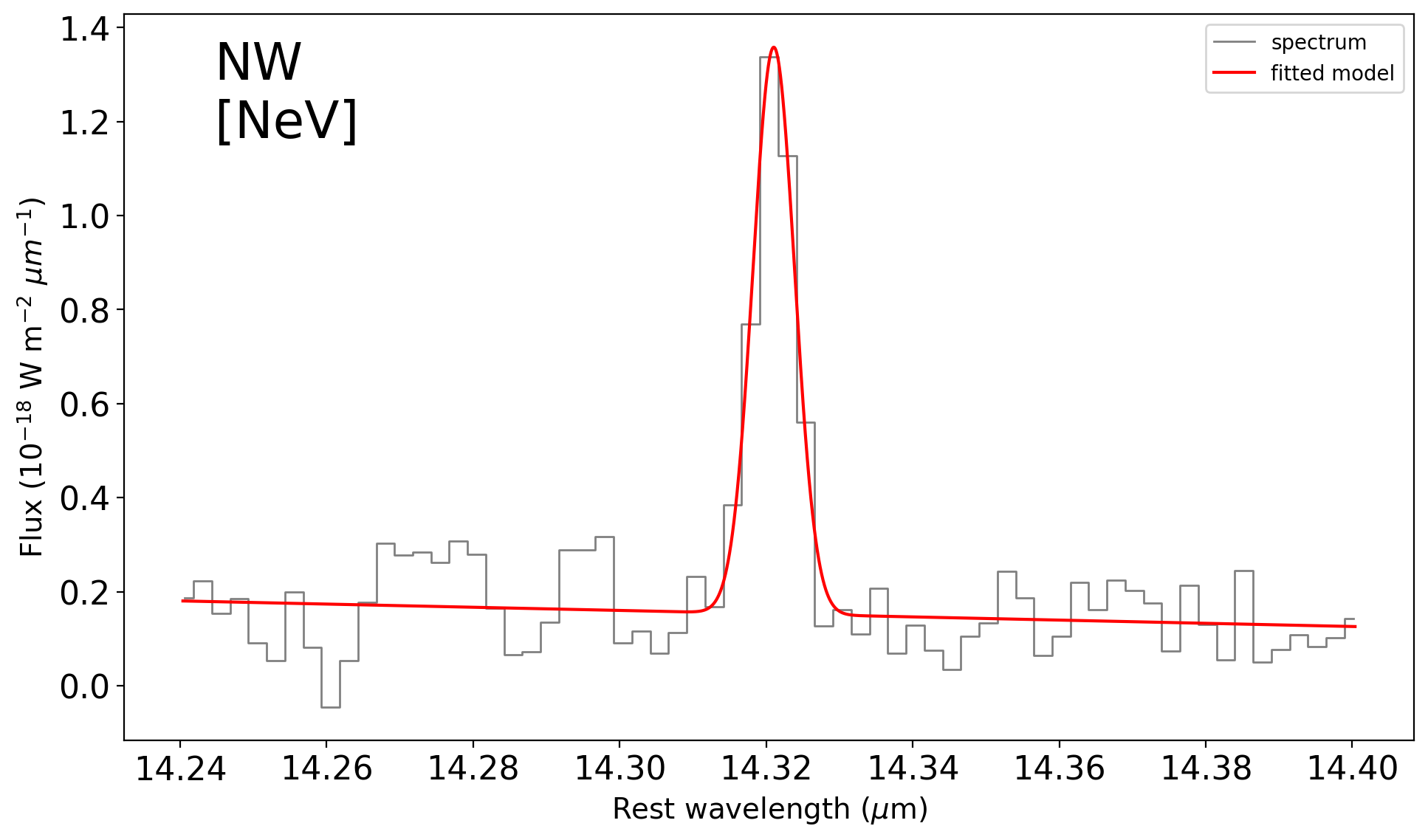}
\includegraphics[width=0.2\linewidth]{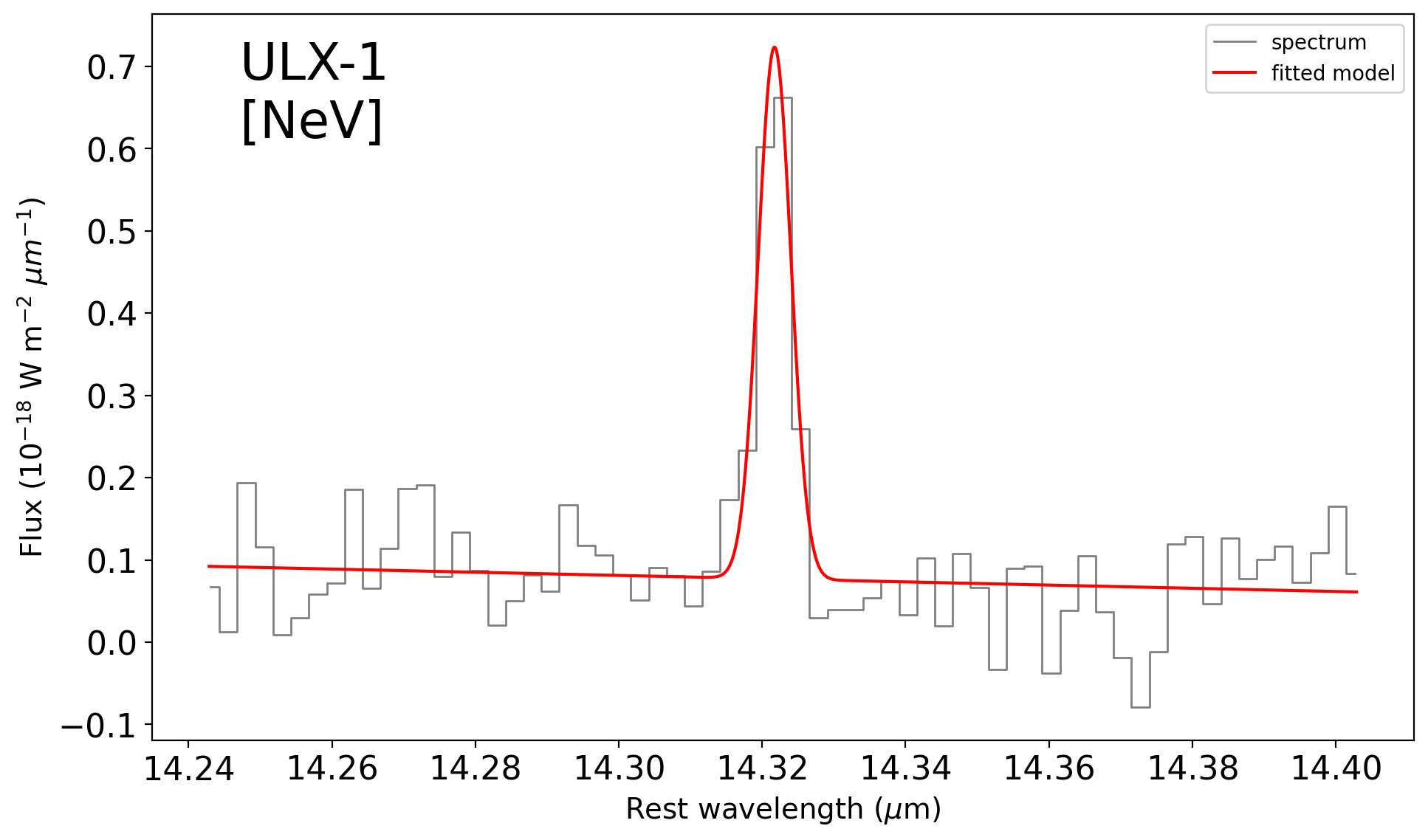}
\includegraphics[width=0.2\linewidth]{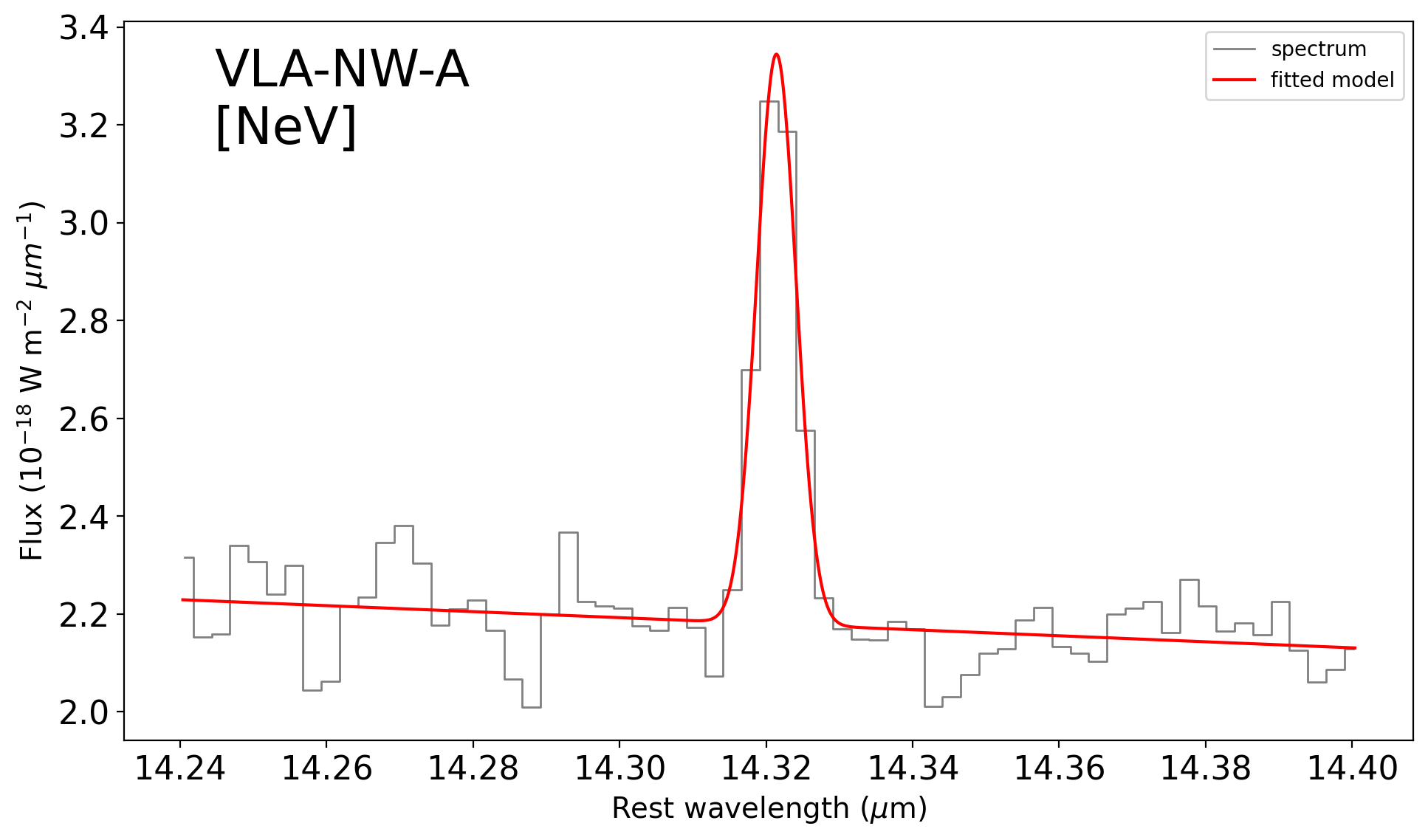} \\
\caption{Gaussian line fits to the detected \nev;
the vertical flux density scale is in units of $10^{-18}$ W\,m\,$^{-2}$\,\micron. 
}
\label{fig:linefits_nev}
\end{figure}

\begin{table*}
\begin{centering}
\caption{Integrated line fluxes in the NW-region apertures with S/N$\,\geq\,3$ \label{tab:nwflux}}
\begin{tabular}{lrcccccc}
\hline
\multicolumn{1}{c}{Line} & \multicolumn{1}{c}{Rest $\lambda$} & NW & VLA-NW-A & VLA-NW-B & VLA-NW-C & CO2-1 & ULX-1\\
& \multicolumn{1}{c}{(\micron)} & \multicolumn{6}{c}{($10^{-21}$\,W\,m$^{-2}$)}\\
\hline
\hline
\multicolumn{8}{c}{\hi\ recombination lines} \\
\hline
\\
HI(10-6) & 5.1287 &  --- & 25.93$\,\pm\,$2.80 &  --- &  --- &  --- &  --- \\
HI(10-7) & 8.7601 & 6.36 $\,\pm\,$0.85 & 16.64$\,\pm\,$0.97 &  --- & 8.01$\,\pm\,$1.40 & 10.33$\,\pm\,$1.10 & 8.49$\,\pm\,$1.09\\
HI(10-8) & 16.2091 & 4.01 $\,\pm\,$0.83 & 10.64$\,\pm\,$0.66 &  --- & 5.69$\,\pm\,$0.62 & 5.37$\,\pm\,$0.63 & 3.67$\,\pm\,$0.82\\
HI(12-9) & 16.8806 & 1.70 $\,\pm\,$0.56 & 4.47$\,\pm\,$0.58 &  --- & 2.31$\,\pm\,$0.56 & 1.87$\,\pm\,$0.57 &  --- \\
HI(13-8) & 9.392 & 10.14 $\,\pm\,$2.59 & 6.12$\,\pm\,$0.93 &  --- & 7.94$\,\pm\,$2.22 & 4.66$\,\pm\,$1.22 & 4.48$\,\pm\,$1.25\\
HI(13-9) & 14.1831 &  --- & 3.27$\,\pm\,$0.77 &  --- &  --- &  --- &  --- \\
HI(15-9) & 11.5395 &  --- & 2.50$\,\pm\,$0.71 &  --- &  --- &  --- &  --- \\
HI(17-7) & 5.3798 &  --- & 18.95$\,\pm\,$4.84 &  --- &  --- &  --- &  --- \\
Pf$\alpha$ & 7.4599 & 60.27 $\,\pm\,$1.68 & 183.10$\,\pm\,$3.23 &  --- &  --- & 89.33$\,\pm\,$2.39 & 68.54$\,\pm\,$2.09\\
Hu$\alpha$ & 12.3719 & 23.95 $\,\pm\,$1.61 & 65.59$\,\pm\,$3.89 & 14.10$\,\pm\,$1.29 & 31.28$\,\pm\,$2.61 & 32.51$\,\pm\,$2.21 & 26.85$\,\pm\,$1.76\\
Hu$\beta$ & 7.5025 & 16.31 $\,\pm\,$1.50 & 52.20$\,\pm\,$3.67 &  --- & 8.58$\,\pm\,$1.01 & 25.12$\,\pm\,$2.04 & 19.26$\,\pm\,$1.83\\
HI(8-7) & 19.0619 &  --- & 25.69$\,\pm\,$2.86 &  --- &  --- & 14.94$\,\pm\,$2.08 &  --- \\
Hu$\gamma$ & 5.9082 & 9.15 $\,\pm\,$2.75 & 29.30$\,\pm\,$4.29 &  --- &  --- &  --- &  --- \\
HI(9-7) & 11.3087 & 7.38 $\,\pm\,$0.74 & 22.73$\,\pm\,$1.39 &  --- & 9.88$\,\pm\,$0.90 & 12.10$\,\pm\,$1.20 & 10.06$\,\pm\,$1.05\\
\\
\hline
\multicolumn{8}{c}{Fine-structure lines} \\
\hline
\\
\ariii\ & 8.9914 & 16.75 $\,\pm\,$1.02 & 54.64$\,\pm\,$1.38 &  --- & 30.42$\,\pm\,$1.11 & 30.72$\,\pm\,$1.31 & 22.27$\,\pm\,$0.81\\
\feii\ & 5.3402 & 11.56 $\,\pm\,$3.38 & 32.28$\,\pm\,$5.76 &  --- &  --- & 19.85$\,\pm\,$4.81 & 16.62$\,\pm\,$4.67\\
\neiii\ & 15.5551 & 177.30 $\,\pm\,$0.88 & 495.50$\,\pm\,$2.21 & 111.50$\,\pm\,$0.98 & 250.30$\,\pm\,$1.45 & 280.60$\,\pm\,$2.24 & 208.40$\,\pm\,$1.62\\
\neii\ & 12.8135 & 10.26 $\,\pm\,$1.08 & 29.34$\,\pm\,$0.51 & 7.79$\,\pm\,$1.01 & 20.01$\,\pm\,$0.89 & 21.06$\,\pm\,$1.02 & 14.52$\,\pm\,$0.98\\
\nev\ & 14.3217 & 8.60 $\,\pm\,$0.82 & 7.84$\,\pm\,$0.77 &  --- &  --- & 3.44$\,\pm\,$0.65 & 3.70$\,\pm\,$0.53\\
\oiv\ & 25.8903 & 227.60 $\,\pm\,$8.65 & 247.80$\,\pm\,$12.21 & 101.10$\,\pm\,$7.33 & 122.60$\,\pm\,$8.58 & 173.60$\,\pm\,$11.61 & 165.60$\,\pm\,$9.96\\
\piii\ & 17.885 &  --- & 2.72$\,\pm\,$0.79 &  --- &  --- &  --- &  --- \\
\siii\ & 18.713 & 66.60 $\,\pm\,$2.27 & 182.80$\,\pm\,$3.16 & 49.55$\,\pm\,$1.85 & 108.20$\,\pm\,$2.74 & 110.10$\,\pm\,$2.38 & 85.19$\,\pm\,$2.51\\
\siv\ & 10.5105 & 240.50 $\,\pm\,$2.72 & 637.20$\,\pm\,$9.28 & 9.36$\,\pm\,$1.23 & 241.00$\,\pm\,$2.77 & 330.30$\,\pm\,$3.95 & 247.80$\,\pm\,$2.90\\
\\
\hline
\hline
\end{tabular}
\end{centering}
\end{table*}

\begin{table*}
\begin{centering}
\caption{Integrated line fluxes in the SE-region apertures with S/N$\,\geq\,3$ \label{tab:seflux} }
\begin{tabular}{lrccccc}
\hline
\hline
\multicolumn{1}{c}{Line} & \multicolumn{1}{c}{Rest $\lambda$} & JWST-SE-1 & JWST-SE-2 & JWST-SE-3 & SE & VLA-SE\\
& \multicolumn{1}{c}{(\micron)} & \multicolumn{5}{c}{($10^{-21}$\,W\,m$^{-2}$)}\\
\hline
\hline
\multicolumn{7}{c}{\hi\ recombination lines} \\
\hline
\\
HI(10-6) & 5.1287 & 11.84 $\,\pm\,$3.39 & 13.32$\,\pm\,$3.12 &  --- & 11.62$\,\pm\,$2.81 & 15.20$\,\pm\,$2.94 \\
HI(10-7) & 8.7601 & 9.23 $\,\pm\,$0.94 & 12.12$\,\pm\,$1.30 & 7.13$\,\pm\,$1.24 & 11.56$\,\pm\,$1.18 & 13.43$\,\pm\,$1.30 \\
HI(10-8) & 16.2091 & 5.37 $\,\pm\,$0.63 & 7.25$\,\pm\,$0.78 & 2.76$\,\pm\,$0.55 & 6.23$\,\pm\,$0.59 & 8.04$\,\pm\,$0.68 \\
HI(12-9) & 16.8806 &  --- & 2.46$\,\pm\,$0.61 &  --- & 2.01$\,\pm\,$0.57 & 2.62$\,\pm\,$0.60 \\
HI(13-8) & 9.392 &  --- &  --- &  --- & 4.67$\,\pm\,$1.10 &  --- \\
HI(13-9) & 14.1831 &  --- & 2.57$\,\pm\,$0.77 &  --- &  --- & 2.81$\,\pm\,$0.75 \\
HI(14-7) & 5.9568 &  --- & 10.20$\,\pm\,$2.42 &  --- &  --- & 11.35$\,\pm\,$2.40 \\
HI(15-9) & 11.5395 &  --- &  --- &  --- & 2.39$\,\pm\,$0.46 & 2.42$\,\pm\,$0.50 \\
Pf\,$\alpha$ & 7.4599 & 92.87 $\,\pm\,$2.26 & 117.90$\,\pm\,$2.40 & 59.04$\,\pm\,$2.14 & 104.50$\,\pm\,$2.38 & 128.30$\,\pm\,$2.51 \\
Hu\,$\alpha$ & 12.3719 & 32.64 $\,\pm\,$1.78 & 42.21$\,\pm\,$2.58 & 18.04$\,\pm\,$1.42 & 37.20$\,\pm\,$2.10 & 48.26$\,\pm\,$2.97 \\
Hu\,$\beta$ & 7.5025 & 24.70 $\,\pm\,$2.02 & 30.99$\,\pm\,$2.73 & 11.25$\,\pm\,$1.78 & 26.17$\,\pm\,$2.40 & 34.78$\,\pm\,$3.15 \\
HI(8-7) & 19.0619 &  --- & 19.27$\,\pm\,$2.01 &  --- & 13.72$\,\pm\,$2.39 & 18.69$\,\pm\,$2.08 \\
Hu\,$\gamma$ & 5.9082 & 14.18 $\,\pm\,$3.12 & 18.44$\,\pm\,$3.23 & 10.95$\,\pm\,$2.10 & 16.43$\,\pm\,$2.43 & 20.25$\,\pm\,$3.12 \\
HI(9-7) & 11.3087 & 13.66 $\,\pm\,$0.94 & 15.55$\,\pm\,$1.19 & 5.87$\,\pm\,$0.88 & 13.66$\,\pm\,$1.01 & 17.04$\,\pm\,$1.03 \\
\\
\hline
\multicolumn{7}{c}{Fine-structure lines} \\
\hline
\\
\ariii\ & 8.9914 & 41.10 $\,\pm\,$1.01 & 45.68$\,\pm\,$1.88 & 21.82$\,\pm\,$1.16 & 44.22$\,\pm\,$1.25 & 52.16$\,\pm\,$1.63 \\
\arii\ & 6.9853 & 12.66 $\,\pm\,$2.13 & 10.94$\,\pm\,$1.62 &  --- & 12.78$\,\pm\,$1.65 & 11.51$\,\pm\,$1.44 \\
\feii\ & 5.3402 & 36.34 $\,\pm\,$3.25 & 42.98$\,\pm\,$6.41 & 21.94$\,\pm\,$3.02 & 40.13$\,\pm\,$3.06 & 44.79$\,\pm\,$3.94 \\
\neiii\ & 15.5551 & 270.90 $\,\pm\,$1.95 & 308.20$\,\pm\,$1.46 & 146.00$\,\pm\,$0.81 & 292.00$\,\pm\,$1.68 & 361.40$\,\pm\,$1.66 \\
\neii\ & 12.8135 & 34.16 $\,\pm\,$0.79 & 39.16$\,\pm\,$0.92 & 20.98$\,\pm\,$0.90 & 36.53$\,\pm\,$0.88 & 39.56$\,\pm\,$0.98 \\
\oiv\ & 25.8903 &  --- & 47.94$\,\pm\,$12.93 &  --- &  --- & 54.93$\,\pm\,$13.93 \\
\siii\ & 18.713 & 132.20 $\,\pm\,$1.95 & 166.70$\,\pm\,$4.52 & 67.44$\,\pm\,$2.78 & 143.20$\,\pm\,$2.57 & 186.90$\,\pm\,$3.91 \\
\siv\ & 10.5105 & 185.00 $\,\pm\,$2.42 & 219.60$\,\pm\,$3.46 & 84.73$\,\pm\,$1.27 & 198.60$\,\pm\,$2.72 & 291.70$\,\pm\,$3.84 \\
\\
\\
\hline
\hline
\end{tabular}
\end{centering}
\end{table*}



\bibliography{izw18}{}

\begin{thebibliography}{}
\expandafter\ifx\csname natexlab\endcsname\relax\def\natexlab#1{#1}\fi
\providecommand{\url}[1]{\href{#1}{#1}}
\providecommand{\dodoi}[1]{doi:~\href{http://doi.org/#1}{\nolinkurl{#1}}}
\providecommand{\doeprint}[1]{\href{http://ascl.net/#1}{\nolinkurl{http://ascl.net/#1}}}
\providecommand{\doarXiv}[1]{\href{https://arxiv.org/abs/#1}{\nolinkurl{https://arxiv.org/abs/#1}}}

\bibitem[{{Alarie} \& {Morisset}(2019)}]{alarie19}
{Alarie}, A., \& {Morisset}, C. 2019, \rmxaa, 55, 377,
  \dodoi{10.22201/ia.01851101p.2019.55.02.21}

\bibitem[{{Aloisi} {et~al.}(2007){Aloisi}, {Clementini}, {Tosi}, {Annibali},
  {Contreras}, {Fiorentino}, {Mack}, {Marconi}, {Musella}, {Saha}, {Sirianni},
  \& {van der Marel}}]{aloisi07}
{Aloisi}, A., {Clementini}, G., {Tosi}, M., {et~al.} 2007, \apjl, 667, L151,
  \dodoi{10.1086/522368}

\bibitem[{{Annibali} {et~al.}(2013){Annibali}, {Cignoni}, {Tosi}, {van der
  Marel}, {Aloisi}, {Clementini}, {Contreras Ramos}, {Fiorentino}, {Marconi},
  \& {Musella}}]{annibali13}
{Annibali}, F., {Cignoni}, M., {Tosi}, M., {et~al.} 2013, \aj, 146, 144,
  \dodoi{10.1088/0004-6256/146/6/144}

\bibitem[{{Argyriou} {et~al.}(2023){Argyriou}, {Glasse}, {Law}, {Labiano},
  {{\'A}lvarez-M{\'a}rquez}, {Patapis}, {Kavanagh}, {Gasman}, {Mueller},
  {Larson}, {Vandenbussche}, {Glauser}, {Royer}, {Dicken}, {Harkett},
  {Sargent}, {Engesser}, {Jones}, {Kendrew}, {Noriega-Crespo}, {Brandl},
  {Rieke}, {Wright}, {Lee}, \& {Wells}}]{argyriou23}
{Argyriou}, I., {Glasse}, A., {Law}, D.~R., {et~al.} 2023, \aap, 675, A111,
  \dodoi{10.1051/0004-6361/202346489}

\bibitem[{{Astropy Collaboration} {et~al.}(2013){Astropy Collaboration},
  {Robitaille}, {Tollerud}, {Greenfield}, {Droettboom}, {Bray}, {Aldcroft},
  {Davis}, {Ginsburg}, {Price-Whelan}, {Kerzendorf}, {Conley}, {Crighton},
  {Barbary}, {Muna}, {Ferguson}, {Grollier}, {Parikh}, {Nair}, {Unther},
  {Deil}, {Woillez}, {Conseil}, {Kramer}, {Turner}, {Singer}, {Fox}, {Weaver},
  {Zabalza}, {Edwards}, {Azalee Bostroem}, {Burke}, {Casey}, {Crawford},
  {Dencheva}, {Ely}, {Jenness}, {Labrie}, {Lim}, {Pierfederici}, {Pontzen},
  {Ptak}, {Refsdal}, {Servillat}, \& {Streicher}}]{2013A&A...558A..33A}
{Astropy Collaboration}, {Robitaille}, T.~P., {Tollerud}, E.~J., {et~al.} 2013,
  \aap, 558, A33, \dodoi{10.1051/0004-6361/201322068}

\bibitem[{{Astropy Collaboration} {et~al.}(2018){Astropy Collaboration},
  {Price-Whelan}, {Sip{\H{o}}cz}, {G{\"u}nther}, {Lim}, {Crawford}, {Conseil},
  {Shupe}, {Craig}, {Dencheva}, {Ginsburg}, {VanderPlas}, {Bradley},
  {P{\'e}rez-Su{\'a}rez}, {de Val-Borro}, {Aldcroft}, {Cruz}, {Robitaille},
  {Tollerud}, {Ardelean}, {Babej}, {Bach}, {Bachetti}, {Bakanov}, {Bamford},
  {Barentsen}, {Barmby}, {Baumbach}, {Berry}, {Biscani}, {Boquien}, {Bostroem},
  {Bouma}, {Brammer}, {Bray}, {Breytenbach}, {Buddelmeijer}, {Burke},
  {Calderone}, {Cano Rodr{\'\i}guez}, {Cara}, {Cardoso}, {Cheedella}, {Copin},
  {Corrales}, {Crichton}, {D'Avella}, {Deil}, {Depagne}, {Dietrich}, {Donath},
  {Droettboom}, {Earl}, {Erben}, {Fabbro}, {Ferreira}, {Finethy}, {Fox},
  {Garrison}, {Gibbons}, {Goldstein}, {Gommers}, {Greco}, {Greenfield},
  {Groener}, {Grollier}, {Hagen}, {Hirst}, {Homeier}, {Horton}, {Hosseinzadeh},
  {Hu}, {Hunkeler}, {Ivezi{\'c}}, {Jain}, {Jenness}, {Kanarek}, {Kendrew},
  {Kern}, {Kerzendorf}, {Khvalko}, {King}, {Kirkby}, {Kulkarni}, {Kumar},
  {Lee}, {Lenz}, {Littlefair}, {Ma}, {Macleod}, {Mastropietro}, {McCully},
  {Montagnac}, {Morris}, {Mueller}, {Mumford}, {Muna}, {Murphy}, {Nelson},
  {Nguyen}, {Ninan}, {N{\"o}the}, {Ogaz}, {Oh}, {Parejko}, {Parley}, {Pascual},
  {Patil}, {Patil}, {Plunkett}, {Prochaska}, {Rastogi}, {Reddy Janga},
  {Sabater}, {Sakurikar}, {Seifert}, {Sherbert}, {Sherwood-Taylor}, {Shih},
  {Sick}, {Silbiger}, {Singanamalla}, {Singer}, {Sladen}, {Sooley},
  {Sornarajah}, {Streicher}, {Teuben}, {Thomas}, {Tremblay}, {Turner},
  {Terr{\'o}n}, {van Kerkwijk}, {de la Vega}, {Watkins}, {Weaver}, {Whitmore},
  {Woillez}, {Zabalza}, \& {Astropy Contributors}}]{2018AJ....156..123A}
{Astropy Collaboration}, {Price-Whelan}, A.~M., {Sip{\H{o}}cz}, B.~M., {et~al.}
  2018, \aj, 156, 123, \dodoi{10.3847/1538-3881/aabc4f}

\bibitem[{{Atek} {et~al.}(2022){Atek}, {Furtak}, {Oesch}, {van Dokkum},
  {Reddy}, {Contini}, {Illingworth}, \& {Wilkins}}]{atek22}
{Atek}, H., {Furtak}, L.~J., {Oesch}, P., {et~al.} 2022, \mnras, 511, 4464,
  \dodoi{10.1093/mnras/stac360}

\bibitem[{{Atek} {et~al.}(2024){Atek}, {Labb{\'e}}, {Furtak}, {Chemerynska},
  {Fujimoto}, {Setton}, {Miller}, {Oesch}, {Bezanson}, {Price}, {Dayal},
  {Zitrin}, {Kokorev}, {Weaver}, {Brammer}, {Dokkum}, {Williams}, {Cutler},
  {Feldmann}, {Fudamoto}, {Greene}, {Leja}, {Maseda}, {Muzzin}, {Pan},
  {Papovich}, {Nelson}, {Nanayakkara}, {Stark}, {Stefanon}, {Suess}, {Wang}, \&
  {Whitaker}}]{atek24}
{Atek}, H., {Labb{\'e}}, I., {Furtak}, L.~J., {et~al.} 2024, \nat, 626, 975,
  \dodoi{10.1038/s41586-024-07043-6}

\bibitem[{{Babul} \& {Ferguson}(1996)}]{babul96}
{Babul}, A., \& {Ferguson}, H.~C. 1996, \apj, 458, 100, \dodoi{10.1086/176795}

\bibitem[{{Behroozi} {et~al.}(2019){Behroozi}, {Wechsler}, {Hearin}, \&
  {Conroy}}]{behroozi19}
{Behroozi}, P., {Wechsler}, R.~H., {Hearin}, A.~P., \& {Conroy}, C. 2019,
  \mnras, 488, 3143, \dodoi{10.1093/mnras/stz1182}

\bibitem[{{Behroozi} {et~al.}(2020){Behroozi}, {Conroy}, {Wechsler}, {Hearin},
  {Williams}, {Moster}, {Yung}, {Somerville}, {Gottl{\"o}ber}, {Yepes}, \&
  {Endsley}}]{behroozi20}
{Behroozi}, P., {Conroy}, C., {Wechsler}, R.~H., {et~al.} 2020, \mnras, 499,
  5702, \dodoi{10.1093/mnras/staa3164}

\bibitem[{{Berghea} {et~al.}(2020){Berghea}, {Johnson}, {Secrest}, {Dudik},
  {Hennessy}, \& {El-khatib}}]{berghea20}
{Berghea}, C.~T., {Johnson}, M.~C., {Secrest}, N.~J., {et~al.} 2020, \apj, 896,
  117, \dodoi{10.3847/1538-4357/ab9108}

\bibitem[{{Bernard-Salas} {et~al.}(2009){Bernard-Salas}, {Spoon},
  {Charmandaris}, {Lebouteiller}, {Farrah}, {Devost}, {Brandl}, {Wu}, {Armus},
  {Hao}, {Sloan}, {Weedman}, \& {Houck}}]{bernardsalas09}
{Bernard-Salas}, J., {Spoon}, H.~W.~W., {Charmandaris}, V., {et~al.} 2009,
  \apjs, 184, 230, \dodoi{10.1088/0067-0049/184/2/230}

\bibitem[{{Bhowmick} {et~al.}(2024){Bhowmick}, {Blecha}, {Torrey}, {Kelley},
  {Weinberger}, {Vogelsberger}, {Hernquist}, {Somerville}, \&
  {Evans}}]{bhowmick24}
{Bhowmick}, A.~K., {Blecha}, L., {Torrey}, P., {et~al.} 2024, \mnras, 531,
  4311, \dodoi{10.1093/mnras/stae1386}

\bibitem[{{Binette} {et~al.}(1985){Binette}, {Dopita}, \& {Tuohy}}]{binette85}
{Binette}, L., {Dopita}, M.~A., \& {Tuohy}, I.~R. 1985, \apj, 297, 476,
  \dodoi{10.1086/163544}

\bibitem[{{Bolatto} {et~al.}(2013){Bolatto}, {Wolfire}, \& {Leroy}}]{bolatto13}
{Bolatto}, A.~D., {Wolfire}, M., \& {Leroy}, A.~K. 2013, \araa, 51, 207,
  \dodoi{10.1146/annurev-astro-082812-140944}

\bibitem[{{Bortolini} {et~al.}(2024){Bortolini}, {{\"O}stlin}, {Habel},
  {Hirschauer}, {Jones}, {Justtanont}, {Meixner}, {Boyer}, {Blommaert},
  {Crouzet}, {Lenki{\'c}}, {Nally}, {Sargent}, {van der Werf}, {G{\"u}del},
  {Henning}, \& {Lagage}}]{bortolini24}
{Bortolini}, G., {{\"O}stlin}, G., {Habel}, N., {et~al.} 2024, \aap, 689, A146,
  \dodoi{10.1051/0004-6361/202450632}

\bibitem[{Bradley {et~al.}(2024)Bradley, Sip{\H o}cz, Robitaille, Tollerud,
  Vin{\'{\i}}cius, Deil, Barbary, Wilson, Busko, Donath, G{\"u}nther, Cara,
  Lim, Me{\ss}linger, Conseil, Burnett, Bostroem, Droettboom, Bray, Bratholm,
  Ginsburg, Jamieson, Barentsen, Craig, Morris, Perrin, Rathi, Pascual, \&
  Georgiev}]{larry_bradley_2024_13989456}
Bradley, L., Sip{\H o}cz, B., Robitaille, T., {et~al.} 2024, astropy/photutils:
  2.0.2, 2.0.2,  Zenodo, \dodoi{10.5281/zenodo.13989456}

\bibitem[{{Bromm} \& {Yoshida}(2011)}]{bromm11}
{Bromm}, V., \& {Yoshida}, N. 2011, \araa, 49, 373,
  \dodoi{10.1146/annurev-astro-081710-102608}

\bibitem[{{Brown} {et~al.}(2002){Brown}, {Heap}, {Hubeny}, {Lanz}, \&
  {Lindler}}]{brown02}
{Brown}, T.~M., {Heap}, S.~R., {Hubeny}, I., {Lanz}, T., \& {Lindler}, D. 2002,
  \apjl, 579, L75, \dodoi{10.1086/345336}

\bibitem[{{Bullock} \& {Boylan-Kolchin}(2017)}]{bullock17}
{Bullock}, J.~S., \& {Boylan-Kolchin}, M. 2017, \araa, 55, 343,
  \dodoi{10.1146/annurev-astro-091916-055313}

\bibitem[{{Bushouse} {et~al.}(2024){Bushouse}, {Eisenhamer}, {Dencheva},
  {Davies}, {Greenfield}, {Morrison}, {Hodge}, {Simon}, {Grumm}, {Droettboom},
  {Slavich}, {Sosey}, {Pauly}, {Miller}, {Jedrzejewski}, {Hack}, {Davis},
  {Crawford}, {Law}, {Gordon}, {Regan}, {Cara}, {MacDonald}, {Bradley},
  {Shanahan}, {Jamieson}, {Teodoro}, {Williams}, \&
  {Pena-Guerrero}}]{bushouse24}
{Bushouse}, H., {Eisenhamer}, J., {Dencheva}, N., {et~al.} 2024, {JWST
  Calibration Pipeline}, 1.15.0,  Zenodo, \dodoi{10.5281/zenodo.12556702}

\bibitem[{{Calzetti} {et~al.}(1994){Calzetti}, {Kinney}, \&
  {Storchi-Bergmann}}]{calzetti94}
{Calzetti}, D., {Kinney}, A.~L., \& {Storchi-Bergmann}, T. 1994, \apj, 429,
  582, \dodoi{10.1086/174346}

\bibitem[{{Campbell} {et~al.}(1986){Campbell}, {Terlevich}, \&
  {Melnick}}]{campbell86}
{Campbell}, A., {Terlevich}, R., \& {Melnick}, J. 1986, \mnras, 223, 811,
  \dodoi{10.1093/mnras/223.4.811}

\bibitem[{{Cannon} {et~al.}(2002){Cannon}, {Skillman}, {Garnett}, \&
  {Dufour}}]{cannon02}
{Cannon}, J.~M., {Skillman}, E.~D., {Garnett}, D.~R., \& {Dufour}, R.~J. 2002,
  \apj, 565, 931, \dodoi{10.1086/324691}

\bibitem[{{Cannon} {et~al.}(2005){Cannon}, {Walter}, {Skillman}, \& {van
  Zee}}]{cannon05}
{Cannon}, J.~M., {Walter}, F., {Skillman}, E.~D., \& {van Zee}, L. 2005, \apjl,
  621, L21, \dodoi{10.1086/428943}

\bibitem[{{Carnall} {et~al.}(2023){Carnall}, {Begley}, {McLeod}, {Hamadouche},
  {Donnan}, {McLure}, {Dunlop}, {Milvang-Jensen}, {Bondestam}, {Cullen},
  {Jewell}, \& {Pollock}}]{carnall23}
{Carnall}, A.~C., {Begley}, R., {McLeod}, D.~J., {et~al.} 2023, \mnras, 518,
  L45, \dodoi{10.1093/mnrasl/slac136}

\bibitem[{{Cervi{\~n}o} {et~al.}(2002){Cervi{\~n}o}, {Mas-Hesse}, \&
  {Kunth}}]{cervino02}
{Cervi{\~n}o}, M., {Mas-Hesse}, J.~M., \& {Kunth}, D. 2002, \aap, 392, 19,
  \dodoi{10.1051/0004-6361:20020785}

\bibitem[{{Chevalier} \& {Clegg}(1985)}]{chevalier85}
{Chevalier}, R.~A., \& {Clegg}, A.~W. 1985, \nat, 317, 44,
  \dodoi{10.1038/317044a0}

\bibitem[{{Chisholm} {et~al.}(2024){Chisholm}, {Berg}, {Endsley}, {Gazagnes},
  {Richardson}, {Lambrides}, {Greene}, {Finkelstein}, {Flury}, {Guseva},
  {Henry}, {Hutchison}, {Izotov}, {Marques-Chaves}, {Oesch}, {Papovich},
  {Saldana-Lopez}, {Schaerer}, \& {Stephenson}}]{chisholm24}
{Chisholm}, J., {Berg}, D.~A., {Endsley}, R., {et~al.} 2024, \mnras, 534, 2633,
  \dodoi{10.1093/mnras/stae2199}

\bibitem[{{Churchwell} {et~al.}(2006){Churchwell}, {Povich}, {Allen}, {Taylor},
  {Meade}, {Babler}, {Indebetouw}, {Watson}, {Whitney}, {Wolfire}, {Bania},
  {Benjamin}, {Clemens}, {Cohen}, {Cyganowski}, {Jackson}, {Kobulnicky},
  {Mathis}, {Mercer}, {Stolovy}, {Uzpen}, {Watson}, \& {Wolff}}]{churchwell06}
{Churchwell}, E., {Povich}, M.~S., {Allen}, D., {et~al.} 2006, \apj, 649, 759,
  \dodoi{10.1086/507015}

\bibitem[{{Conselice} {et~al.}(2025){Conselice}, {Adams}, {Harvey}, {Austin},
  {Ferreira}, {Ormerod}, {Duan}, {Trussler}, {Li}, {Juod{\v{z}}balis},
  {Westcott}, {Harris}, {Seeyave}, {Bluck}, {Windhorst}, {Bhatawdekar}, {Coe},
  {Cohen}, {Cheng}, {Driver}, {Frye}, {Furtak}, {Grogin}, {Hathi}, {Holwerda},
  {Jansen}, {Koekemoer}, {Marshall}, {Nonino}, {Robotham}, {Summers},
  {Wilkins}, {Willmer}, {Yan}, \& {Zitrin}}]{conselice25}
{Conselice}, C.~J., {Adams}, N., {Harvey}, T., {et~al.} 2025, \apj, 983, 30,
  \dodoi{10.3847/1538-4357/ada608}

\bibitem[{{Cormier} {et~al.}(2015){Cormier}, {Madden}, {Lebouteiller}, {Abel},
  {Hony}, {Galliano}, {R{\'e}my-Ruyer}, {Bigiel}, {Baes}, {Boselli},
  {Chevance}, {Cooray}, {De Looze}, {Doublier}, {Galametz}, {Hughes},
  {Karczewski}, {Lee}, {Lu}, \& {Spinoglio}}]{cormier15}
{Cormier}, D., {Madden}, S.~C., {Lebouteiller}, V., {et~al.} 2015, \aap, 578,
  A53, \dodoi{10.1051/0004-6361/201425207}

\bibitem[{{Curti} {et~al.}(2023){Curti}, {D'Eugenio}, {Carniani}, {Maiolino},
  {Sandles}, {Witstok}, {Baker}, {Bennett}, {Piotrowska}, {Tacchella},
  {Charlot}, {Nakajima}, {Maheson}, {Mannucci}, {Amiri}, {Arribas}, {Belfiore},
  {Bonaventura}, {Bunker}, {Chevallard}, {Cresci}, {Curtis-Lake},
  {Hayden-Pawson}, {Jones}, {Kumari}, {Laseter}, {Looser}, {Marconi}, {Maseda},
  {Scholtz}, {Smit}, {{\"U}bler}, \& {Wallace}}]{curti23}
{Curti}, M., {D'Eugenio}, F., {Carniani}, S., {et~al.} 2023, \mnras, 518, 425,
  \dodoi{10.1093/mnras/stac2737}

\bibitem[{{Curti} {et~al.}(2024){Curti}, {Maiolino}, {Curtis-Lake},
  {Chevallard}, {Carniani}, {D'Eugenio}, {Looser}, {Scholtz}, {Charlot},
  {Cameron}, {{\"U}bler}, {Witstok}, {Boyett}, {Laseter}, {Sandles}, {Arribas},
  {Bunker}, {Giardino}, {Maseda}, {Rawle}, {Rodr{\'\i}guez Del Pino}, {Smit},
  {Willott}, {Eisenstein}, {Hausen}, {Johnson}, {Rieke}, {Robertson},
  {Tacchella}, {Williams}, {Willmer}, {Baker}, {Bhatawdekar}, {Egami},
  {Helton}, {Ji}, {Kumari}, {Perna}, {Shivaei}, \& {Sun}}]{curti24}
{Curti}, M., {Maiolino}, R., {Curtis-Lake}, E., {et~al.} 2024, \aap, 684, A75,
  \dodoi{10.1051/0004-6361/202346698}

\bibitem[{{Dale} {et~al.}(2009){Dale}, {Smith}, {Schlawin}, {Armus},
  {Buckalew}, {Cohen}, {Helou}, {Jarrett}, {Johnson}, {Moustakas}, {Murphy},
  {Roussel}, {Sheth}, {Staudaher}, {Bot}, {Calzetti}, {Engelbracht}, {Gordon},
  {Hollenbach}, {Kennicutt}, \& {Malhotra}}]{dale09}
{Dale}, D.~A., {Smith}, J.~D.~T., {Schlawin}, E.~A., {et~al.} 2009, \apj, 693,
  1821, \dodoi{10.1088/0004-637X/693/2/1821}

\bibitem[{{Davidzon} {et~al.}(2017){Davidzon}, {Ilbert}, {Laigle}, {Coupon},
  {McCracken}, {Delvecchio}, {Masters}, {Capak}, {Hsieh}, {Le F{\`e}vre},
  {Tresse}, {Bethermin}, {Chang}, {Faisst}, {Le Floc'h}, {Steinhardt}, {Toft},
  {Aussel}, {Dubois}, {Hasinger}, {Salvato}, {Sanders}, {Scoville}, \&
  {Silverman}}]{davidzon17}
{Davidzon}, I., {Ilbert}, O., {Laigle}, C., {et~al.} 2017, \aap, 605, A70,
  \dodoi{10.1051/0004-6361/201730419}

\bibitem[{{Decleir} {et~al.}(2022){Decleir}, {Gordon}, {Andrews}, {Clayton},
  {Cushing}, {Misselt}, {Pendleton}, {Rayner}, {Vacca}, \&
  {Whittet}}]{decleir22}
{Decleir}, M., {Gordon}, K.~D., {Andrews}, J.~E., {et~al.} 2022, \apj, 930, 15,
  \dodoi{10.3847/1538-4357/ac5dbe}

\bibitem[{{Della Bruna} {et~al.}(2022){Della Bruna}, {Adamo}, {McLeod},
  {Smith}, {Savard}, {Robert}, {Sun}, {Amram}, {Bik}, {Blair}, {Long},
  {Renaud}, {Walterbos}, \& {Usher}}]{dellabruna22}
{Della Bruna}, L., {Adamo}, A., {McLeod}, A.~F., {et~al.} 2022, \aap, 666, A29,
  \dodoi{10.1051/0004-6361/202243395}

\bibitem[{{Deshmukh} {et~al.}(2024){Deshmukh}, {Linden}, {Calzetti}, {Adamo},
  {Messa}, {Grasha}, {Sabbi}, {Smith}, \& {Johnson}}]{deshmukh24}
{Deshmukh}, S., {Linden}, S.~T., {Calzetti}, D., {et~al.} 2024, \apjl, 974,
  L24, \dodoi{10.3847/2041-8213/ad7ba9}

\bibitem[{{Draine}(2011)}]{draine11}
{Draine}, B.~T. 2011, {Physics of the Interstellar and Intergalactic Medium}

\bibitem[{{Draine} \& {Bertoldi}(1996)}]{draine96}
{Draine}, B.~T., \& {Bertoldi}, F. 1996, \apj, 468, 269, \dodoi{10.1086/177689}

\bibitem[{{Draine} \& {McKee}(1993)}]{draine93}
{Draine}, B.~T., \& {McKee}, C.~F. 1993, \araa, 31, 373,
  \dodoi{10.1146/annurev.aa.31.090193.002105}

\bibitem[{{Earl} {et~al.}(2020){Earl}, {Tollerud}, {Jones}, {Kerzendorf},
  {Shaileshahuja}, {D'Avella}, {Robitaille}, {Ginsburg}, {Sip{\H{o}}cz},
  {Busko}, {Ogaz}, {G{\"u}nther}, {Rosteen}, {Barbary}, {Foster}, {Torres},
  {Droettboom}, {Bray}, {Davies}, {Casey}, {Ferguson}, {Crawford}, {Teuben},
  {Homeier}, {Cruz}, {Pickering}, {Dencheva}, {Ninan}, {Gmduvvuri}, \&
  {Deil}}]{specutils}
{Earl}, N., {Tollerud}, E., {Jones}, C., {et~al.} 2020, {astropy/specutils:
  v1.0}, v1.0,  Zenodo, \dodoi{10.5281/zenodo.3718589}

\bibitem[{{Eldridge} {et~al.}(2017){Eldridge}, {Stanway}, {Xiao}, {McClelland},
  {Taylor}, {Ng}, {Greis}, \& {Bray}}]{eldridge17}
{Eldridge}, J.~J., {Stanway}, E.~R., {Xiao}, L., {et~al.} 2017, \pasa, 34,
  e058, \dodoi{10.1017/pasa.2017.51}

\bibitem[{{Endsley} {et~al.}(2024){Endsley}, {Stark}, {Whitler}, {Topping},
  {Johnson}, {Robertson}, {Tacchella}, {Alberts}, {Baker}, {Bhatawdekar},
  {Boyett}, {Bunker}, {Cameron}, {Carniani}, {Charlot}, {Chen}, {Chevallard},
  {Curtis-Lake}, {Danhaive}, {Egami}, {Eisenstein}, {Hainline}, {Helton}, {Ji},
  {Looser}, {Maiolino}, {Nelson}, {Pusk{\'a}s}, {Rieke}, {Rieke}, {Rix},
  {Sandles}, {Saxena}, {Simmonds}, {Smit}, {Sun}, {Williams}, {Willmer},
  {Willott}, \& {Witstok}}]{endsley24}
{Endsley}, R., {Stark}, D.~P., {Whitler}, L., {et~al.} 2024, \mnras, 533, 1111,
  \dodoi{10.1093/mnras/stae1857}

\bibitem[{{Engelbracht} {et~al.}(2005){Engelbracht}, {Gordon}, {Rieke},
  {Werner}, {Dale}, \& {Latter}}]{engelbracht05}
{Engelbracht}, C.~W., {Gordon}, K.~D., {Rieke}, G.~H., {et~al.} 2005, \apjl,
  628, L29, \dodoi{10.1086/432613}

\bibitem[{{Feltre} {et~al.}(2016){Feltre}, {Charlot}, \& {Gutkin}}]{feltre16}
{Feltre}, A., {Charlot}, S., \& {Gutkin}, J. 2016, \mnras, 456, 3354,
  \dodoi{10.1093/mnras/stv2794}

\bibitem[{{Fisher} {et~al.}(2014){Fisher}, {Bolatto}, {Herrera-Camus},
  {Draine}, {Donaldson}, {Walter}, {Sandstrom}, {Leroy}, {Cannon}, \&
  {Gordon}}]{fisher14}
{Fisher}, D.~B., {Bolatto}, A.~D., {Herrera-Camus}, R., {et~al.} 2014, \nat,
  505, 186, \dodoi{10.1038/nature12765}

\bibitem[{{Fitzpatrick} {et~al.}(2019){Fitzpatrick}, {Massa}, {Gordon},
  {Bohlin}, \& {Clayton}}]{fitzpatrick19}
{Fitzpatrick}, E.~L., {Massa}, D., {Gordon}, K.~D., {Bohlin}, R., \& {Clayton},
  G.~C. 2019, \apj, 886, 108, \dodoi{10.3847/1538-4357/ab4c3a}

\bibitem[{{Flury} {et~al.}(2024){Flury}, {Arellano-C{\'o}rdova}, {Moran}, \&
  {Einsig}}]{flury24}
{Flury}, S.~R., {Arellano-C{\'o}rdova}, K.~Z., {Moran}, E.~C., \& {Einsig}, A.
  2024, arXiv e-prints, arXiv:2412.06763, \dodoi{10.48550/arXiv.2412.06763}

\bibitem[{{Franeck} {et~al.}(2022){Franeck}, {W{\"u}nsch},
  {Mart{\'\i}nez-Gonz{\'a}lez}, {Orlitov{\'a}}, {Boorman}, {Svoboda},
  {Sz{\'e}csi}, \& {Douna}}]{franeck22}
{Franeck}, A., {W{\"u}nsch}, R., {Mart{\'\i}nez-Gonz{\'a}lez}, S., {et~al.}
  2022, \apj, 927, 212, \dodoi{10.3847/1538-4357/ac4fc2}

\bibitem[{{Fumagalli} {et~al.}(2010){Fumagalli}, {Krumholz}, \&
  {Hunt}}]{fumagalli10}
{Fumagalli}, M., {Krumholz}, M.~R., \& {Hunt}, L.~K. 2010, \apj, 722, 919,
  \dodoi{10.1088/0004-637X/722/1/919}

\bibitem[{{Furtak} {et~al.}(2023){Furtak}, {Shuntov}, {Atek}, {Zitrin},
  {Richard}, {Lehnert}, \& {Chevallard}}]{furtak23}
{Furtak}, L.~J., {Shuntov}, M., {Atek}, H., {et~al.} 2023, \mnras, 519, 3064,
  \dodoi{10.1093/mnras/stac3717}

\bibitem[{{Gaia Collaboration} {et~al.}(2021){Gaia Collaboration}, {Brown},
  {Vallenari}, {Prusti}, {de Bruijne}, {Babusiaux}, {Biermann}, {Creevey},
  {Evans}, {Eyer}, {Hutton}, {Jansen}, {Jordi}, {Klioner}, {Lammers},
  {Lindegren}, {Luri}, {Mignard}, {Panem}, {Pourbaix}, {Randich}, {Sartoretti},
  {Soubiran}, {Walton}, {Arenou}, {Bailer-Jones}, {Bastian}, {Cropper},
  {Drimmel}, {Katz}, {Lattanzi}, {van Leeuwen}, {Bakker}, {Cacciari},
  {Casta{\~n}eda}, {De Angeli}, {Ducourant}, {Fabricius}, {Fouesneau},
  {Fr{\'e}mat}, {Guerra}, {Guerrier}, {Guiraud}, {Jean-Antoine Piccolo},
  {Masana}, {Messineo}, {Mowlavi}, {Nicolas}, {Nienartowicz}, {Pailler},
  {Panuzzo}, {Riclet}, {Roux}, {Seabroke}, {Sordo}, {Tanga}, {Th{\'e}venin},
  {Gracia-Abril}, {Portell}, {Teyssier}, {Altmann}, {Andrae}, {Bellas-Velidis},
  {Benson}, {Berthier}, {Blomme}, {Brugaletta}, {Burgess}, {Busso}, {Carry},
  {Cellino}, {Cheek}, {Clementini}, {Damerdji}, {Davidson}, {Delchambre},
  {Dell'Oro}, {Fern{\'a}ndez-Hern{\'a}ndez}, {Galluccio}, {Garc{\'\i}a-Lario},
  {Garcia-Reinaldos}, {Gonz{\'a}lez-N{\'u}{\~n}ez}, {Gosset}, {Haigron},
  {Halbwachs}, {Hambly}, {Harrison}, {Hatzidimitriou}, {Heiter},
  {Hern{\'a}ndez}, {Hestroffer}, {Hodgkin}, {Holl}, {Jan{\ss}en}, {Jevardat de
  Fombelle}, {Jordan}, {Krone-Martins}, {Lanzafame}, {L{\"o}ffler}, {Lorca},
  {Manteiga}, {Marchal}, {Marrese}, {Moitinho}, {Mora}, {Muinonen}, {Osborne},
  {Pancino}, {Pauwels}, {Petit}, {Recio-Blanco}, {Richards}, {Riello},
  {Rimoldini}, {Robin}, {Roegiers}, {Rybizki}, {Sarro}, {Siopis}, {Smith},
  {Sozzetti}, {Ulla}, {Utrilla}, {van Leeuwen}, {van Reeven}, {Abbas}, {Abreu
  Aramburu}, {Accart}, {Aerts}, {Aguado}, {Ajaj}, {Altavilla}, {{\'A}lvarez},
  {{\'A}lvarez Cid-Fuentes}, {Alves}, {Anderson}, {Anglada Varela}, {Antoja},
  {Audard}, {Baines}, {Baker}, {Balaguer-N{\'u}{\~n}ez}, {Balbinot}, {Balog},
  {Barache}, {Barbato}, {Barros}, {Barstow}, {Bartolom{\'e}}, {Bassilana},
  {Bauchet}, {Baudesson-Stella}, {Becciani}, {Bellazzini}, {Bernet}, {Bertone},
  {Bianchi}, {Blanco-Cuaresma}, {Boch}, {Bombrun}, {Bossini}, {Bouquillon},
  {Bragaglia}, {Bramante}, {Breedt}, {Bressan}, {Brouillet}, {Bucciarelli},
  {Burlacu}, {Busonero}, {Butkevich}, {Buzzi}, {Caffau}, {Cancelliere},
  {C{\'a}novas}, {Cantat-Gaudin}, {Carballo}, {Carlucci}, {Carnerero},
  {Carrasco}, {Casamiquela}, {Castellani}, {Castro-Ginard}, {Castro Sampol},
  {Chaoul}, {Charlot}, {Chemin}, {Chiavassa}, {Cioni}, {Comoretto}, {Cooper},
  {Cornez}, {Cowell}, {Crifo}, {Crosta}, {Crowley}, {Dafonte}, {Dapergolas},
  {David}, \& {David}}]{gaiadr3}
{Gaia Collaboration}, {Brown}, A.~G.~A., {Vallenari}, A., {et~al.} 2021, \aap,
  649, A1, \dodoi{10.1051/0004-6361/202039657}

\bibitem[{{Galliano} {et~al.}(2021){Galliano}, {Nersesian}, {Bianchi}, {De
  Looze}, {Roychowdhury}, {Baes}, {Casasola}, {Cassar{\'a}}, {Dobbels},
  {Fritz}, {Galametz}, {Jones}, {Madden}, {Mosenkov}, {Xilouris}, \&
  {Ysard}}]{galliano21}
{Galliano}, F., {Nersesian}, A., {Bianchi}, S., {et~al.} 2021, \aap, 649, A18,
  \dodoi{10.1051/0004-6361/202039701}

\bibitem[{{Garofali} {et~al.}(2024){Garofali}, {Basu-Zych}, {Johnson},
  {Tzanavaris}, {Jaskot}, {Richardson}, {Lehmer}, {Yukita}, {Hodges-Kluck},
  {Hornschemeier}, {Ptak}, \& {Vulic}}]{garofali24}
{Garofali}, K., {Basu-Zych}, A.~R., {Johnson}, B.~D., {et~al.} 2024, \apj, 960,
  13, \dodoi{10.3847/1538-4357/ad0a6a}

\bibitem[{{Gil de Paz} {et~al.}(2003){Gil de Paz}, {Madore}, \&
  {Pevunova}}]{gildepaz03}
{Gil de Paz}, A., {Madore}, B.~F., \& {Pevunova}, O. 2003, \apjs, 147, 29,
  \dodoi{10.1086/374737}

\bibitem[{{Gordon}(2024)}]{gordon24}
{Gordon}, K. 2024, The Journal of Open Source Software, 9, 7023,
  \dodoi{10.21105/joss.07023}

\bibitem[{{Gordon} {et~al.}(2009){Gordon}, {Cartledge}, \&
  {Clayton}}]{gordon09}
{Gordon}, K.~D., {Cartledge}, S., \& {Clayton}, G.~C. 2009, \apj, 705, 1320,
  \dodoi{10.1088/0004-637X/705/2/1320}

\bibitem[{{Gordon} {et~al.}(2023){Gordon}, {Clayton}, {Decleir}, {Fitzpatrick},
  {Massa}, {Misselt}, \& {Tollerud}}]{gordon23}
{Gordon}, K.~D., {Clayton}, G.~C., {Decleir}, M., {et~al.} 2023, \apj, 950, 86,
  \dodoi{10.3847/1538-4357/accb59}

\bibitem[{{Gordon} {et~al.}(2021){Gordon}, {Misselt}, {Bouwman}, {Clayton},
  {Decleir}, {Hines}, {Pendleton}, {Rieke}, {Smith}, \& {Whittet}}]{gordon21}
{Gordon}, K.~D., {Misselt}, K.~A., {Bouwman}, J., {et~al.} 2021, \apj, 916, 33,
  \dodoi{10.3847/1538-4357/ac00b7}

\bibitem[{{Green}(2018)}]{green18}
{Green}, G. 2018, The Journal of Open Source Software, 3, 695,
  \dodoi{10.21105/joss.00695}

\bibitem[{{G{\'u}rpide} {et~al.}(2022){G{\'u}rpide}, {Parra}, {Godet},
  {Contini}, \& {Olive}}]{gurpide22}
{G{\'u}rpide}, A., {Parra}, M., {Godet}, O., {Contini}, T., \& {Olive}, J.~F.
  2022, \aap, 666, A100, \dodoi{10.1051/0004-6361/202142229}

\bibitem[{{Hao} {et~al.}(2009){Hao}, {Wu}, {Charmandaris}, {Spoon},
  {Bernard-Salas}, {Devost}, {Lebouteiller}, \& {Houck}}]{hao09}
{Hao}, L., {Wu}, Y., {Charmandaris}, V., {et~al.} 2009, \apj, 704, 1159,
  \dodoi{10.1088/0004-637X/704/2/1159}

\bibitem[{Harris {et~al.}(2020)Harris, Millman, van~der Walt, Gommers,
  Virtanen, Cournapeau, Wieser, Taylor, Berg, Smith, Kern, Picus, Hoyer, van
  Kerkwijk, Brett, Haldane, del R{\'{i}}o, Wiebe, Peterson,
  G{\'{e}}rard-Marchant, Sheppard, Reddy, Weckesser, Abbasi, Gohlke, \&
  Oliphant}]{harris2020array}
Harris, C.~R., Millman, K.~J., van~der Walt, S.~J., {et~al.} 2020, Nature, 585,
  357, \dodoi{10.1038/s41586-020-2649-2}

\bibitem[{{Hassani} {et~al.}(2023){Hassani}, {Rosolowsky}, {Leroy}, {Boquien},
  {Lee}, {Barnes}, {Belfiore}, {Bigiel}, {Cao}, {Chevance}, {Dale}, {Egorov},
  {Emsellem}, {Faesi}, {Grasha}, {Kim}, {Klessen}, {Kreckel}, {Kruijssen},
  {Larson}, {Meidt}, {Sandstrom}, {Schinnerer}, {Thilker}, {Watkins},
  {Whitmore}, \& {Williams}}]{hassani23}
{Hassani}, H., {Rosolowsky}, E., {Leroy}, A.~K., {et~al.} 2023, \apjl, 944,
  L21, \dodoi{10.3847/2041-8213/aca8ab}

\bibitem[{{Heintz} {et~al.}(2023){Heintz}, {Gim{\'e}nez-Arteaga}, {Fujimoto},
  {Brammer}, {Espada}, {Gillman}, {Gonz{\'a}lez-L{\'o}pez}, {Greve},
  {Harikane}, {Hatsukade}, {Knudsen}, {Koekemoer}, {Kohno}, {Kokorev}, {Lee},
  {Magdis}, {Nelson}, {Rizzo}, {Sanders}, {Schaerer}, {Shapley}, {Strait},
  {Toft}, {Valentino}, {van der Wel}, {Vijayan}, {Watson}, {Bauer},
  {Christiansen}, \& {Wilson}}]{heintz23}
{Heintz}, K.~E., {Gim{\'e}nez-Arteaga}, C., {Fujimoto}, S., {et~al.} 2023,
  \apjl, 944, L30, \dodoi{10.3847/2041-8213/acb2cf}

\bibitem[{{Hernandez} {et~al.}(2025){Hernandez}, {Smith}, {Jones}, {Togi},
  {Melendez}, {Abril-Melgarejo}, {Adamo}, {Alonso Herrero}, {Diaz-Santos},
  {Fischer}, {Garcia-Burillo}, {Hirschauer}, {Hunt}, {James}, {Lebouteiller},
  {Long}, {Mingozzi}, {Ramambason}, \& {Ramos Almeida}}]{hernandez25}
{Hernandez}, S., {Smith}, L.~J., {Jones}, L.~H., {et~al.} 2025, arXiv e-prints,
  arXiv:2502.17621, \dodoi{10.48550/arXiv.2502.17621}

\bibitem[{{Hirschauer} {et~al.}(2024){Hirschauer}, {Crouzet}, {Habel},
  {Lenki{\'c}}, {Nally}, {Jones}, {Bortolini}, {Boyer}, {Justtanont},
  {Meixner}, {{\"O}stlin}, {Wright}, {Azzollini}, {Blommaert}, {Brandl},
  {Decin}, {Nayak}, {Royer}, {Sargent}, \& {van der Werf}}]{hirschauer24}
{Hirschauer}, A.~S., {Crouzet}, N., {Habel}, N., {et~al.} 2024, \aj, 168, 23,
  \dodoi{10.3847/1538-3881/ad4967}

\bibitem[{{Hood} {et~al.}(2017){Hood}, {Barth}, {Ho}, \& {Greene}}]{hood17}
{Hood}, C.~E., {Barth}, A.~J., {Ho}, L.~C., \& {Greene}, J.~E. 2017, \apj, 838,
  26, \dodoi{10.3847/1538-4357/aa60c9}

\bibitem[{{Hunt} {et~al.}(2005){Hunt}, {Dyer}, \& {Thuan}}]{hunt05}
{Hunt}, L.~K., {Dyer}, K.~K., \& {Thuan}, T.~X. 2005, \aap, 436, 837,
  \dodoi{10.1051/0004-6361:20052915}

\bibitem[{{Hunt} {et~al.}(2010){Hunt}, {Thuan}, {Izotov}, \&
  {Sauvage}}]{hunt10}
{Hunt}, L.~K., {Thuan}, T.~X., {Izotov}, Y.~I., \& {Sauvage}, M. 2010, \apj,
  712, 164, \dodoi{10.1088/0004-637X/712/1/164}

\bibitem[{{Hunt} {et~al.}(2001){Hunt}, {Vanzi}, \& {Thuan}}]{hunt01}
{Hunt}, L.~K., {Vanzi}, L., \& {Thuan}, T.~X. 2001, \aap, 377, 66,
  \dodoi{10.1051/0004-6361:20011088}

\bibitem[{{Hunt} {et~al.}(2014){Hunt}, {Testi}, {Casasola},
  {Garc{\'\i}a-Burillo}, {Combes}, {Nikutta}, {Caselli}, {Henkel}, {Maiolino},
  {Menten}, {Sauvage}, \& {Weiss}}]{hunt14}
{Hunt}, L.~K., {Testi}, L., {Casasola}, V., {et~al.} 2014, \aap, 561, A49,
  \dodoi{10.1051/0004-6361/201322739}

\bibitem[{{Hunter} \& {Thronson}(1995)}]{hunter95}
{Hunter}, D.~A., \& {Thronson}, Jr., H.~A. 1995, \apj, 452, 238,
  \dodoi{10.1086/176295}

\bibitem[{{Inami} {et~al.}(2013){Inami}, {Armus}, {Charmandaris}, {Groves},
  {Kewley}, {Petric}, {Stierwalt}, {D{\'\i}az-Santos}, {Surace}, {Rich},
  {Haan}, {Howell}, {Evans}, {Mazzarella}, {Marshall}, {Appleton}, {Lord},
  {Spoon}, {Frayer}, {Matsuhara}, \& {Veilleux}}]{inami13}
{Inami}, H., {Armus}, L., {Charmandaris}, V., {et~al.} 2013, \apj, 777, 156,
  \dodoi{10.1088/0004-637X/777/2/156}

\bibitem[{{Izotov} {et~al.}(1999){Izotov}, {Chaffee}, {Foltz}, {Green},
  {Guseva}, \& {Thuan}}]{izotov99}
{Izotov}, Y.~I., {Chaffee}, F.~H., {Foltz}, C.~B., {et~al.} 1999, \apj, 527,
  757, \dodoi{10.1086/308119}

\bibitem[{{Izotov} {et~al.}(1997){Izotov}, {Foltz}, {Green}, {Guseva}, \&
  {Thuan}}]{izotov97}
{Izotov}, Y.~I., {Foltz}, C.~B., {Green}, R.~F., {Guseva}, N.~G., \& {Thuan},
  T.~X. 1997, \apjl, 487, L37, \dodoi{10.1086/310872}

\bibitem[{{Izotov} \& {Thuan}(2016)}]{izotov16}
{Izotov}, Y.~I., \& {Thuan}, T.~X. 2016, \mnras, 457, 64,
  \dodoi{10.1093/mnras/stv2957}

\bibitem[{{Izotov} {et~al.}(2021){Izotov}, {Thuan}, \& {Guseva}}]{izotov21}
{Izotov}, Y.~I., {Thuan}, T.~X., \& {Guseva}, N.~G. 2021, \mnras, 508, 2556,
  \dodoi{10.1093/mnras/stab2798}

\bibitem[{{Izotov} {et~al.}(2012){Izotov}, {Thuan}, \& {Privon}}]{izotov12}
{Izotov}, Y.~I., {Thuan}, T.~X., \& {Privon}, G. 2012, \mnras, 427, 1229,
  \dodoi{10.1111/j.1365-2966.2012.22051.x}

\bibitem[{{Janowiecki} {et~al.}(2017){Janowiecki}, {Salzer}, {van Zee},
  {Rosenberg}, \& {Skillman}}]{jano17}
{Janowiecki}, S., {Salzer}, J.~J., {van Zee}, L., {Rosenberg}, J.~L., \&
  {Skillman}, E. 2017, \apj, 836, 128, \dodoi{10.3847/1538-4357/836/1/128}

\bibitem[{{Kaaret} \& {Feng}(2013)}]{kaaret13}
{Kaaret}, P., \& {Feng}, H. 2013, \apj, 770, 20,
  \dodoi{10.1088/0004-637X/770/1/20}

\bibitem[{{Kaaret} {et~al.}(2017){Kaaret}, {Feng}, \& {Roberts}}]{kaaret17}
{Kaaret}, P., {Feng}, H., \& {Roberts}, T.~P. 2017, \araa, 55, 303,
  \dodoi{10.1146/annurev-astro-091916-055259}

\bibitem[{{Kehrig} {et~al.}(2021){Kehrig}, {Guerrero}, {V{\'\i}lchez}, \&
  {Ramos-Larios}}]{kehrig21}
{Kehrig}, C., {Guerrero}, M.~A., {V{\'\i}lchez}, J.~M., \& {Ramos-Larios}, G.
  2021, \apjl, 908, L54, \dodoi{10.3847/2041-8213/abe41b}

\bibitem[{{Kehrig} {et~al.}(2015){Kehrig}, {V{\'\i}lchez}, {P{\'e}rez-Montero},
  {Iglesias-P{\'a}ramo}, {Brinchmann}, {Kunth}, {Durret}, \& {Bayo}}]{kehrig15}
{Kehrig}, C., {V{\'\i}lchez}, J.~M., {P{\'e}rez-Montero}, E., {et~al.} 2015,
  \apjl, 801, L28, \dodoi{10.1088/2041-8205/801/2/L28}

\bibitem[{{Kehrig} {et~al.}(2016){Kehrig}, {V{\'\i}lchez}, {P{\'e}rez-Montero},
  {Iglesias-P{\'a}ramo}, {Hern{\'a}ndez-Fern{\'a}ndez}, {Duarte Puertas},
  {Brinchmann}, {Durret}, \& {Kunth}}]{kehrig16}
---. 2016, \mnras, 459, 2992, \dodoi{10.1093/mnras/stw806}

\bibitem[{{Klessen} \& {Glover}(2023)}]{klessen23}
{Klessen}, R.~S., \& {Glover}, S. C.~O. 2023, \araa, 61, 65,
  \dodoi{10.1146/annurev-astro-071221-053453}

\bibitem[{{Krumholz} {et~al.}(2019){Krumholz}, {McKee}, \&
  {Bland-Hawthorn}}]{krumholz19}
{Krumholz}, M.~R., {McKee}, C.~F., \& {Bland-Hawthorn}, J. 2019, \araa, 57,
  227, \dodoi{10.1146/annurev-astro-091918-104430}

\bibitem[{{Law} {et~al.}(2023){Law}, {E. Morrison}, {Argyriou}, {Patapis},
  {{\'A}lvarez-M{\'a}rquez}, {Labiano}, \& {Vandenbussche}}]{law23}
{Law}, D.~R., {E. Morrison}, J., {Argyriou}, I., {et~al.} 2023, \aj, 166, 45,
  \dodoi{10.3847/1538-3881/acdddc}

\bibitem[{{Lebouteiller} {et~al.}(2013){Lebouteiller}, {Heap}, {Hubeny}, \&
  {Kunth}}]{lebout13}
{Lebouteiller}, V., {Heap}, S., {Hubeny}, I., \& {Kunth}, D. 2013, \aap, 553,
  A16, \dodoi{10.1051/0004-6361/201220948}

\bibitem[{{Lebouteiller} {et~al.}(2017){Lebouteiller}, {P{\'e}quignot},
  {Cormier}, {Madden}, {Pakull}, {Kunth}, {Galliano}, {Chevance}, {Heap},
  {Lee}, \& {Polles}}]{lebout17}
{Lebouteiller}, V., {P{\'e}quignot}, D., {Cormier}, D., {et~al.} 2017, \aap,
  602, A45, \dodoi{10.1051/0004-6361/201629675}

\bibitem[{{Legrand} {et~al.}(1997){Legrand}, {Kunth}, {Roy}, {Mas-Hesse}, \&
  {Walsh}}]{legrand97}
{Legrand}, F., {Kunth}, D., {Roy}, J.~R., {Mas-Hesse}, J.~M., \& {Walsh}, J.~R.
  1997, \aap, 326, L17, \dodoi{10.48550/arXiv.astro-ph/9707279}

\bibitem[{{Leitherer} {et~al.}(2014){Leitherer}, {Ekstr{\"o}m}, {Meynet},
  {Schaerer}, {Agienko}, \& {Levesque}}]{leitherer14}
{Leitherer}, C., {Ekstr{\"o}m}, S., {Meynet}, G., {et~al.} 2014, \apjs, 212,
  14, \dodoi{10.1088/0067-0049/212/1/14}

\bibitem[{{Lelli} {et~al.}(2012){Lelli}, {Verheijen}, {Fraternali}, \&
  {Sancisi}}]{lelli12}
{Lelli}, F., {Verheijen}, M., {Fraternali}, F., \& {Sancisi}, R. 2012, \aap,
  537, A72, \dodoi{10.1051/0004-6361/201117867}

\bibitem[{{Leroy} {et~al.}(2018){Leroy}, {Bolatto}, {Ostriker}, {Walter},
  {Gorski}, {Ginsburg}, {Krieger}, {Levy}, {Meier}, {Mills}, {Ott},
  {Rosolowsky}, {Thompson}, {Veilleux}, \& {Zschaechner}}]{leroy18}
{Leroy}, A.~K., {Bolatto}, A.~D., {Ostriker}, E.~C., {et~al.} 2018, \apj, 869,
  126, \dodoi{10.3847/1538-4357/aaecd1}

\bibitem[{{Leslie} {et~al.}(2020){Leslie}, {Schinnerer}, {Liu}, {Magnelli},
  {Algera}, {Karim}, {Davidzon}, {Gozaliasl}, {Jim{\'e}nez-Andrade}, {Lang},
  {Sargent}, {Novak}, {Groves}, {Smol{\v{c}}i{\'c}}, {Zamorani}, {Vaccari},
  {Battisti}, {Vardoulaki}, {Peng}, \& {Kartaltepe}}]{leslie20}
{Leslie}, S.~K., {Schinnerer}, E., {Liu}, D., {et~al.} 2020, \apj, 899, 58,
  \dodoi{10.3847/1538-4357/aba044}

\bibitem[{{Levy} {et~al.}(2024){Levy}, {Bolatto}, {Mayya}, {Cuevas-Otahola},
  {Tarantino}, {Boyer}, {Boogaard}, {B{\"o}ker}, {Cronin}, {Dale}, {Donaghue},
  {Emig}, {Fisher}, {Glover}, {Herrera-Camus}, {Jim{\'e}nez-Donaire},
  {Klessen}, {Lenki{\'c}}, {Leroy}, {De Looze}, {Meier}, {Mills}, {Ott},
  {Rela{\~n}o}, {Veilleux}, {Villanueva}, {Walter}, \& {van der Werf}}]{levy24}
{Levy}, R.~C., {Bolatto}, A.~D., {Mayya}, D., {et~al.} 2024, \apjl, 973, L55,
  \dodoi{10.3847/2041-8213/ad7af3}

\bibitem[{{Linden} {et~al.}(2024){Linden}, {Lai}, {Evans}, {Armus}, {Larson},
  {Rich}, {U}, {Privon}, {Inami}, {Song}, {Bianchin}, {Bohn}, {Buiten},
  {Sanchez-Garc{\'\i}a}, {Kader}, {Lenki{\'c}}, {Medling}, {B{\"o}ker},
  {D{\'\i}az-Santos}, {Charmandaris}, {Barcos-Mu{\~n}oz}, {van der Werf},
  {Stierwalt}, {Aalto}, {Appleton}, {Hayward}, {Howell}, {Malkan},
  {Mazzarella}, {Murphy}, \& {Surace}}]{linden24}
{Linden}, S.~T., {Lai}, T., {Evans}, A.~S., {et~al.} 2024, \apjl, 974, L27,
  \dodoi{10.3847/2041-8213/ad7eae}

\bibitem[{{Looser} {et~al.}(2025){Looser}, {D'Eugenio}, {Maiolino},
  {Tacchella}, {Curti}, {Arribas}, {Baker}, {Baum}, {Bonaventura}, {Boyett},
  {Bunker}, {Carniani}, {Charlot}, {Chevallard}, {Curtis-Lake}, {Lola
  Danhaive}, {Eisenstein}, {de Graaff}, {Hainline}, {Ji}, {Johnson}, {Kumari},
  {Nelson}, {Parlanti}, {Rix}, {Robertson}, {Del Pino}, {Sandles}, {Scholtz},
  {Smit}, {Stark}, {{\"U}bler}, {Williams}, {Willott}, \& {Witstok}}]{looser25}
{Looser}, T.~J., {D'Eugenio}, F., {Maiolino}, R., {et~al.} 2025, \aap, 697,
  A88, \dodoi{10.1051/0004-6361/202347102}

\bibitem[{{Luridiana} {et~al.}(2015){Luridiana}, {Morisset}, \&
  {Shaw}}]{luridiana15}
{Luridiana}, V., {Morisset}, C., \& {Shaw}, R.~A. 2015, \aap, 573, A42,
  \dodoi{10.1051/0004-6361/201323152}

\bibitem[{{Madden} {et~al.}(2006){Madden}, {Galliano}, {Jones}, \&
  {Sauvage}}]{madden06}
{Madden}, S.~C., {Galliano}, F., {Jones}, A.~P., \& {Sauvage}, M. 2006, \aap,
  446, 877, \dodoi{10.1051/0004-6361:20053890}

\bibitem[{{Madden} {et~al.}(2014){Madden}, {R{\'e}my-Ruyer}, {Galametz},
  {Cormier}, {Lebouteiller}, {Galliano}, {Hony}, {Bendo}, {Smith}, {Pohlen},
  {Roussel}, {Sauvage}, {Wu}, {Sturm}, {Poglitsch}, {Contursi}, {Doublier},
  {Baes}, {Barlow}, {Boselli}, {Boquien}, {Carlson}, {Ciesla}, {Cooray},
  {Cortese}, {De Looze}, {Irwin}, {Isaak}, {Kamenetzky}, {Karczewski}, {Lu},
  {MacHattie}, {O'Halloran}, {Parkin}, {Rangwala}, {Schirm}, {Schulz},
  {Spinoglio}, {Vaccari}, {Wilson}, \& {Wozniak}}]{madden14}
{Madden}, S.~C., {R{\'e}my-Ruyer}, A., {Galametz}, M., {et~al.} 2014, {An
  Overview of the Dwarf Galaxy Survey (PASP, 125, 600,
  [2013]){\textemdash}Corrigendum},  IOP, \dodoi{10.1086/679312}

\bibitem[{{Martin}(1996)}]{martin96}
{Martin}, C.~L. 1996, \apj, 465, 680, \dodoi{10.1086/177453}

\bibitem[{{McQuaid} {et~al.}(2024){McQuaid}, {Calzetti}, {Linden}, {Messa},
  {Adamo}, {Elmegreen}, {Grasha}, {Johnson}, {Smith}, \& {Bajaj}}]{mcquaid24}
{McQuaid}, T., {Calzetti}, D., {Linden}, S.~T., {et~al.} 2024, \apj, 967, 102,
  \dodoi{10.3847/1538-4357/ad3e64}

\bibitem[{{Mingozzi} {et~al.}(2025){Mingozzi}, {Garcia del Valle-Espinosa},
  {James}, {Rickards Vaught}, {Hayes}, {Amor{\'\i}n}, {Leitherer}, {Aloisi},
  {Hunt}, {Law}, {Richardson}, {Pidgeon}, {Arellano-C{\'o}rdova}, {Berg},
  {Chisholm}, {Hernandez}, {Jones}, {Kumari}, {Martin}, {Ravindranath},
  {Vallini}, \& {Xu}}]{mingozzi25}
{Mingozzi}, M., {Garcia del Valle-Espinosa}, M., {James}, B.~L., {et~al.} 2025,
  \apj, 985, 253, \dodoi{10.3847/1538-4357/adc996}

\bibitem[{{Moon} {et~al.}(2011){Moon}, {Harrison}, {Cenko}, \&
  {Shariff}}]{moon11}
{Moon}, D.-S., {Harrison}, F.~A., {Cenko}, S.~B., \& {Shariff}, J.~A. 2011,
  \apjl, 731, L32, \dodoi{10.1088/2041-8205/731/2/L32}

\bibitem[{{Morishita} {et~al.}(2024){Morishita}, {Stiavelli}, {Grillo},
  {Rosati}, {Schuldt}, {Trenti}, {Bergamini}, {Boyett}, {Chary},
  {Leethochawalit}, {Roberts-Borsani}, {Treu}, \& {Vanzella}}]{morishita24}
{Morishita}, T., {Stiavelli}, M., {Grillo}, C., {et~al.} 2024, \apj, 971, 43,
  \dodoi{10.3847/1538-4357/ad5290}

\bibitem[{{Nakajima} {et~al.}(2023){Nakajima}, {Ouchi}, {Isobe}, {Harikane},
  {Zhang}, {Ono}, {Umeda}, \& {Oguri}}]{nakajima23}
{Nakajima}, K., {Ouchi}, M., {Isobe}, Y., {et~al.} 2023, \apjs, 269, 33,
  \dodoi{10.3847/1538-4365/acd556}

\bibitem[{{Nanni} {et~al.}(2020){Nanni}, {Burgarella}, {Theul{\'e}},
  {C{\^o}t{\'e}}, \& {Hirashita}}]{nanni20}
{Nanni}, A., {Burgarella}, D., {Theul{\'e}}, P., {C{\^o}t{\'e}}, B., \&
  {Hirashita}, H. 2020, \aap, 641, A168, \dodoi{10.1051/0004-6361/202037833}

\bibitem[{{Naz{\'e}} {et~al.}(2003){Naz{\'e}}, {Rauw}, {Manfroid}, {Chu}, \&
  {Vreux}}]{naze03}
{Naz{\'e}}, Y., {Rauw}, G., {Manfroid}, J., {Chu}, Y.~H., \& {Vreux}, J.~M.
  2003, \aap, 408, 171, \dodoi{10.1051/0004-6361:20030847}

\bibitem[{{Noeske} {et~al.}(2007){Noeske}, {Weiner}, {Faber}, {Papovich},
  {Koo}, {Somerville}, {Bundy}, {Conselice}, {Newman}, {Schiminovich}, {Le
  Floc'h}, {Coil}, {Rieke}, {Lotz}, {Primack}, {Barmby}, {Cooper}, {Davis},
  {Ellis}, {Fazio}, {Guhathakurta}, {Huang}, {Kassin}, {Martin}, {Phillips},
  {Rich}, {Small}, {Willmer}, \& {Wilson}}]{noeske07}
{Noeske}, K.~G., {Weiner}, B.~J., {Faber}, S.~M., {et~al.} 2007, \apjl, 660,
  L43, \dodoi{10.1086/517926}

\bibitem[{{Oskinova} \& {Schaerer}(2022)}]{oskinova22}
{Oskinova}, L.~M., \& {Schaerer}, D. 2022, \aap, 661, A67,
  \dodoi{10.1051/0004-6361/202142520}

\bibitem[{{Ott} {et~al.}(2005){Ott}, {Walter}, \& {Brinks}}]{ott05}
{Ott}, J., {Walter}, F., \& {Brinks}, E. 2005, \mnras, 358, 1423,
  \dodoi{10.1111/j.1365-2966.2005.08862.x}

\bibitem[{{Patapis} {et~al.}(2024){Patapis}, {Argyriou}, {Law}, {Glauser},
  {Glasse}, {Labiano}, {{\'A}lvarez-M{\'a}rquez}, {Kavanagh}, {Gasman},
  {Mueller}, {Larson}, {Vandenbussche}, {Lee}, {Klaassen}, {Guillard}, \&
  {Wright}}]{patapis24}
{Patapis}, P., {Argyriou}, I., {Law}, D.~R., {et~al.} 2024, \aap, 682, A53,
  \dodoi{10.1051/0004-6361/202347339}

\bibitem[{{Planck Collaboration} {et~al.}(2016){Planck Collaboration},
  {Aghanim}, {Ashdown}, {Aumont}, {Baccigalupi}, {Ballardini}, {Banday},
  {Barreiro}, {Bartolo}, {Basak}, {Benabed}, {Bernard}, {Bersanelli},
  {Bielewicz}, {Bonavera}, {Bond}, {Borrill}, {Bouchet}, {Boulanger},
  {Burigana}, {Calabrese}, {Cardoso}, {Carron}, {Chiang}, {Colombo}, {Comis},
  {Couchot}, {Coulais}, {Crill}, {Curto}, {Cuttaia}, {de Bernardis}, {de
  Zotti}, {Delabrouille}, {Di Valentino}, {Dickinson}, {Diego}, {Dor{\'e}},
  {Douspis}, {Ducout}, {Dupac}, {Dusini}, {Elsner}, {En{\ss}lin}, {Eriksen},
  {Falgarone}, {Fantaye}, {Finelli}, {Forastieri}, {Frailis}, {Fraisse},
  {Franceschi}, {Frolov}, {Galeotta}, {Galli}, {Ganga}, {G{\'e}nova-Santos},
  {Gerbino}, {Ghosh}, {Giraud-H{\'e}raud}, {Gonz{\'a}lez-Nuevo}, {G{\'o}rski},
  {Gruppuso}, {Gudmundsson}, {Hansen}, {Helou}, {Henrot-Versill{\'e}},
  {Herranz}, {Hivon}, {Huang}, {Jaffe}, {Jones}, {Keih{\"a}nen}, {Keskitalo},
  {Kiiveri}, {Kisner}, {Krachmalnicoff}, {Kunz}, {Kurki-Suonio}, {Lamarre},
  {Langer}, {Lasenby}, {Lattanzi}, {Lawrence}, {Le Jeune}, {Levrier}, {Lilje},
  {Lilley}, {Lindholm}, {L{\'o}pez-Caniego}, {Ma}, {Mac{\'\i}as-P{\'e}rez},
  {Maggio}, {Maino}, {Mandolesi}, {Mangilli}, {Maris}, {Martin},
  {Mart{\'\i}nez-Gonz{\'a}lez}, {Matarrese}, {Mauri}, {McEwen}, {Melchiorri},
  {Mennella}, {Migliaccio}, {Miville-Desch{\^e}nes}, {Molinari}, {Moneti},
  {Montier}, {Morgante}, {Moss}, {Natoli}, {Oxborrow}, {Pagano}, {Paoletti},
  {Patanchon}, {Perdereau}, {Perotto}, {Pettorino}, {Piacentini},
  {Plaszczynski}, {Polastri}, {Polenta}, {Puget}, {Rachen}, {Racine},
  {Reinecke}, {Remazeilles}, {Renzi}, {Rocha}, {Rosset}, {Rossetti}, {Roudier},
  {Rubi{\~n}o-Mart{\'\i}n}, {Ruiz-Granados}, {Salvati}, {Sandri}, {Savelainen},
  {Scott}, {Sirignano}, {Sirri}, {Soler}, {Spencer}, {Suur-Uski}, {Tauber},
  {Tavagnacco}, {Tenti}, {Toffolatti}, {Tomasi}, {Tristram}, {Trombetti},
  {Valiviita}, {Van Tent}, {Vielva}, {Villa}, {Vittorio}, {Wandelt}, {Wehus},
  {Zacchei}, \& {Zonca}}]{planck16}
{Planck Collaboration}, {Aghanim}, N., {Ashdown}, M., {et~al.} 2016, \aap, 596,
  A109, \dodoi{10.1051/0004-6361/201629022}

\bibitem[{{R{\'e}my-Ruyer} {et~al.}(2014){R{\'e}my-Ruyer}, {Madden},
  {Galliano}, {Galametz}, {Takeuchi}, {Asano}, {Zhukovska}, {Lebouteiller},
  {Cormier}, {Jones}, {Bocchio}, {Baes}, {Bendo}, {Boquien}, {Boselli},
  {DeLooze}, {Doublier-Pritchard}, {Hughes}, {Karczewski}, \&
  {Spinoglio}}]{remy14}
{R{\'e}my-Ruyer}, A., {Madden}, S.~C., {Galliano}, F., {et~al.} 2014, \aap,
  563, A31, \dodoi{10.1051/0004-6361/201322803}

\bibitem[{{Rhoads} {et~al.}(2023){Rhoads}, {Wold}, {Harish}, {Kim}, {Pharo},
  {Malhotra}, {Gabrielpillai}, {Jiang}, \& {Yang}}]{rhoads23}
{Rhoads}, J.~E., {Wold}, I. G.~B., {Harish}, S., {et~al.} 2023, \apjl, 942,
  L14, \dodoi{10.3847/2041-8213/acaaaf}

\bibitem[{{Richardson} {et~al.}(2022){Richardson}, {Simpson}, {Polimera},
  {Kannappan}, {Bellovary}, {Greene}, \& {Jenkins}}]{richardson22}
{Richardson}, C.~T., {Simpson}, C., {Polimera}, M.~S., {et~al.} 2022, \apj,
  927, 165, \dodoi{10.3847/1538-4357/ac510c}

\bibitem[{{Richardson} {et~al.}(2025){Richardson}, {Wels}, {Garofali},
  {Levanti}, {Lebouteiller}, {Lehmer}, {Basu-Zych}, {Berg}, {Bellovary},
  {Chisholm}, {Kannappan}, {Lambrides}, {Polimera}, {Ramambason}, {Varese}, \&
  {Vivona}}]{richardson25}
{Richardson}, C.~T., {Wels}, J., {Garofali}, K., {et~al.} 2025, arXiv e-prints,
  arXiv:2505.07749, \dodoi{10.48550/arXiv.2505.07749}

\bibitem[{{Rickards Vaught} {et~al.}(2021){Rickards Vaught}, {Sandstrom}, \&
  {Hunt}}]{rickards21}
{Rickards Vaught}, R.~J., {Sandstrom}, K.~M., \& {Hunt}, L.~K. 2021, \apjl,
  911, L17, \dodoi{10.3847/2041-8213/abf09b}

\bibitem[{{Rickards Vaught} {et~al.}(2025, ApJ, in press){Rickards Vaught},
  {Hunt}, {Aloisi}, {Navarro-Ovando}, {Mingozzi}, {James}, {del
  Valle-Espinosa}, {Sandstrom}, {Adamo}, {Annibali}, {Calzetti}, {Draine},
  {Hernandez}, {Hirschauer}, {Meixner}, {Rigopoulou}, \& {Tosi}}]{rickards25}
{Rickards Vaught}, R.~J., {Hunt}, L.~K., {Aloisi}, A., {et~al.} 2025, ApJ, in
  press, arXiv e-prints, arXiv:2507.12222, \dodoi{10.48550/arXiv.2507.12222}

\bibitem[{{Rigby} {et~al.}(2023){Rigby}, {Perrin}, {McElwain}, {Kimble},
  {Friedman}, {Lallo}, {Doyon}, {Feinberg}, {Ferruit}, {Glasse}, {Rieke},
  {Rieke}, {Wright}, {Willott}, {Colon}, {Milam}, {Neff}, {Stark}, {Valenti},
  {Abell}, {Abney}, {Abul-Huda}, {Acton}, {Adams}, {Adler}, {Aguilar}, {Ahmed},
  {Albert}, {Alberts}, {Aldridge}, {Allen}, {Altenburg},
  {{\'A}lvarez-M{\'a}rquez}, {Alves de Oliveira}, {Andersen}, {Anderson},
  {Anderson}, {Argyriou}, {Armstrong}, {Arribas}, {Artigau}, {Arvai},
  {Atkinson}, {Bacon}, {Bair}, {Banks}, {Barrientes}, {Barringer}, {Bartosik},
  {Bast}, {Baudoz}, {Beatty}, {Bechtold}, {Beck}, {Bergeron}, {Bergkoetter},
  {Bhatawdekar}, {Birkmann}, {Blazek}, {Blome}, {Boccaletti}, {B{\"o}ker},
  {Boia}, {Bonaventura}, {Bond}, {Bosley}, {Boucarut}, {Bourque}, {Bouwman},
  {Bower}, {Bowers}, {Boyer}, {Bradley}, {Brady}, {Braun}, {Breda},
  {Bresnahan}, {Bright}, {Britt}, {Bromenschenkel}, {Brooks}, {Brooks},
  {Brown}, {Brown}, {Brown}, {Bunker}, {Burger}, {Bushouse}, {Cale}, {Cameron},
  {Cameron}, {Canipe}, {Caplinger}, {Caputo}, {Cara}, {Carey}, {Carniani},
  {Carrasquilla}, {Carruthers}, {Case}, {Catherine}, {Chance}, {Chapman},
  {Charlot}, {Charlow}, {Chayer}, {Chen}, {Cherinka}, {Chichester}, {Chilton},
  {Chonis}, {Clampin}, {Clark}, {Clark}, {Coe}, {Coleman}, {Comber}, {Comeau},
  {Connolly}, {Cooper}, {Cooper}, {Coppock}, {Correnti}, {Cossou}, {Coulais},
  {Coyle}, {Cracraft}, {Curti}, {Cuturic}, {Davis}, {Davis}, {Dean}, {DeLisa},
  {deMeester}, {Dencheva}, {Dencheva}, {DePasquale}, {Deschenes}, {Hunor
  Detre}, {Diaz}, {Dicken}, {DiFelice}, {Dillman}, {Dixon}, {Doggett},
  {Donaldson}, {Douglas}, {DuPrie}, {Dupuis}, {Durning}, {Easmin}, {Eck},
  {Edeani}, {Egami}, {Ehrenwinkler}, {Eisenhamer}, {Eisenhower}, {Elie},
  {Elliott}, {Elliott}, {Ellis}, {Engesser}, {Espinoza}, {Etienne}, {Etxaluze},
  {Falini}, {Feeney}, {Ferry}, {Filippazzo}, {Fincham}, {Fix}, {Flagey},
  {Florian}, {Flynn}, {Fontanella}, {Ford}, {Forshay}, {Fox}, {Franz}, {Fu},
  {Fullerton}, {Galkin}, {Galyer}, {Garc{\'\i}a Mar{\'\i}n}, {Gardner},
  {Gardner}, {Garland}, {Garrett}, {Gasman}, {Gaspar}, {Gaudreau}, {Gauthier},
  {Geers}, {Geithner}, {Gennaro}, {Giardino}, {Girard}, {Giuliano},
  {Glassmire}, \& {Glauser}}]{rigby23}
{Rigby}, J., {Perrin}, M., {McElwain}, M., {et~al.} 2023, \pasp, 135, 048001,
  \dodoi{10.1088/1538-3873/acb293}

\bibitem[{{Rodr{\'\i}guez} {et~al.}(2025){Rodr{\'\i}guez}, {Lee}, {Indebetouw},
  {Whitmore}, {Maschmann}, {Williams}, {Chandar}, {Barnes}, {Gnedin},
  {Sandstrom}, {Rosolowsky}, {Leroy}, {Thilker}, {Kim}, {Sun}, {Klessen},
  {Groves}, {Wofford}, {Boquien}, {Dale}, {{\'U}beda}, {Larson}, {Grasha},
  {Johnson}, {Levy}, {Bigiel}, {Hassani}, \& {Sarbadhicary}}]{rodriguez25}
{Rodr{\'\i}guez}, M.~J., {Lee}, J.~C., {Indebetouw}, R., {et~al.} 2025, \apj,
  983, 137, \dodoi{10.3847/1538-4357/adbb69}

\bibitem[{{Satyapal} {et~al.}(2008){Satyapal}, {Vega}, {Dudik}, {Abel}, \&
  {Heckman}}]{satyapal08}
{Satyapal}, S., {Vega}, D., {Dudik}, R.~P., {Abel}, N.~P., \& {Heckman}, T.
  2008, \apj, 677, 926, \dodoi{10.1086/529014}

\bibitem[{{Schaerer} {et~al.}(1999){Schaerer}, {Contini}, \&
  {Pindao}}]{schaerer99}
{Schaerer}, D., {Contini}, T., \& {Pindao}, M. 1999, \aaps, 136, 35,
  \dodoi{10.1051/aas:1999197}

\bibitem[{{Schaerer} {et~al.}(2019){Schaerer}, {Fragos}, \&
  {Izotov}}]{schaerer19}
{Schaerer}, D., {Fragos}, T., \& {Izotov}, Y.~I. 2019, \aap, 622, L10,
  \dodoi{10.1051/0004-6361/201935005}

\bibitem[{{Schaerer} \& {Stasi{\'n}ska}(1999)}]{schaererstasinska99}
{Schaerer}, D., \& {Stasi{\'n}ska}, G. 1999, \aap, 345, L17,
  \dodoi{10.48550/arXiv.astro-ph/9903430}

\bibitem[{{Schaerer} \& {Vacca}(1998)}]{schaerer98}
{Schaerer}, D., \& {Vacca}, W.~D. 1998, \apj, 497, 618, \dodoi{10.1086/305487}

\bibitem[{{Schlafly} \& {Finkbeiner}(2011)}]{schlafly11}
{Schlafly}, E.~F., \& {Finkbeiner}, D.~P. 2011, \apj, 737, 103,
  \dodoi{10.1088/0004-637X/737/2/103}

\bibitem[{{Schlegel} {et~al.}(1998){Schlegel}, {Finkbeiner}, \&
  {Davis}}]{schlegel98}
{Schlegel}, D.~J., {Finkbeiner}, D.~P., \& {Davis}, M. 1998, \apj, 500, 525,
  \dodoi{10.1086/305772}

\bibitem[{{Sharma} {et~al.}(2011){Sharma}, {Corbelli}, {Giovanardi}, {Hunt}, \&
  {Palla}}]{sharma11}
{Sharma}, S., {Corbelli}, E., {Giovanardi}, C., {Hunt}, L.~K., \& {Palla}, F.
  2011, \aap, 534, A96, \dodoi{10.1051/0004-6361/201117812}

\bibitem[{{Shenar} {et~al.}(2016){Shenar}, {Hainich}, {Todt}, {Sander},
  {Hamann}, {Moffat}, {Eldridge}, {Pablo}, {Oskinova}, \&
  {Richardson}}]{shenar16}
{Shenar}, T., {Hainich}, R., {Todt}, H., {et~al.} 2016, \aap, 591, A22,
  \dodoi{10.1051/0004-6361/201527916}

\bibitem[{{Smith} {et~al.}(2020){Smith}, {Bajaj}, {Ryon}, \& {Sabbi}}]{smith20}
{Smith}, L.~J., {Bajaj}, V., {Ryon}, J., \& {Sabbi}, E. 2020, \apj, 896, 84,
  \dodoi{10.3847/1538-4357/ab8f94}

\bibitem[{{Soria} {et~al.}(2021){Soria}, {Pakull}, {Motch}, {Miller-Jones},
  {Schwope}, {Urquhart}, \& {Ryan}}]{soria21}
{Soria}, R., {Pakull}, M.~W., {Motch}, C., {et~al.} 2021, \mnras, 501, 1644,
  \dodoi{10.1093/mnras/staa3784}

\bibitem[{{Speagle} {et~al.}(2014){Speagle}, {Steinhardt}, {Capak}, \&
  {Silverman}}]{speagle14}
{Speagle}, J.~S., {Steinhardt}, C.~L., {Capak}, P.~L., \& {Silverman}, J.~D.
  2014, \apjs, 214, 15, \dodoi{10.1088/0067-0049/214/2/15}

\bibitem[{{Stanway} {et~al.}(2016){Stanway}, {Eldridge}, \&
  {Becker}}]{stanway16}
{Stanway}, E.~R., {Eldridge}, J.~J., \& {Becker}, G.~D. 2016, \mnras, 456, 485,
  \dodoi{10.1093/mnras/stv2661}

\bibitem[{{Stefanon} {et~al.}(2021){Stefanon}, {Bouwens}, {Labb{\'e}},
  {Illingworth}, {Gonzalez}, \& {Oesch}}]{stefanon21}
{Stefanon}, M., {Bouwens}, R.~J., {Labb{\'e}}, I., {et~al.} 2021, \apj, 922,
  29, \dodoi{10.3847/1538-4357/ac1bb6}

\bibitem[{{Sturm} {et~al.}(2002){Sturm}, {Lutz}, {Verma}, {Netzer},
  {Sternberg}, {Moorwood}, {Oliva}, \& {Genzel}}]{sturm02}
{Sturm}, E., {Lutz}, D., {Verma}, A., {et~al.} 2002, \aap, 393, 821,
  \dodoi{10.1051/0004-6361:20021043}

\bibitem[{{Sun} {et~al.}(2024){Sun}, {He}, {Batschkun}, {Levy}, {Emig},
  {Rodr{\'\i}guez}, {Hassani}, {Leroy}, {Schinnerer}, {Ostriker}, {Wilson},
  {Bolatto}, {Mills}, {Rosolowsky}, {Lee}, {Dale}, {Larson}, {Thilker},
  {Ubeda}, {Whitmore}, {Williams}, {Barnes}, {Bigiel}, {Chevance}, {Glover},
  {Grasha}, {Groves}, {Henshaw}, {Indebetouw}, {Jim{\'e}nez-Donaire},
  {Klessen}, {Koch}, {Liu}, {Mathur}, {Meidt}, {Menon}, {Neumann}, {Pinna},
  {Querejeta}, {Sormani}, \& {Tress}}]{sun24}
{Sun}, J., {He}, H., {Batschkun}, K., {et~al.} 2024, \apj, 967, 133,
  \dodoi{10.3847/1538-4357/ad3de6}

\bibitem[{{Sutherland} {et~al.}(2018){Sutherland}, {Dopita}, {Binette}, \&
  {Groves}}]{sutherland18}
{Sutherland}, R., {Dopita}, M., {Binette}, L., \& {Groves}, B. 2018, {MAPPINGS
  V: Astrophysical plasma modeling code}, Astrophysics Source Code Library,
  record ascl:1807.005

\bibitem[{{Sutherland} \& {Dopita}(2017)}]{sutherland17}
{Sutherland}, R.~S., \& {Dopita}, M.~A. 2017, \apjs, 229, 34,
  \dodoi{10.3847/1538-4365/aa6541}

\bibitem[{{Tarantino} {et~al.}(2024){Tarantino}, {Bolatto}, {Indebetouw},
  {Rubio}, {Sandstrom}, {Smith}, {Stapleton}, \& {Wolfire}}]{tarantino24}
{Tarantino}, E., {Bolatto}, A.~D., {Indebetouw}, R., {et~al.} 2024, \apj, 969,
  101, \dodoi{10.3847/1538-4357/ad3def}

\bibitem[{{Thornley} {et~al.}(2000){Thornley}, {F{\"o}rster Schreiber}, {Lutz},
  {Genzel}, {Spoon}, {Kunze}, \& {Sternberg}}]{thornley00}
{Thornley}, M.~D., {F{\"o}rster Schreiber}, N.~M., {Lutz}, D., {et~al.} 2000,
  \apj, 539, 641, \dodoi{10.1086/309261}

\bibitem[{{Thuan} {et~al.}(2004){Thuan}, {Bauer}, {Papaderos}, \&
  {Izotov}}]{thuan04}
{Thuan}, T.~X., {Bauer}, F.~E., {Papaderos}, P., \& {Izotov}, Y.~I. 2004, \apj,
  606, 213, \dodoi{10.1086/382949}

\bibitem[{{Thuan} \& {Izotov}(2005)}]{thuan05}
{Thuan}, T.~X., \& {Izotov}, Y.~I. 2005, \apjs, 161, 240,
  \dodoi{10.1086/491657}

\bibitem[{{Trump} {et~al.}(2023){Trump}, {Arrabal Haro}, {Simons}, {Backhaus},
  {Amor{\'\i}n}, {Dickinson}, {Fern{\'a}ndez}, {Papovich}, {Nicholls},
  {Kewley}, {Brunker}, {Salzer}, {Wilkins}, {Almaini}, {Bagley}, {Berg},
  {Bhatawdekar}, {Bisigello}, {Buat}, {Burgarella}, {Calabr{\`o}}, {Casey},
  {Ciesla}, {Cleri}, {Cole}, {Cooper}, {Cooray}, {Costantin}, {Croton},
  {Ferguson}, {Finkelstein}, {Fujimoto}, {Gardner}, {Gawiser}, {Giavalisco},
  {Grazian}, {Grogin}, {Hathi}, {Hirschmann}, {Holwerda}, {Huertas-Company},
  {Hutchison}, {Jogee}, {Juneau}, {Jung}, {Kartaltepe}, {Kirkpatrick},
  {Kocevski}, {Koekemoer}, {Lotz}, {Lucas}, {Magnelli}, {Matharu},
  {P{\'e}rez-Gonz{\'a}lez}, {Pirzkal}, {Rafelski}, {Rose}, {Seill{\'e}},
  {Somerville}, {Straughn}, {Tacchella}, {Vanderhoof}, {Weiner}, {Wuyts},
  {Yung}, \& {Zavala}}]{trump23}
{Trump}, J.~R., {Arrabal Haro}, P., {Simons}, R.~C., {et~al.} 2023, \apj, 945,
  35, \dodoi{10.3847/1538-4357/acba8a}

\bibitem[{{van Dishoeck} \& {Black}(1988)}]{vandishoeck88}
{van Dishoeck}, E.~F., \& {Black}, J.~H. 1988, \apj, 334, 771,
  \dodoi{10.1086/166877}

\bibitem[{{van Dokkum}(2001)}]{vandokkum01}
{van Dokkum}, P.~G. 2001, \pasp, 113, 1420, \dodoi{10.1086/323894}

\bibitem[{{van Hoof}(2018)}]{vanhoof18}
{van Hoof}, P. A.~M. 2018, Galaxies, 6, 63, \dodoi{10.3390/galaxies6020063}

\bibitem[{{van Zee} {et~al.}(1998){van Zee}, {Westpfahl}, {Haynes}, \&
  {Salzer}}]{vanzee98}
{van Zee}, L., {Westpfahl}, D., {Haynes}, M.~P., \& {Salzer}, J.~J. 1998, \aj,
  115, 1000, \dodoi{10.1086/300251}

\bibitem[{{Verma} {et~al.}(2003){Verma}, {Lutz}, {Sturm}, {Sternberg},
  {Genzel}, \& {Vacca}}]{verma03}
{Verma}, A., {Lutz}, D., {Sturm}, E., {et~al.} 2003, \aap, 403, 829,
  \dodoi{10.1051/0004-6361:20030408}

\bibitem[{{Vilchez} \& {Pagel}(1988)}]{vilchez98}
{Vilchez}, J.~M., \& {Pagel}, B.~E.~J. 1988, \mnras, 231, 257,
  \dodoi{10.1093/mnras/231.2.257}

\bibitem[{Virtanen {et~al.}(2020)Virtanen, Gommers, Oliphant, Haberland, Reddy,
  Cournapeau, Burovski, Peterson, Weckesser, Bright, {van der Walt}, Brett,
  Wilson, Millman, Mayorov, Nelson, Jones, Kern, Larson, Carey, Polat, Feng,
  Moore, {VanderPlas}, Laxalde, Perktold, Cimrman, Henriksen, Quintero, Harris,
  Archibald, Ribeiro, Pedregosa, {van Mulbregt}, \& {SciPy 1.0
  Contributors}}]{2020SciPy-NMeth}
Virtanen, P., Gommers, R., Oliphant, T.~E., {et~al.} 2020, Nature Methods, 17,
  261, \dodoi{10.1038/s41592-019-0686-2}

\bibitem[{{Weedman} {et~al.}(2005){Weedman}, {Hao}, {Higdon}, {Devost}, {Wu},
  {Charmandaris}, {Brandl}, {Bass}, \& {Houck}}]{weedman05}
{Weedman}, D.~W., {Hao}, L., {Higdon}, S.~J.~U., {et~al.} 2005, \apj, 633, 706,
  \dodoi{10.1086/466520}

\bibitem[{{Weingartner} \& {Draine}(2001)}]{wd01}
{Weingartner}, J.~C., \& {Draine}, B.~T. 2001, \apj, 548, 296,
  \dodoi{10.1086/318651}

\bibitem[{{Whitaker} {et~al.}(2014){Whitaker}, {Franx}, {Leja}, {van Dokkum},
  {Henry}, {Skelton}, {Fumagalli}, {Momcheva}, {Brammer}, {Labb{\'e}},
  {Nelson}, \& {Rigby}}]{whitaker14}
{Whitaker}, K.~E., {Franx}, M., {Leja}, J., {et~al.} 2014, \apj, 795, 104,
  \dodoi{10.1088/0004-637X/795/2/104}

\bibitem[{{White} \& {Frenk}(1991)}]{white91}
{White}, S. D.~M., \& {Frenk}, C.~S. 1991, \apj, 379, 52,
  \dodoi{10.1086/170483}

\bibitem[{{Whitmore} {et~al.}(2025){Whitmore}, {Chandar}, {Lee}, {Henny},
  {Rodr{\'\i}guez}, {Baron}, {Bigiel}, {Boquien}, {Chevance}, {Chown}, {Dale},
  {Floyd}, {Grasha}, {Glover}, {Gnedin}, {Hassani}, {Indebetouw}, {Kapoor},
  {Larson}, {Leroy}, {Maschmann}, {Scheuermann}, {Sutter}, {Schinnerer},
  {Sarbadhicary}, {Thilker}, {Williams}, \& {Wofford}}]{whitmore25}
{Whitmore}, B.~C., {Chandar}, R., {Lee}, J.~C., {et~al.} 2025, \apj, 982, 50,
  \dodoi{10.3847/1538-4357/adb3a2}

\bibitem[{{Wise} \& {Cen}(2009)}]{wise09}
{Wise}, J.~H., \& {Cen}, R. 2009, \apj, 693, 984,
  \dodoi{10.1088/0004-637X/693/1/984}

\bibitem[{{Wu} {et~al.}(2007){Wu}, {Charmandaris}, {Hunt}, {Bernard-Salas},
  {Brandl}, {Marshall}, {Lebouteiller}, {Hao}, \& {Houck}}]{wu07}
{Wu}, Y., {Charmandaris}, V., {Hunt}, L.~K., {et~al.} 2007, \apj, 662, 952,
  \dodoi{10.1086/517988}

\bibitem[{{Yoshimoto} {et~al.}(2024){Yoshimoto}, {Yoneyama}, {Noda}, {Odaka},
  \& {Matsumoto}}]{yoshimoto24}
{Yoshimoto}, M., {Yoneyama}, T., {Noda}, H., {Odaka}, H., \& {Matsumoto}, H.
  2024, \apj, 970, 8, \dodoi{10.3847/1538-4357/ad4e34}

\bibitem[{{Zhou} {et~al.}(2021){Zhou}, {Shi}, {Zhang}, \& {Wang}}]{zhou21}
{Zhou}, L., {Shi}, Y., {Zhang}, Z.-Y., \& {Wang}, J. 2021, \aap, 653, L10,
  \dodoi{10.1051/0004-6361/202039033}

\end{thebibliography}
\bibliographystyle{aasjournal}



\end{document}